\documentclass[10pt,aps,prc,floatfix,twocolumn,nofootinbib]{revtex4-1}
\usepackage{graphicx,amsmath,amssymb,bm}
\usepackage{amsfonts}
\usepackage[utf8]{inputenc}
\usepackage{verbatim}
\usepackage{float}
\usepackage[labelfont={small},subrefformat=parens,caption=false,labelformat=empty]{subfig}
\usepackage{cancel}
\usepackage{multirow}
\usepackage{array}
\usepackage{xparse}

\usepackage{color}

\usepackage[pdfpagelabels]{hyperref}

\usepackage{cellspace}
\newcommand{\beq}{\begin{equation}}
\newcommand{\eeq}{\end{equation}}

\newcommand{\bseq}{\begin{subequations}}
\newcommand{\eseq}{\end{subequations}}

\newcommand{\ds}{\displaystyle}

\newcommand{\la}{\langle}
\newcommand{\ra}{\rangle}

\newcommand{\NLO}{\ensuremath{{\rm NLO}}}
\newcommand{\NNLO}{\ensuremath{{\rm N}{}^2{\rm LO}}}
\newcommand{\NNNLO}{\ensuremath{{\rm N}{}^3{\rm LO}}}
\newcommand{\NNNNLO}{\ensuremath{{\rm N}{}^4{\rm LO}}}

\newcommand{\npr}{\ensuremath{np}}

\newcommand{\boldvec}[1]{\bm{#1}}

\newcommand{\nvec}{\boldvec{n}}

\newcommand{\svec}{\boldvec{s}}

\newcommand{\ellvec}{\boldvec{\ell}}
\newcommand{\mvec}{\boldvec{m}}
\newcommand{\kvec}{\boldvec{k}}
\newcommand{\kpvec}{\boldvec{k}'}
\newcommand{\kprvec}{\boldvec{k}_{R_1}'}
\newcommand{\spvec}{\boldvec{s}'}
\newcommand{\sprvec}{\boldvec{s}_{R_1}'}
\newcommand{\kppvec}{\boldvec{k}''}
\newcommand{\kpprvec}{\boldvec{k}_{R_2}''}
\newcommand{\sppvec}{\boldvec{s}''}
\newcommand{\spprvec}{\boldvec{s}_{R_2}''}
\newcommand{\ckvec}{\mathbf{c}_k}
\newcommand{\ckvecsq}{\mathbf{c}_k^{2}}
\newcommand{\bkvec}{\mathbf{b}_k}

\newcommand{\kinparvec}{\boldvec{\alpha}}

\newcommand{\bi}{\begin{itemize}}
\newcommand{\ei}{\end{itemize}}
\newcommand{\I}{\item}
\newcommand{\be}{\begin{enumerate}}
\newcommand{\ee}{\end{enumerate}}
\newcommand{\bc}{\begin{center}}
\newcommand{\ec}{\end{center}}

\DeclareMathOperator{\pr}{pr}

\newcommand{\cbar}{\bar{c}}
\newcommand{\cbarsubk}[1]{\cbar_{(#1)}}
\newcommand{\cbark}{\cbarsubk{k}}
\newcommand{\cbarmin}{\cbar_{<}}
\newcommand{\cbarmax}{\cbar_{>}}

\newcommand{\Q}{Q}

\newcommand{\Elab}{E_{\rm lab}}

\newcommand{\prel}{p_{\rm rel}}

\newcommand{\Aeps}{\ensuremath{{\rm A}_{\epsilon}}}
\newcommand{\Ceps}{\ensuremath{{\rm C}_{\epsilon}}}
\newcommand{\Aepsone}{\Aeps^{(1)}}

\newcommand{\Aone}{\ensuremath{{\rm A}^{(1)}}}

\newcommand{\Af}[2]{\ensuremath{{\rm A}_{#1\text{-}#2}}}
\newcommand{\Cf}[2]{\ensuremath{{\rm C}_{#1\text{-}#2}}}

\newcommand{\Cfone}[2]{\Cf{#1}{#2}^{(1)}}

\newcommand{\Xref}{\ensuremath{X_{\text{ref}}}}

\newcommand{\ppercent}{(100*p)\%}

\newcommand{\phase}[3]{\ensuremath{\bar\delta^{#1 #3}_{#2}}}
\newcommand{\blattphase}[3]{\ensuremath{\delta^{#1 #3}_{#2}}}
\newcommand{\epsmix}[1]{\ensuremath{\bar\epsilon_{#1}}}
\newcommand{\blattepsmix}[1]{\ensuremath{\epsilon_{#1}}}
\newcommand{\gammix}[1]{\ensuremath{\bar\gamma_{#1}}}

\newcommand{\Ron}[4]{R^{#1 #4}_{#2 #3}}

\newcommand{\ls}{l.s.}
\newcommand{\cms}{c.m.}

\def\diffd{\mathrm{d}}  

\DeclareDocumentCommand\differential{ o g d() }{ 
    \IfNoValueTF{#2}{
        \IfNoValueTF{#3}
            {\diffd\IfNoValueTF{#1}{}{^{#1}}}
            {\mathinner{\diffd\IfNoValueTF{#1}{}{^{#1}}\argopen(#3\argclose)}}
        }
        {\mathinner{\diffd\IfNoValueTF{#1}{}{^{#1}}#2} \IfNoValueTF{#3}{}{(#3)}}
    }
\DeclareDocumentCommand\dd{}{\differential} 

\makeatletter
\def\dsigma{\@ifstar\@@dsigmawithstar\@dsigma}
\def\@dsigma{\frac{\dd\sigma}{\dd\Omega}}  
\def\@@dsigmawithstar{\dd\sigma/\dd\Omega} 
\makeatother

\begin{document}

\title{Bayesian truncation errors in chiral effective field theory: nucleon-nucleon observables}

\author{J.~A. Melendez}
\email{melendez.27@osu.edu}
\affiliation{Department of Physics, The Ohio State University, Columbus, OH 43210, USA}

\author{S.~Wesolowski}
\email{wesolowski.14@osu.edu}
\affiliation{Department of Physics, The Ohio State University, Columbus, OH 43210, USA}

\author{R.~J. Furnstahl}
\email{furnstahl.1@osu.edu}
\affiliation{Department of Physics, The Ohio State University, Columbus, OH 43210, USA}

\date{\today}

\begin{abstract}
Chiral effective field theory (EFT) predictions are necessarily truncated at
some order in the EFT expansion, which induces an error that must be quantified 
for robust statistical comparisons to experiment.
In previous work, a Bayesian model for truncation errors of perturbative expansions 
was adapted to EFTs. 
The model yields posterior probability distribution 
functions (pdfs) for these errors based on expectations of naturalness encoded
in Bayesian priors and the observed order-by-order convergence pattern of the EFT\@.
A first application was made to chiral EFT for neutron-proton scattering using 
the semi-local potentials of Epelbaum, Krebs, and Mei{\ss}ner (EKM).
Here we extend this application to consider a larger set of regulator parameters,
energies, and observables as a general example of a statistical approach to truncation errors. 
The Bayesian approach allows for statistical validations of the assumptions 
and enables the calculation of posterior pdfs for the EFT breakdown scale.
The statistical model is validated for EKM potentials whose convergence behavior
is not distorted by regulator artifacts.
For these cases, the posterior for the breakdown scale is consistent with
EKM assumptions. 
\end{abstract}

\maketitle

\newpage


\section{Introduction}
\label{sec:introduction}

The scope of \emph{ab initio} nuclear structure and reactions has increased dramatically
due to recent advances in many-body methods~\cite{Barrett:2013nh,Hagen:2013nca,Carlson:2014vla,Hergert:2016iju,Barbieri:2016uib,Lee:2016fhn}, 
continued growth in computational power, 
and new developments in chiral effective field theory 
(EFT)~\cite{Epelbaum:2008ga,Machleidt:2011zz,Gezerlis:2014zia,Epelbaum:2014sza,Piarulli:2014bda,Entem:2017gor}.
To properly judge the successes and predictive power of \emph{ab initio} nuclear theory,
however, it is necessary that theory errors be understood.
Thus, quantifying the theoretical uncertainties of nuclear calculations has now become a 
critical task for confronting experiment and theory and for extrapolating
to unmeasured phenomena~\cite{Dobaczewski:2014jga,0954-3899-42-3-030301,Carlsson:2015vda}.

Uncertainties in chiral EFT predictions arise from three sources~\cite{Furnstahl:2014xsa}:
uncertainty in the input data to which the EFT parameters are fit, errors in the Hamiltonian,
and numerical approximations.
Here we focus on quantifying the Hamiltonian truncation error as part of the larger
BUQEYE program~\cite{Furnstahl:2014xsa} of quantifying all uncertainties for EFT predictions.
Despite the promise of systematic expansions, uncertainties from truncation have been difficult
to estimate and, when provided, generally lack a well-defined statistical interpretation.

In Ref.~\cite{Furnstahl:2015rha},
a Bayesian model for truncation errors originally applied to perturbative
expansions in quantum chromodynamics~\cite{Cacciari:2011ze,Bagnaschi:2014eva}
was adapted to EFTs.
The generic assumption is that the EFT provides us
with a dimensionless expansion parameter $Q$, 
which is a ratio of scales, and an
associated expansion for quantities $X$
(usually observables):
\beq
  X= \Xref \sum_{n=0}^{\infty} c_n \Q^n  \;.
  \label{eq:obsexp}
\eeq
Here, $\Xref$ is the natural size of $X$, which could be the leading-order estimate $X_0$, and 
the $c_n$s are dimensionless coefficients.  For chiral EFT $c_1$ is zero by symmetry, and
we have a double expansion in $Q = \{p, m_\pi\}/\Lambda_b$, where $p$ is the relative
momentum of two scattering nucleons, $m_\pi$ is the pion mass, and 
$\Lambda_b$ is the EFT breakdown scale.
The goal is to estimate the error incurred in the observable by truncating the expansion at order $k$.
Note that this does not exclude an asymptotic expansion, but assumes that we truncate
while the result is still improving.

In some cases the expansion in Eq.~\eqref{eq:obsexp} may follow directly from 
a perturbative EFT expansion of a Lagrangian, i.e., through a sum of Feynman diagrams
with powers of $Q$ coming from a simple power-law dependence on momentum or a mass
(such as the pion mass in a chiral perturbation theory expansion).
There will also be implicit $Q$ dependence, often in the form of logarithms,
which vary much more slowly. 
But in other cases, such as EFT for more than one nucleon, 
the calculations are nonperturbative, and
the dependence on momentum or energy will be complicated and nonlinear in general.
Nevertheless, if the EFT is working we expect the calculation of $X$ to improve systematically as we go to higher orders.  

Equation~\eqref{eq:obsexp} can be interpreted as a summary of that
expected systematic improvement.
Namely, that the correction term with each successive order is on average a factor $Q$ smaller
than the previous order.
For this to be the case, we need the $c_n$ coefficients to be roughly the same size.
Because the coefficients are unknown \emph{a priori}, we treat them as drawn from
a random distribution with a characteristic size.
This is a realization of the underlying assumption
that the naturalness of the low-energy constants (LECs) in the EFT Lagrangian 
propagates to the expansion for any observable.
We have no general proof of this assumption, so we aim to validate it in each application.

In Ref.~\cite{Furnstahl:2015rha} we made a first pass at formalizing and testing
the assumptions behind the expansion in Eq.~\eqref{eq:obsexp}, 
building on an analogous Bayesian analysis applied to perturbative QCD calculations~\cite{Cacciari:2011ze,Bagnaschi:2014eva}.
We considered various priors for the $c_n$s, made an application to a small
subset of results from Epelbaum, Krebs, and Mei{\ss}ner (here EKM)
for neutron-proton (\npr) scattering cross sections using their new
semi-local potentials~\cite{Epelbaum:2014efa,Epelbaum:2014sza}, and tested
the consistency of assumed expansion parameters, which are associated
with the expected breakdown scales of the EFT implementation.
Here we revisit the EKM application to further test and generalize those investigations,
which will set the stage for extending our model of EFT truncation errors.

We seek to address the following questions:
\be
  \I 
  Coefficients $c_0$--$c_5$ of the total cross section $\sigma$ given at four energies in
  Refs.~\cite{Epelbaum:2014efa,Epelbaum:2014sza} were examined in Ref.~\cite{Furnstahl:2015rha}.
  Can we validate \emph{a posteriori} our assumption that the observable coefficients 
  follow some bounded random distribution about zero for \emph{all} energies?

  \I The truncation error model 
  of Ref.~\cite{Furnstahl:2015rha} has not yet been applied to other
  nucleon-nucleon (NN) observables calculated in chiral EFT, such as the
  differential cross section and various spin observables.
  How do the coefficient patterns compare for different
  NN scattering observables, considered both as functions of energy and scattering angle?
  Are the naturalness assumptions validated for these observables and for all
  values of the EKM regulator parameter $R$? 

  \I An appropriately assigned $\ppercent$ error band should capture the true value of an
  observable $\ppercent$ of the time.
  How can we utilize known order-by-order results to verify that the error band
  prescriptions work as advertised?
  Can information from different observables be treated as
  independent, and if not, how can we account for their relationships in our
  analysis?
  Is there a well-defined ``correlation length'' in energy or scattering
  angle beyond which expansion coefficients
  may be treated as independent of one another? 

  \I The identification of the expansion parameter $Q$, which in turn is based on
   identifying the scale $\Lambda_b$, is a key element in determining the
   convergence pattern. Is it consistent to take 
   $\Lambda_b$ to be the same scale for every observable? 
   To what extent can we \emph{extract} $\Lambda_b$, given order-by-order expansions
    and our naturalness assumptions encoded in a Bayesian model?
\ee
In the present work we make progress on all these questions.

In Sec.~\ref{sec:formulas}, we summarize and extend the relevant formulas from 
Ref.~\cite{Furnstahl:2015rha}.
We refer the reader to that article for background on the
use of Bayesian statistics in this context, derivations of the formulas
we summarize here, and more general references.
New results and analysis for the total cross section are given in
Sec.~\ref{sec:total_sigma}, and other observables are considered in
Sec.~\ref{sec:other_observables}.
In Sec.~\ref{sec:model_checking} we perform Bayesian model checking~\cite{gelman2013bayesian}
by applying a consistency check used in Ref.~\cite{Furnstahl:2015rha}
and by calculating posterior probability distribution functions
(pdfs) for $\Lambda_b$.
Section~\ref{sec:summary} has our summary and outlook.
For completeness and convenience, we show explicit formulas regarding
the Bayesian model for truncation errors in Appendix~\ref{app:derivations_of_posteriors}
and summarize the notation and formulas used
for NN observables in Appendix~\ref{app:observables}.
The Supplemental Material~\cite{supplmaterial} displays extra figures and data that helped inform our conclusions.


\section{Formulas}
\label{sec:formulas}

If the EFT expansion in Eq.~\eqref{eq:obsexp} 
is truncated at order $k$, then the error induced is $\Xref \Delta_k$,
where the scaled, dimensionless parameter that determines the truncation error is
\beq
  \Delta_k \equiv \sum_{n=k+1}^\infty c_n \Q^n
  \;.
  \label{eq:Deltak_def}
\eeq
Generally it is only practical to approximate
$\Delta_k^{(h)}$,
the error due to the first $h$ omitted higher-order terms.
For sufficiently small values of $\Q$, the first omitted term $\Delta_k^{(1)} = c_{k+1} \Q^{k+1}$ 
is a good estimate for $\Delta_k$, but we do not assume this in general.

We use the notation $\pr(x|I)$ to denote the probability density of
$x$ given information $I$.
Our pdf of interest is $\pr_h(\Delta|\ckvec)$:
the probability distribution for $\Delta_k$ given the vector of relevant lower-order coefficients
$\ckvec$ that have been calculated, assuming that only $h$ higher-order terms contribute to the error and that $\Lambda_b$ is to
be given from other considerations.
The pdf $\pr_h(\Delta|\ckvec)$ is normalized in terms of the dummy variable $\Delta$, 
which is implied to be an estimate of $\Delta_k$ contingent on lower-order coefficients $\ckvec$.

In contrast to Ref.~\cite{Furnstahl:2015rha}, here $c_0 \notin \ckvec$ because it does not provide insight into the convergence pattern of the observables; rather, the LO calculation provides
scaling information.
Again, $c_1 \notin \ckvec$ because $c_1 = 0$ in chiral EFT.
Thus the relevant lower-order coefficients in the determination of $\Delta_k$ in chiral EFT are
\begin{align} \label{eq:ckvec}
  \ckvec = (c_2,c_3, \ldots,c_k)\;.
\end{align}

In the Bayesian framework, the posterior $\pr_h(\Delta | \ckvec)$
contains the complete information we claim to have about the dimensionless 
residual $\Delta_k$.
In general, a posterior pdf can have complex structures such as multiple modes, heavy tails, 
large skewness, etc.
Here we can capture most of the information with a small number of
degree-of-belief (DoB) intervals.%
\footnote{These are also called ``credibility'' or ``credible'' intervals, or ``Bayesian
confidence intervals''.}
We use the highest posterior density (HPD) definition of DoB, which is the shortest interval that contains $\ppercent$ of the area~\cite{kruschke2011doing,gelman2013bayesian,Liu:2015:SSP:2799322.2799342}.
This ensures that the probability density within the DoB is never lower than the density outside.
The HPD definition is particularly well suited for skewed posteriors, as we will encounter in Sec.~\ref{sec:model_checking}.
Because the $\pr_h(\Delta|\ckvec)$ that we consider here are unimodal and symmetric about $\Delta=0$, 
finding the DoB interval reduces to the inversion problem for $d_k^{(p)}$, where
\beq
    p = \int_{-d_k^{(p)}}^{d_k^{(p)}} \dd{\Delta} \pr_h(\Delta|\ckvec)
    \;.
    \label{eq:integralford}
\eeq 
Hence, one believes with $\ppercent$ certainty the true value of the observable $X$
lies within $\pm \Xref\, d_k^{(p)}$ of the N$^k$LO prediction.
In general, Eq.~\eqref{eq:integralford} must be inverted numerically, 
but simplified results for certain
priors and approximations (e.g., that the first omitted term dominates) 
are possible~\cite{Furnstahl:2015rha}.

\begin{figure}[tb]
  \includegraphics[width=0.9\columnwidth]{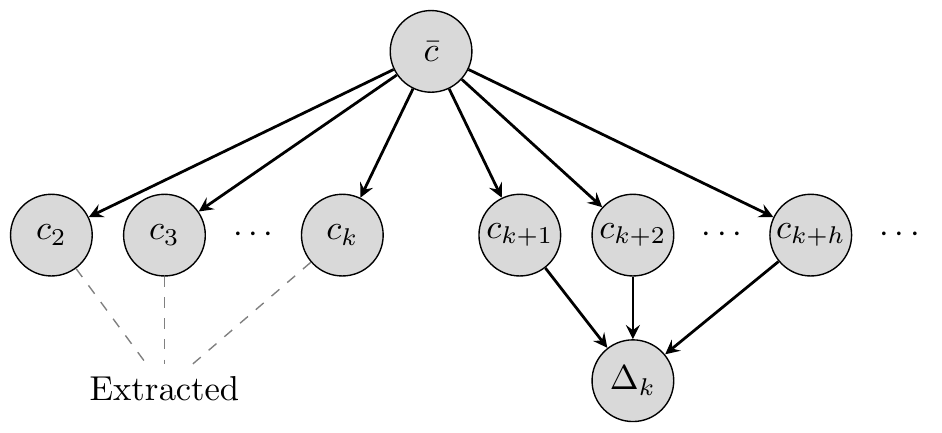}
  \caption{A Bayesian network~\cite{ben2007bayesian,citeulike:13925803} for the 
  $\Delta_k$ truncation error model outlined in~\cite{Furnstahl:2015rha}.}
  \label{fig:Bayesian_Network}
\end{figure}

In Ref.~\cite{Furnstahl:2015rha}, a statistical model for $\Delta_k$ in terms 
of the order-by-order coefficients of the EFT expansion was developed.
It was assumed that naturalness could be implemented by treating the $c_n$s as random
variables drawn from a shared distribution centered at zero with a characteristic size
or upper bound $\cbar$.
The coefficients at each order are treated as independent of one another---the
value of $\cbar$
is the only way that information propagates between orders.
These relationships can be encapsulated in a Bayesian network~\cite{ben2007bayesian,citeulike:13925803}, as shown in Fig.~\ref{fig:Bayesian_Network}.
The nodes of the graph are random variables and the arrows denote causal relationships between them.

While the topology of Fig.~\ref{fig:Bayesian_Network} outlines the 
logic of our model, prescriptions in the 
form of priors $\pr(c_n|\cbar)$ and $\pr(\cbar)$ must be given to make 
quantitative statistical inferences of $\Delta_k$.
When all we know is that there is an upper bound to the coefficients, an application of 
maximum entropy~\cite{Gull:98,Sivia:2006} dictates that the least-informative distribution $\pr(c_n|\cbar)$ is uniform for $|c_n| < \cbar$ and zero otherwise.
Such uniformity is additionally appealing because it can lead to simple, analytic results.
This uniform prior was the initial choice of Ref.~\cite{Cacciari:2011ze}. 
We employ it in priors we denote as ``set A'' and ``set B'' (see Table~\ref{tab:priors}).
The analogous prior of ``set C'' in Table~\ref{tab:priors} corresponds to the ensemble
naturalness assumption of Ref.~\cite{Schindler:2008fh}. This Gaussian prior follows
from the maximum-entropy principle assuming knowledge of testable information on the mean 
and standard deviation of the $c_n$s~\cite{Schindler:2008fh}:
\beq
  \label{eq:ensnat_conditions}
   \left\langle \ckvecsq \right\rangle = (k-1)\cbar^2, \quad \left\langle c_n \right\rangle = 0
   \;.
\eeq

In addition we require a prior for $\cbar$: $\pr(\cbar)$.
Sets A and C of Table~\ref{tab:priors} use a log-uniform prior for $\cbar$ to reflect 
unbiased expectations regarding the scale of $\cbar$~\cite{Jeffreys:1939}
(this was the choice in Ref.~\cite{Cacciari:2011ze} and Ref.~\cite{Schindler:2008fh}).
Such a prior cannot be normalized for $\cbar$ in $(0,\infty)$ and
is therefore termed an ``improper prior''. 
Limiting the range of $\cbar$ through the use of $\theta$ functions 
permits an examination of the otherwise ill-defined limiting behavior.
When marginalizing (i.e.\ integrating) over $\cbar$, we can express complete ignorance of
the scale of $\cbar$ by considering the limit of infinite range ($\Aeps$ or $\Ceps$, see~\cite{Furnstahl:2015rha}),
or render the prior more informative through the use of a finite range $[a,b]$
($\Af{a}{b}$ or $\Cf{a}{b}$).
Alternatively, set~B employs a log-normal distribution about
zero~\cite{Bagnaschi:2014wea,Bagnaschi:2014eva}, which sets the
scale of $\cbar$ with the hyperparameter $\sigma$.

\begin{table}[tb]
\caption{Candidates for prior pdfs~\cite{Furnstahl:2015rha}.}
\label{tab:priors}
\begin{ruledtabular}
\begin{tabular}{ScScScSc}
  Set  &   $\pr(c_n| \cbar)$   &   $\pr(\cbar)$ \\
  \colrule
  A
  &
  $\ds\frac{1}{2\cbar}\,\theta(\cbar-|c_n|)$
  &
  $\ds\frac{1}{\ln \cbarmax/\cbarmin}
   \frac{1}{\cbar} \,
   \theta ( \cbar - \cbarmin) \theta( \cbarmax - \cbar)$
  \\
  B
  &
  $\ds\frac{1}{2\cbar}\,\theta(\cbar-|c_n|)$
  &
  $\ds\frac{1}{\sqrt{2\pi}\cbar\sigma} e^{-(\ln\cbar)^2/2\sigma^2}$
  \\
  C
  &
  $\ds\frac{1}{\sqrt{2\pi}\cbar} e^{-c_n^2/2\cbar^2}$
  &
  $\ds\frac{1}{\ln \cbarmax/\cbarmin}
   \frac{1}{\cbar} \,
   \theta ( \cbar - \cbarmin) \theta( \cbarmax - \cbar)$
\end{tabular}
\end{ruledtabular}
\end{table}

\begin{figure*}[tb!]
  \subfloat{%
    \label{fig:cross_sections_EKM_R0p9_Lambdab_600_Xzero_coeffs}%
    \includegraphics[width=0.42\textwidth]{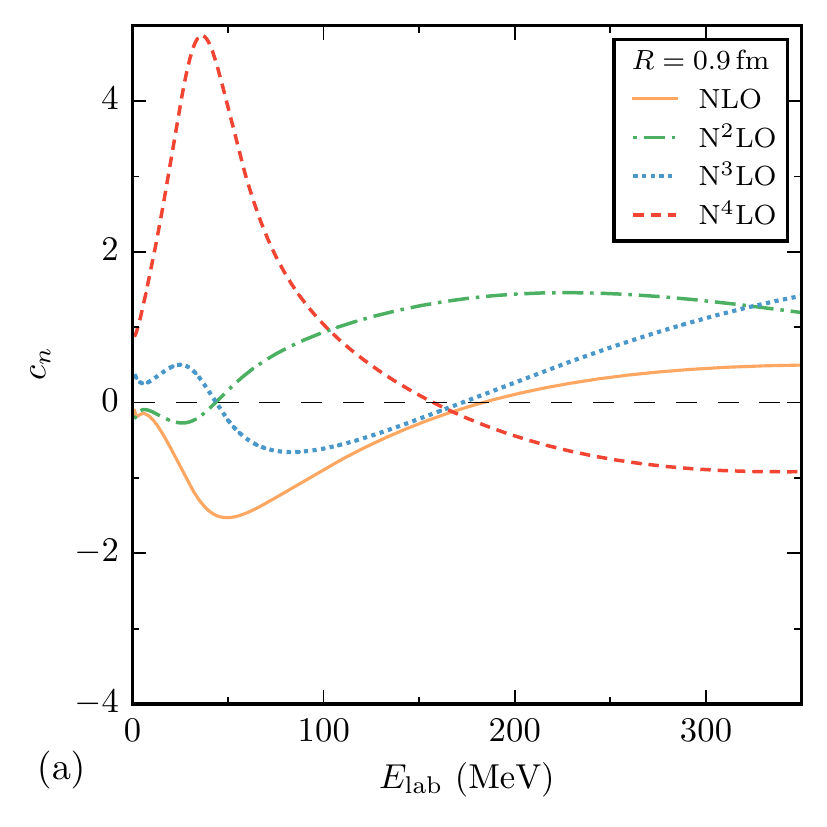}%
  } 
  \hspace*{0.05\textwidth}  
  \subfloat{%
    \label{fig:cross_sections_EKM_R0p9_Lambdab_600_Xzero_coeffs_p_expansion_only}%
  \includegraphics[width=0.42\textwidth]{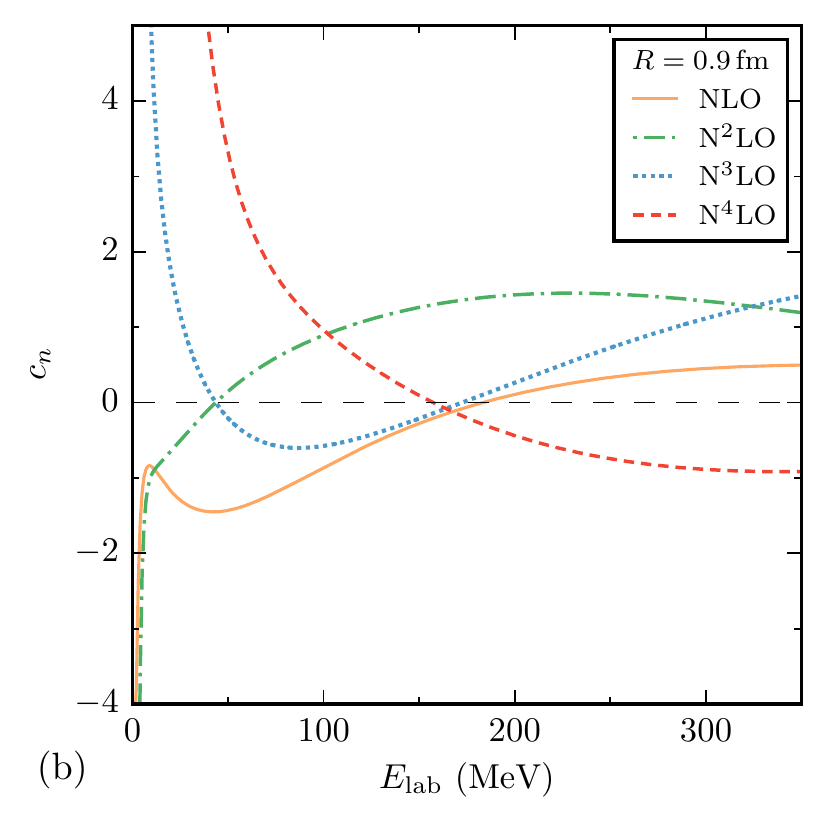}%
  }
  \caption{(a) Dimensionless coefficients as in Eq.~\eqref{eq:obsexp} 
  at each EFT order 
  for the \npr\ total cross section 
  as a function of lab energy for EKM potentials with
  $R=0.9\,$fm, and $\Lambda_b = 600\,$MeV\@.
  Plot (b) uses the $p/\Lambda_b$ expansion only.}
  \label{fig:cross_sections_EKM_R0p9_R0p9_Lambdab_600_600_Xzero_coeffs}
\end{figure*}

The general result for $\pr_h(\Delta|\ckvec)$ implied by
Fig.~\ref{fig:Bayesian_Network} was derived as%
\footnote{We have corrected and simplified here the 
corresponding equation from Ref.~\cite{Furnstahl:2015rha}.}
\begin{align} \label{eq:BayesEpsFull}
  \pr_h(\Delta|\ckvec) = \frac
    {\ds\int_0^\infty \dd{\cbar} \pr_h(\Delta|\cbar) \pr(\cbar) \prod_{n=2}^k \pr(c_n|\cbar)}
    {\ds\int_0^\infty \dd{\cbar} \pr(\cbar) \prod_{n=2}^k \pr(c_n|\cbar)}\;,
\end{align}
where
\begin{align} \label{eq:higher_orders_pdf}
  \pr_h(\Delta|\cbar) \equiv {\left[\prod_{i=k+1}^{k+h} \int_{-\infty}^\infty \! \dd{c_i} \pr(c_i | \cbar) \right]} \delta{\left(\Delta - \Delta_k^{(h)}\right)} \;.
\end{align}
If we assume that the first omitted term dominates the truncation error, 
then Eq.~\eqref{eq:higher_orders_pdf} is easily evaluated by the $\delta$ function for any prior, 
and Eq.~\eqref{eq:BayesEpsFull} reduces to
\begin{align}
  \pr_1(\Delta|\ckvec) =
   \frac{\ds\int_0^\infty \dd{\cbar} \pr(c_{k+1}|\cbar) \pr(\cbar) \prod_{n=2}^k \pr(c_n|\cbar)}
        {\ds Q^{k+1}\int_0^\infty \dd{\cbar} \pr(\cbar) \prod_{n=2}^k \pr(c_n|\cbar)}
        \;,
   \label{eq:BayesEps2}
\end{align}
where $c_{k+1} = \Delta/Q^{k+1}$ as enforced by the $\delta$ function.
Error bands made under this assumption are denoted by the prior with a superscript (1), e.g., $\Aone$.
Further progress, with or without the first-omitted-term approximation,
requires an explicit choice of priors.
The relevant equations for this work, such as posteriors and DoB intervals, are contained Appendix~\ref{app:derivations_of_posteriors}.


\section{Total NN cross section}
\label{sec:total_sigma}

Truncation error DoBs were estimated in Ref.~\cite{Furnstahl:2015rha}
for the \npr\ total cross section at laboratory energies of 50, 96, 143, and 200\,MeV 
from results given explicitly in Refs.~\cite{Epelbaum:2014efa,Epelbaum:2014sza} 
using the new R = 0.9\,fm EKM potential.
In this work we extend these calculations to all of the new potentials
and for all energies up to 350\,MeV\@.
The first step is to extract the $c_n$ coefficients, defined by Eq.~\eqref{eq:obsexp},
from the order-by-order calculations of the \npr\ total cross section $X = \sigma(\Elab)$.
Here we choose our reference scale to be the leading-order calculation
$\Xref = \sigma_0$, so $c_0 \equiv 1$ by construction.
Other reasonable choices of $\Xref$, such as a higher-order result or the experimental value, do not substantially change the convergence pattern for this observable.

\begin{figure*}[tbh!]
   \subfloat{%
     \label{fig:cross_sections_EKM_R0p8_Lambdab_600_Xzero_coeffs}%
   \includegraphics[width=0.42\textwidth]{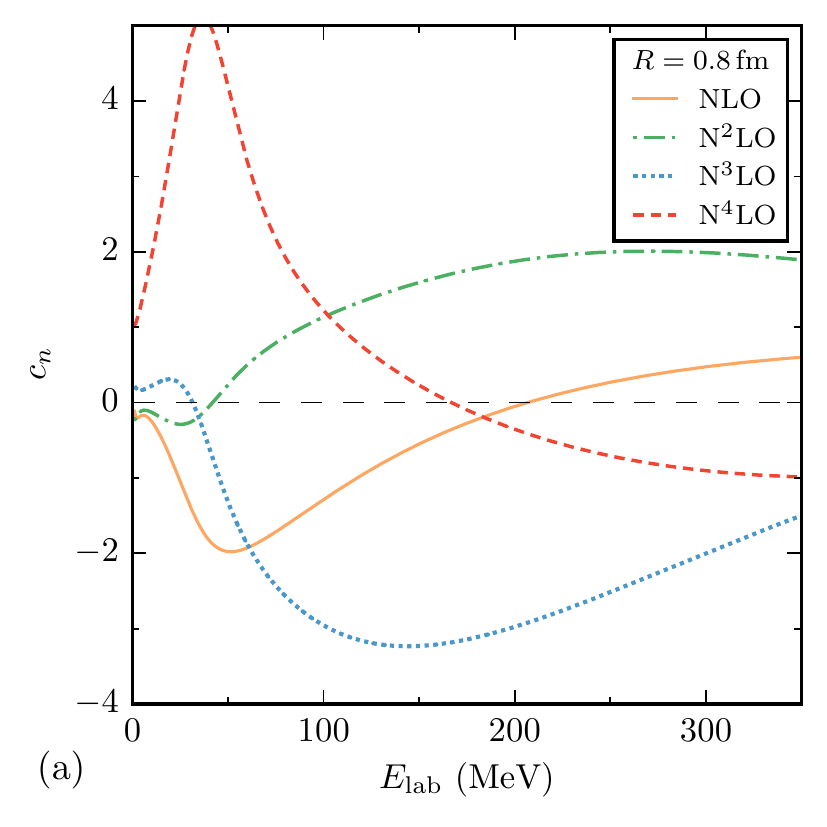}%
   }
  \hspace*{0.05\textwidth}  
  \subfloat{%
    \label{fig:cross_sections_EKM_R1p0_Lambdab_600_Xzero_coeffs}%
    \includegraphics[width=0.42\textwidth]{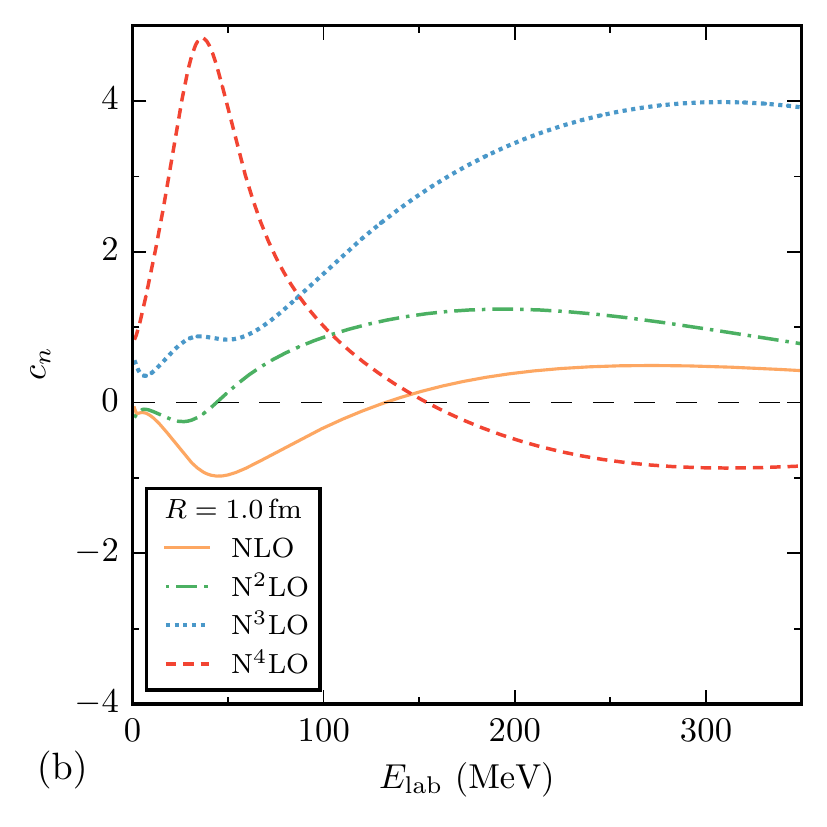}%
  }
  \caption{Dimensionless coefficients as in Eq.~\eqref{eq:obsexp} at each EFT order 
  for the \npr\ total cross section 
  as a function of lab energy for (a) $R=0.8\,$fm and (b) $R=1.0\,$fm EKM potentials, both with $\Lambda_b = 600\,$MeV.}
  \label{fig:cross_sections_EKM_R0p8_R1p0_Lambdab_600_600_Xzero_coeffs}
\end{figure*}

\begin{figure*}[tbh!]
  \subfloat{%
    \label{fig:cross_sections_EKM_R1p1_Lambdab_500_Xzero_coeffs}%
    \includegraphics[width=0.42\textwidth]{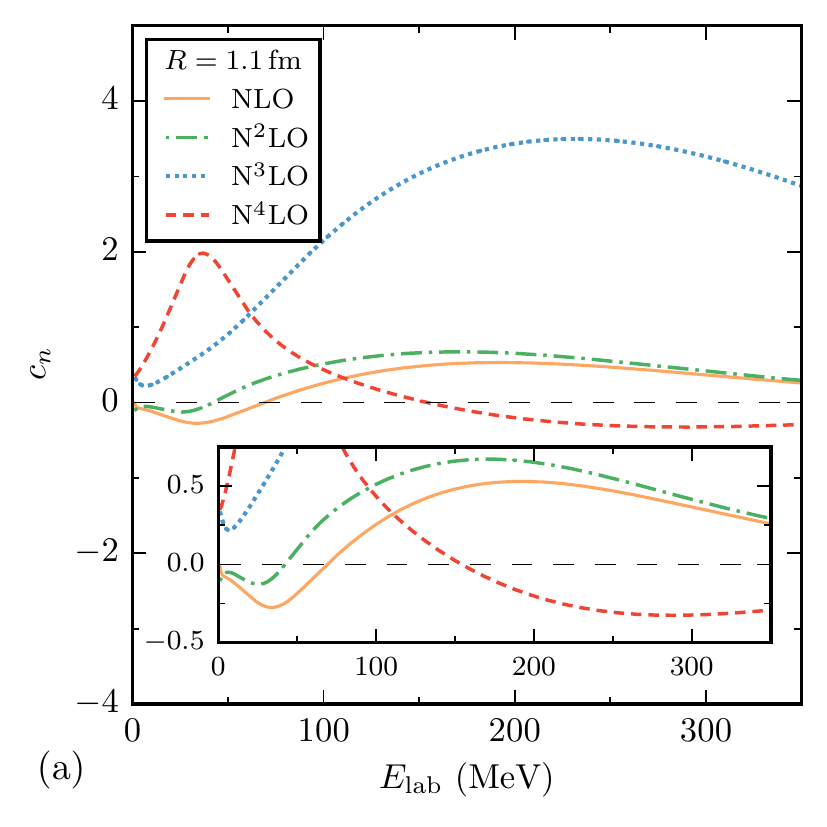}%
  } 
  \hspace*{0.05\textwidth}  
  \subfloat{%
    \label{fig:cross_sections_EKM_R1p2_Lambdab_400_Xzero_coeffs}%
    \includegraphics[width=0.42\textwidth]{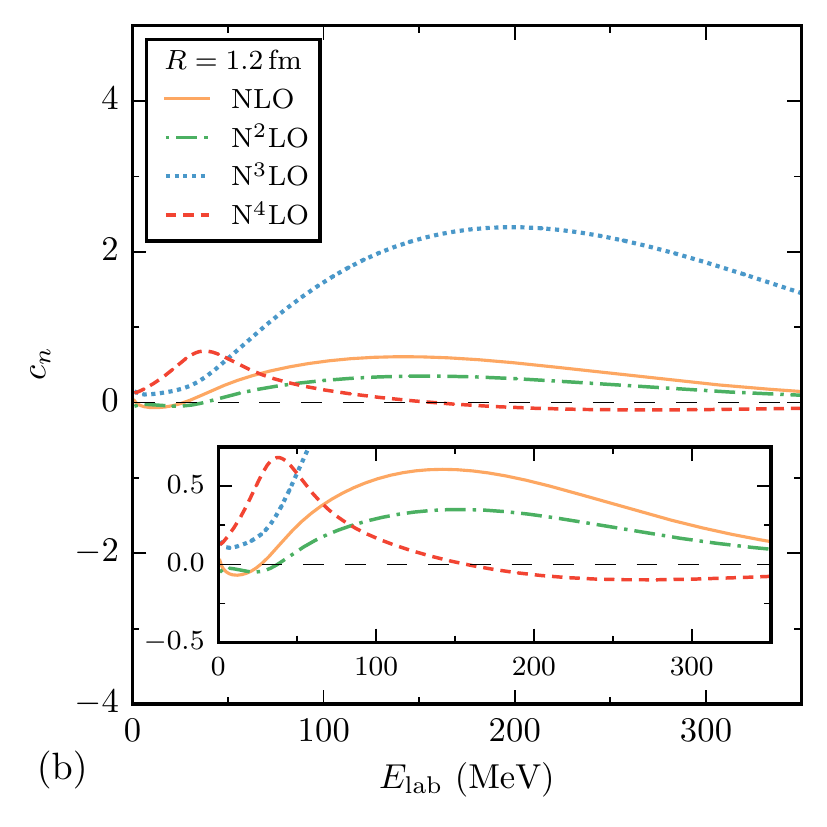}%
  }
  \caption{Same as Fig.~\ref{fig:cross_sections_EKM_R0p8_R1p0_Lambdab_600_600_Xzero_coeffs}
   but for EKM potentials with 
  (a) $R=1.1\,$fm and $\Lambda_b = 500\,$MeV, and (b) $R=1.2\,$fm and $\Lambda_b = 400\,$MeV.}
  \label{fig:cross_sections_EKM_R1p1_R1p2_Lambdab_500_400_Xzero_coeffs}
\end{figure*}

We also need to specify the high-momentum scale $\Lambda_b$.
Here we assume that $\Lambda_b$ is a given quantity, and adopt the values assumed by
EKM\@.  Their choice of
$\Lambda_b \approx 400\mbox{--}600$\,MeV (the particular value depending on a regulator parameter $R$) 
was based on a rough analysis of residual error plots 
(``Lepage plots''), validated by the observation that their choices resulted in natural coefficients in 
the EFT series for \npr\ scattering cross sections~\cite{Epelbaum:2014efa,Epelbaum:2014sza}.
In Sec.~\ref{sec:model_checking}, we make a statistical analysis of whether the EKM
choices of $\Lambda_b$ (or nearby values) lead to self-consistent convergence patterns for
observables, and explore directly determining a posterior probability distribution for $\Lambda_b$.

Because we have a double expansion in $p/\Lambda_b$ and $m_\pi/\Lambda_b$, 
we must develop a prescription to define $Q$.
In Ref.~\cite{Furnstahl:2015rha} we took $Q$ to be
\beq
  Q = \frac{\text{max}\{p,m_\pi\}}{\Lambda_b}
  \label{eq:Qmax}
  \;.
\eeq
We expect that at low momenta the expansion will be dominated by powers
of $m_\pi/\Lambda_b$ and at momenta much higher than $m_\pi$ it will be dominated
by powers of $p/\Lambda$, so the appropriate choice of $Q$ in each region follows correctly
from Eq.~\eqref{eq:Qmax}.  
However, it is not clear how we should parameterize the crossover region.
To avoid cusps at $p = m_\pi$, we choose to replace Eq.~\eqref{eq:Qmax} by a smooth
interpolation function for $Q$:
\beq
  Q_{\text{interp}}(p) = \frac{m_\pi^n + p^n}{m_\pi^{n-1} + p^{n-1}} \frac{1}{\Lambda_b}
  \;,
  \label{eq:Qinterp}
\eeq
where $n$ is a sufficiently high integer (we take $n=8$ here).
But we need to examine the behavior at low energies to assess whether the implicit
equal weighting of the expansions is justified.

Coefficients $c_2$--$c_5$ for the total cross section, calculated with the $R=0.9\,$fm potential
and $\Lambda_b = 600\,$MeV,
are shown as functions of energy in the left panel of
Fig.~\ref{fig:cross_sections_EKM_R0p9_R0p9_Lambdab_600_600_Xzero_coeffs}.
These include the results for four individual energies from \cite{Furnstahl:2015rha},
but now we can see the global pattern.
Except for the \NNNNLO\ coefficient around $\Elab \approx 50$\,MeV,
the coefficients at any fixed energy follow a distribution with a
characteristic size of about one. 
If $Q= p/\Lambda_b$ were used for all energies instead of Eq.~\eqref{eq:Qinterp}, 
then the coefficients would grow very large as $\Elab$ gets small 
(i.e., as $p\rightarrow 0$), as shown in the right panel of
Fig.~\ref{fig:cross_sections_EKM_R0p9_R0p9_Lambdab_600_600_Xzero_coeffs}.
The onset of this behavior in $\Elab$ increases
with chiral order and, for \NNNNLO, is the source
of the large coefficient near $\Elab \approx 50\,$MeV\@.
This reflects the increasing sensitivity at large order to the relative contribution
of the two expansions in the crossover region.  
We do not yet have a model to address this behavior.
If we exclude the crossover region, 
the underlying assumption of the priors $\pr(c_n|\cbar)$ in 
Table~\ref{tab:priors} that the coefficients at a given energy are distributed
with a characteristic size $\cbar$ is validated. 

The observable coefficients for the other EKM
potentials are shown in 
Figs.~\ref{fig:cross_sections_EKM_R0p8_R1p0_Lambdab_600_600_Xzero_coeffs} 
and \ref{fig:cross_sections_EKM_R1p1_R1p2_Lambdab_500_400_Xzero_coeffs}.
For each potential we have adopted the value of $\Lambda_b$ advocated by EKM:
$\Lambda_b$ equal to 600\,MeV for $R=0.8\,$fm, $0.9\,$fm, and $1.0\,$fm,
500\,MeV for $R = 1.1\,$fm, and 400\,MeV for $R = 1.2\,$fm.
We return in Sec.~\ref{sec:model_checking} to consider different choices of $\Lambda_b$.

The general assumption made in constructing a posterior for the truncation error in 
Ref.~\cite{Furnstahl:2015rha}, that the 
coefficients have a characteristic magnitude or upper bound $\cbar$,
is based on the expectation that a well-formulated EFT will have a certain
uniformity in the convergence pattern of observables.  That is, with each 
successive order there is a steady convergence implied by the value of the 
expansion parameter (as shown below,
this corresponds with
a steady improvement
of the prediction for the cross section).
For an integrated observable such as the total cross
section, we expect this to be particularly manifested.  This justifies the use of
lower-order results to inform our expectations for higher-order contributions.

The pattern of coefficients for $R=0.9\,$fm shows this uniformity,
which is mostly still present for $R=0.8\,$fm and $R=1.0\,$fm.  
In particular, we see evidence for a characteristic size for the 
$c_n$s of order unity (in practice about three).
However, the uniformity deteriorates
significantly as one progresses to $R=1.1$, and $1.2\,$fm.
This is a consequence of the growing cutoff artifacts at larger values of $R$.
As the artifacts become more prevalent, there is a decreased contribution
from mid-range pion physics at \NNLO\ and \NNNNLO, which is counteracted by an
increase in the contact terms at \NLO\ and particularly {\NNNLO}\@.
This reflects a partial integrating-out of pion physics, which
takes us closer to a pionless EFT convergence pattern with the dominant contributions
at even orders in the expansion.

In Ref.~\cite{Furnstahl:2015rha}, we analyzed results only for 
$R=0.9\,$fm and $R=1.2\,$fm (as reported by EKM) and only at four energies.
From this limited sample we concluded that the distribution of $c_n$s at these energies 
was consistent with a common $\cbar$ for $R=0.9\,$fm,
but this was not the case for $R=1.2\,$fm.  
In particular, the latter case had 
uniformly small coefficients for N$^2$LO and N$^4$LO, 
consistent with there being no new short-range contributions
at those orders and the regulator greatly reducing the pion tensor-range 
contribution.
Now looking globally at the $R=1.2\,$fm coefficients,
we see that 
N$^2$LO and N$^4$LO stay small for the full range of $\Elab$. 
If we focus on N$^2$LO in each graph, we see that the 
trend of the coefficients with energy is quite similar as $R$ increases
(softening the interaction), 
but the overall scale decreases monotonically.  
The situation with N$^4$LO is similar.

Turning to 
the N$^3$LO coefficients for successive values of $R$, we find a
transition from negative and fairly large (order $-3$)
at $R=0.8\,$fm to positive and fairly large (order $+3$) at $R=1.0\,$fm and above.
$R=0.9\,$fm is the middle of this transition. 
This is not unnatural, but may reflect a tendency toward overfitting at
N$^3$LO (see Ref.~\cite{Wesolowski:2017aa}).
Taking all the orders together, the coefficients imply that 
the convergence pattern for $R=1.1\,$fm and $1.2\,$fm, for which regulator artifacts are
significant, is not consistent with our statistical model.

\begin{figure}[tb!]
\includegraphics[width=0.48\textwidth]{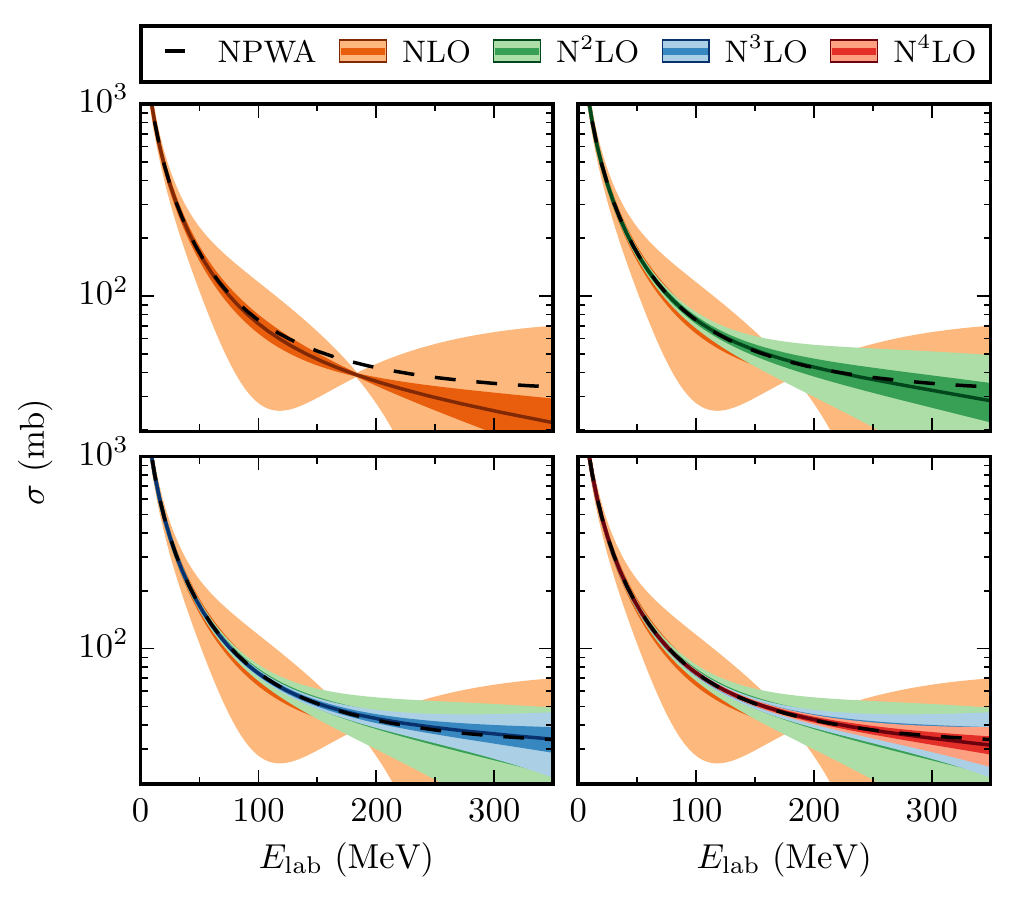}
  \vspace*{-.15in}
\caption{DoB intervals for the \npr\ total cross section for $R=0.9\,$fm
  at each of the orders, using prior set $\Ceps$.
  }
\label{fig:cross_section_DoBs}
\end{figure}

\begin{figure}[tbh!]
  \includegraphics[width=0.48\textwidth]{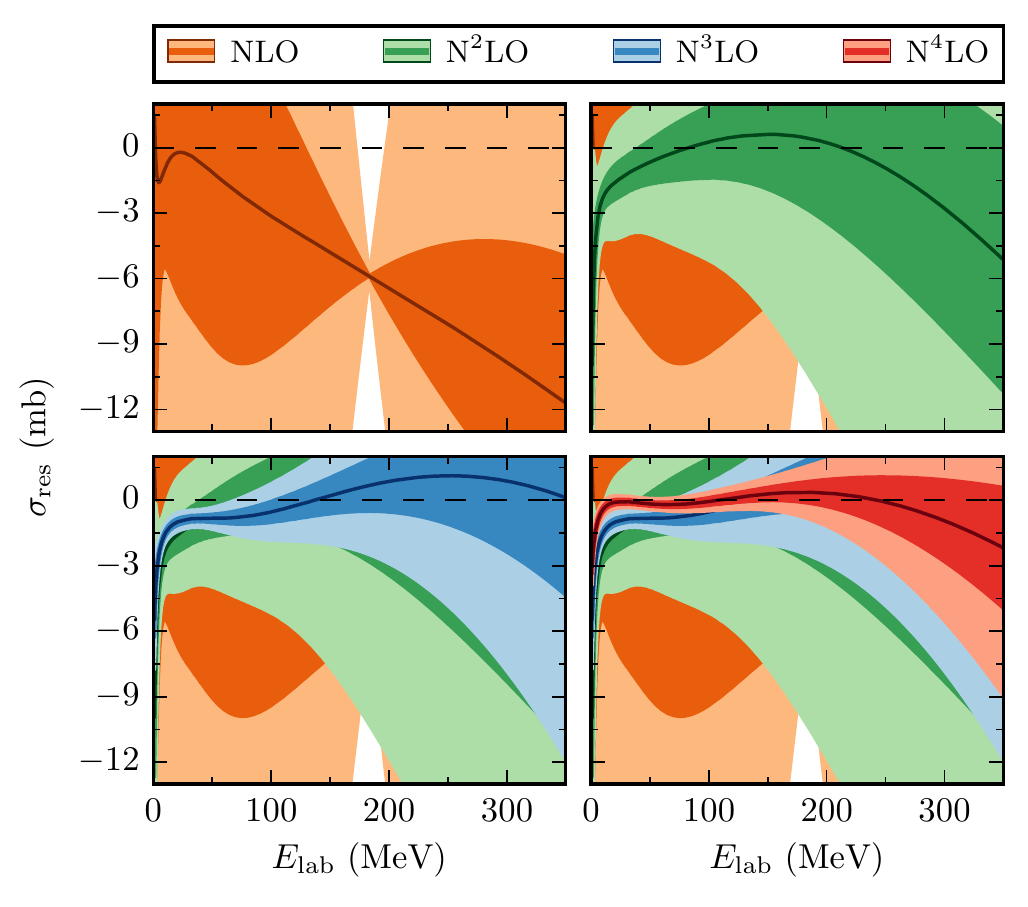}
  \vspace*{-.15in}
\caption{Residuals defined in Eq.~\eqref{eq:residual_def} at each order
for the \npr\ total cross section for $R=0.9\,$fm, 
with DoB intervals calculated using prior set $\Ceps$.
  }
\label{fig:cross_section_residual_DoBs}
\end{figure}

\begin{figure}[tbh!]
\includegraphics[width=0.48\textwidth]{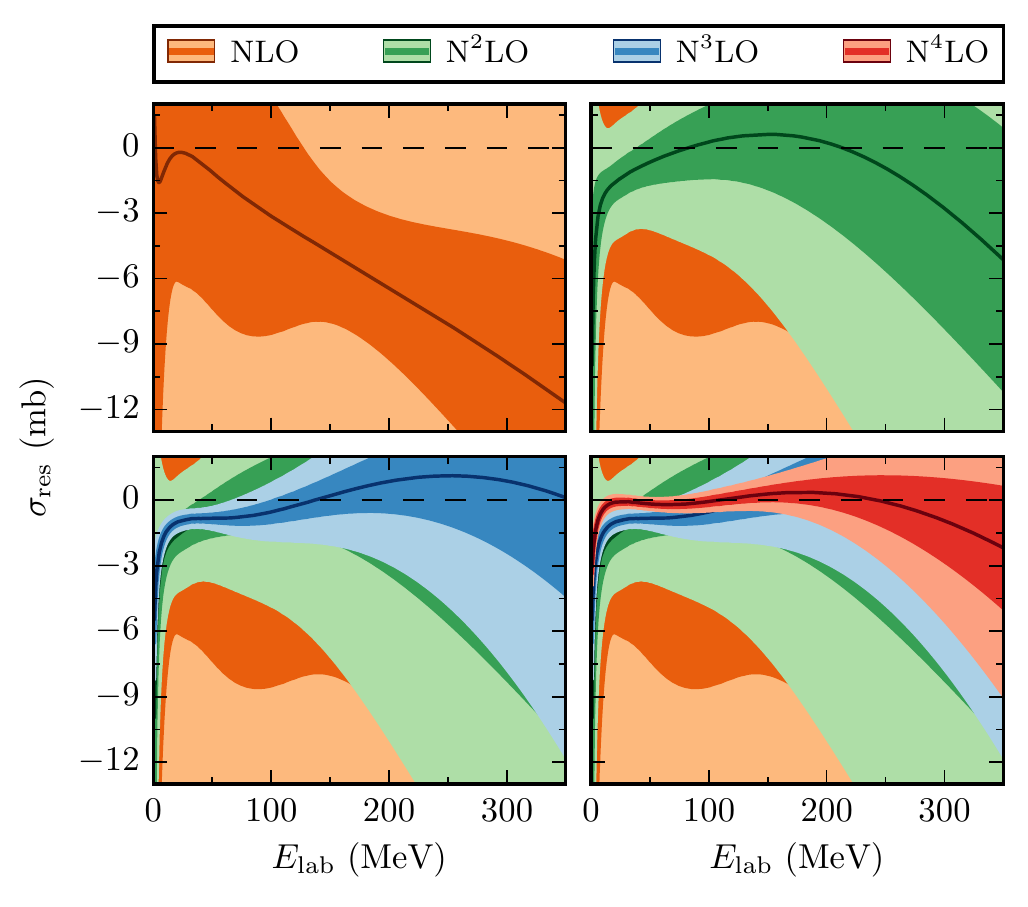}
  \vspace*{-.15in}
\caption{Residuals defined in Eq.~\eqref{eq:residual_def} at each order
for the \npr\ total cross section for $R=0.9\,$fm, 
with DoB intervals calculated using prior set $\Cf{0.25}{10}$.
  }
\label{fig:cross_section_residual_DoBs_Cp25-10}
\end{figure}

Next we estimate DoB intervals for EFT truncation errors using the extracted
coefficients.
We apply at each energy the formulas from Sec.~\ref{sec:formulas} to the
coefficients from that energy only.
The order-by-order results for the total cross section with $R=0.9\,$fm and
prior set $\Ceps$ (using Eq.~\eqref{eq:posterior_setC_analytic_eps} or \eqref{eq:Ceps_d_transcendental}) are shown in 
Fig.~\ref{fig:cross_section_DoBs}.
To amplify the patterns, in Fig.~\ref{fig:cross_section_residual_DoBs} and
below we plot the residuals with respect to the Nijmegen partial-wave analysis (NPWA)~\cite{PhysRevC.48.792}, where the residual for a calculated observable $X$ is defined as
\begin{align} \label{eq:residual_def}
  X_{\text{res}} \equiv X - X_{\text{NPWA}}\;.
\end{align}
All plots of observables and residuals are shown with
solid lines for the calculation at each order, with
dark and light shaded bands denoting the 68\% and 95\% DoB interval for the truncation
error at each kinematic point.  
Note that the errors are not Gaussian, as the 95\% bands are not twice the
size of the corresponding 68\% bands.

The order-by-order convergence of the calculations in Fig.~\ref{fig:cross_section_residual_DoBs}
to the NPWA result is clear, but not surprising---the potential
was fit to reproduce the NPWA in each partial wave.
The pattern of DoB intervals shown in both Figs.~\ref{fig:cross_section_DoBs}
and~\ref{fig:cross_section_residual_DoBs}
is mostly systematic: the widths tend to increase with $Q$,
decrease with order, and overlap with preceding order DoBs. 
Although we used set~C for these truncation error estimates on the cross section,
the results using set~A are similar.
See the Supplemental Material~\cite{supplmaterial} for plots displaying various error band prescriptions.
We return to quantify the effects of prior choice on the success rate of the error bands in Sec.~\ref{sec:model_checking}.

An exception to the systematic and intuitive DoB intervals is for the NLO calculation 
near 200\,MeV, where the intervals vanish.
This is because the prior set $\Ceps$ makes no assumption on the minimum (or maximum) 
size of $\cbar$, so the only information for the DoB at NLO is the NLO coefficient,
which vanishes in that energy range.  In  Ref.~\cite{Furnstahl:2015rha} we included
$c_0 =1$ at NLO, which effectively set a lower limit of $\cbar = 1$.  
Considering the NLO coefficients over the full energy range, as well as the other
coefficients, it is clear that we should use a prior with a non-zero $\cbarmin$.
A more informative, but not too restrictive, choice of $\cbarmin = 0.25$ (and $\cbarmax = 10$) is used
in Fig.~\ref{fig:cross_section_residual_DoBs_Cp25-10}.
The DoB intervals at low order are now more plausible while there is no significant
difference at the two highest orders.

Of course, it is not enough that the DoB bands are plausible;
they should be statistically valid.
If our DoB intervals are consistent, we might expect the NPWA line
in Fig.~\ref{fig:cross_section_DoBs} or the zero line in
Figs.~\ref{fig:cross_section_residual_DoBs} or~\ref{fig:cross_section_residual_DoBs_Cp25-10}
to lie outside the 68\% region roughly 1/3 of the time and outside
the 95\% region roughly 1/20 of the time.
With this in mind, a rough examination of
Fig.~\ref{fig:cross_section_residual_DoBs_Cp25-10} shows
the bands are not ideal: 
the 95\% DoBs appear too large for \NLO\ and \NNLO, while they are too small for \NNNLO;
the 68\% DoBs underestimate the error on the \NLO\ plot;
and the \NNNLO\ DoBs do not perform well at low energies.
In Sec.~\ref{sec:model_checking}, we perform a systematic analysis
using Bayesian model checking, where
we evaluate how well the DoBs predict the subsequent order correction.
Section~\ref{sec:summary}  concludes
with a reassuring proof of concept (Fig.~\ref{fig:cons_sigma_npwa_Cp25-10_orders}), which shows 
that, on average, our DoBs accurately assess the error of the order-by-order results
when compared to NPWA data.


\section{Other NN scattering observables}  \label{sec:other_observables}

In this section we extend our analysis to other NN scattering observables.
For convenience we have collected in the Appendix the relevant notation
and formulas we have used, as well as a brief comparison to other notations
in the literature.
We focus on the differential cross section and
a set of the most commonly considered spin observables,
namely the analyzing power $A_y$, polarization transfer coefficients
$A$ and $D$, and the spin correlation parameters $A_{xx}$ and $A_{yy}$.
Each observable has been generated from LO through N$^4$LO,
primarily using the $R= 0.9\,$fm potential, which we have seen demonstrates
the best convergence pattern.
We consider the observables both at fixed energy as a 
function of angle and at fixed angle as a function of energy.

\begin{figure}[tbh!]
  \centering
  \includegraphics[width=.483\textwidth]{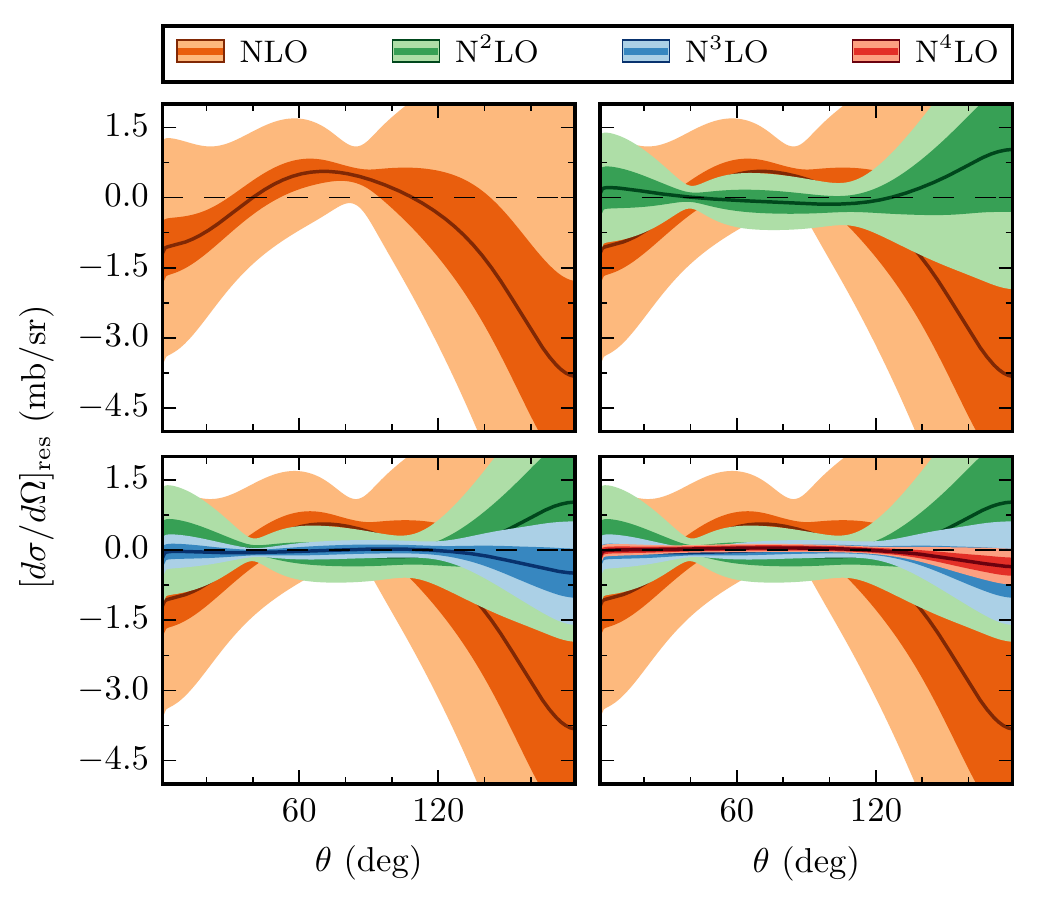}
  \vspace*{-.1in}
  \caption{Residuals defined in Eq.~\eqref{eq:residual_def} at each order for $[\protect\dsigma*]_{\text{res}}$ vs \cms\ angle $\theta$ with $R=0.9$\,fm, $\Elab = 96$\,MeV, and error bands generated using $\Cf{0.25}{10}$.
  }
  \label{fig:residual_diff_cross_section_R0p9_E96MeV}
\end{figure}

In Fig.~\ref{fig:residual_diff_cross_section_R0p9_E96MeV}, residuals for the differential cross section 
as a function of scattering angle at fixed $\Elab = 96\,$MeV 
for the $R=0.9\,$fm potential are shown as a characteristic example of this observable;
other energies display similar characteristics.
The detailed order-by-order convergence pattern does not seem to depend on angle,
suggesting that it is plausible to describe the convergence statistically.  

\begin{figure}[tbh!]
  \includegraphics[width=.40\textwidth]{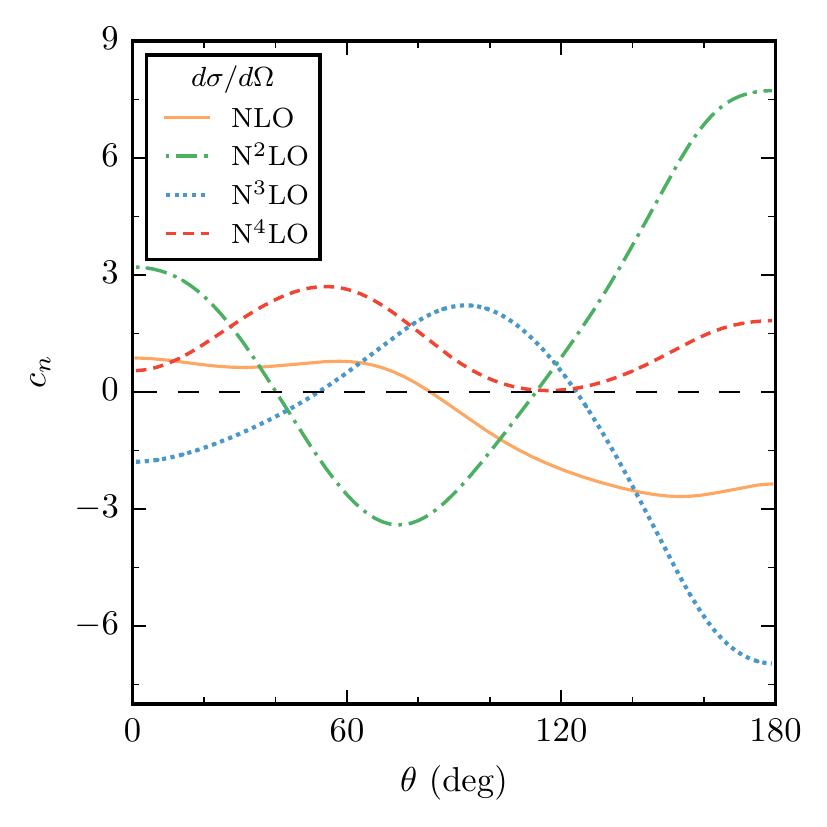}
  \vspace*{-.1in}
  \caption{Dimensionless coefficients as in Eq.~\eqref{eq:obsexp}
  extracted at each $\theta$ from the differential cross section at $\Elab=96$\,MeV\@.
  $\Xref$ is chosen to be $X_0$, the leading order result, which is consistent with natural coefficients.}
  \label{fig:diff_cross_section_R0p9_E96MeV_coeff}
\end{figure}

These observations are supported by the plot of coefficients as a function of angle, 
shown in Fig.~\ref{fig:diff_cross_section_R0p9_E96MeV_coeff}, for which
each order takes a turn at being the largest in magnitude. 
For this extraction, the leading-order result $X_0$ was taken for $\Xref$; the results
are not sensitive to this choice.
The scale of the dimensionless coefficients is roughly uniform at angles
less than 150$^\circ$ (about four), which is several times larger than the
scale for the integrated cross section at this energy.
The scale is significantly larger at back angles, where the momentum transfer
becomes twice the relative nucleon momentum, which may require a reexamination
of the expansion in this region.  
However, our overall conclusion is that the naturalness assumption, 
in the form of a characteristic size for the coefficient variations, still holds without 
integrating over all angles.


\begin{figure*}[p]
\centering
\subfloat[\label{fig:observable_coeffs_X0_R0p9_Elab_250_MeV}]{%
\includegraphics[width=\textwidth]{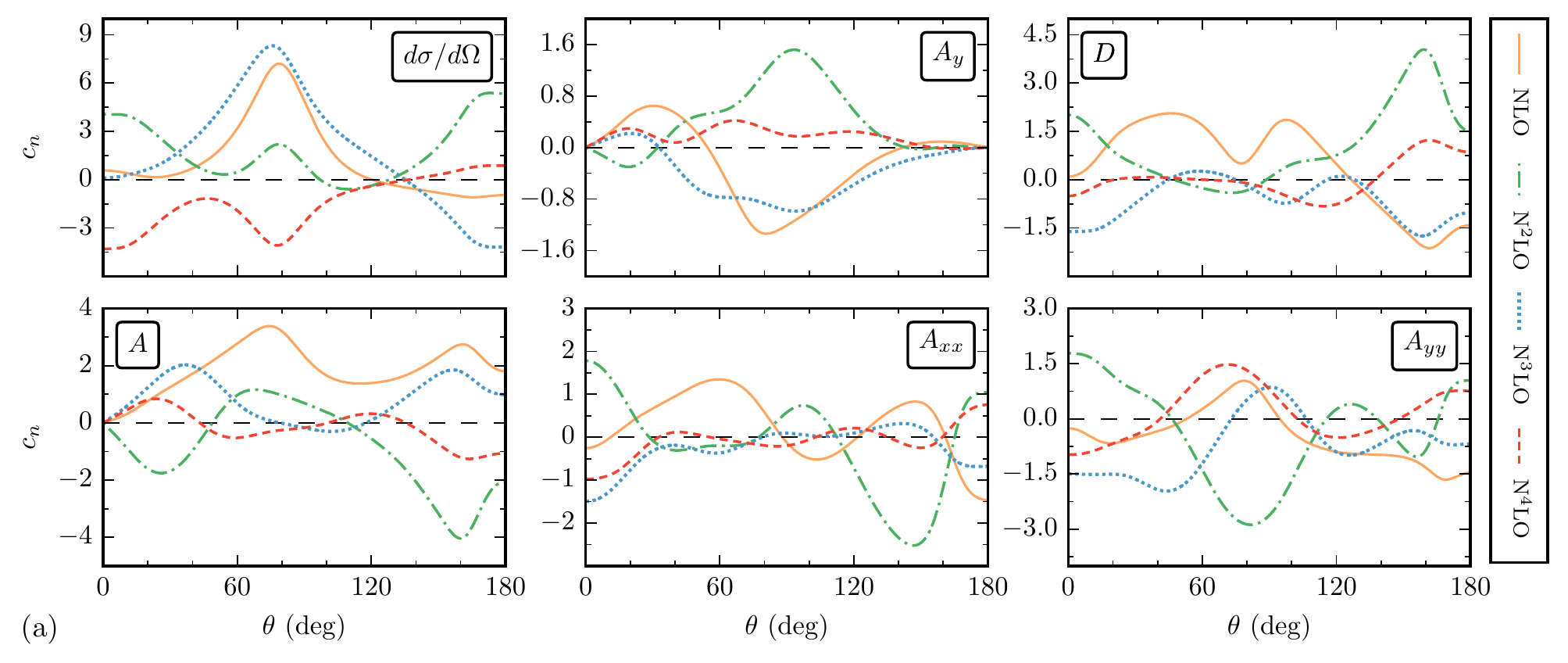}
}
\vspace*{-.42in}\\
\subfloat[\label{fig:spin_observable_residuals_X0_R0p9_Elab_250_MeV}]{%
\includegraphics[width=\textwidth]{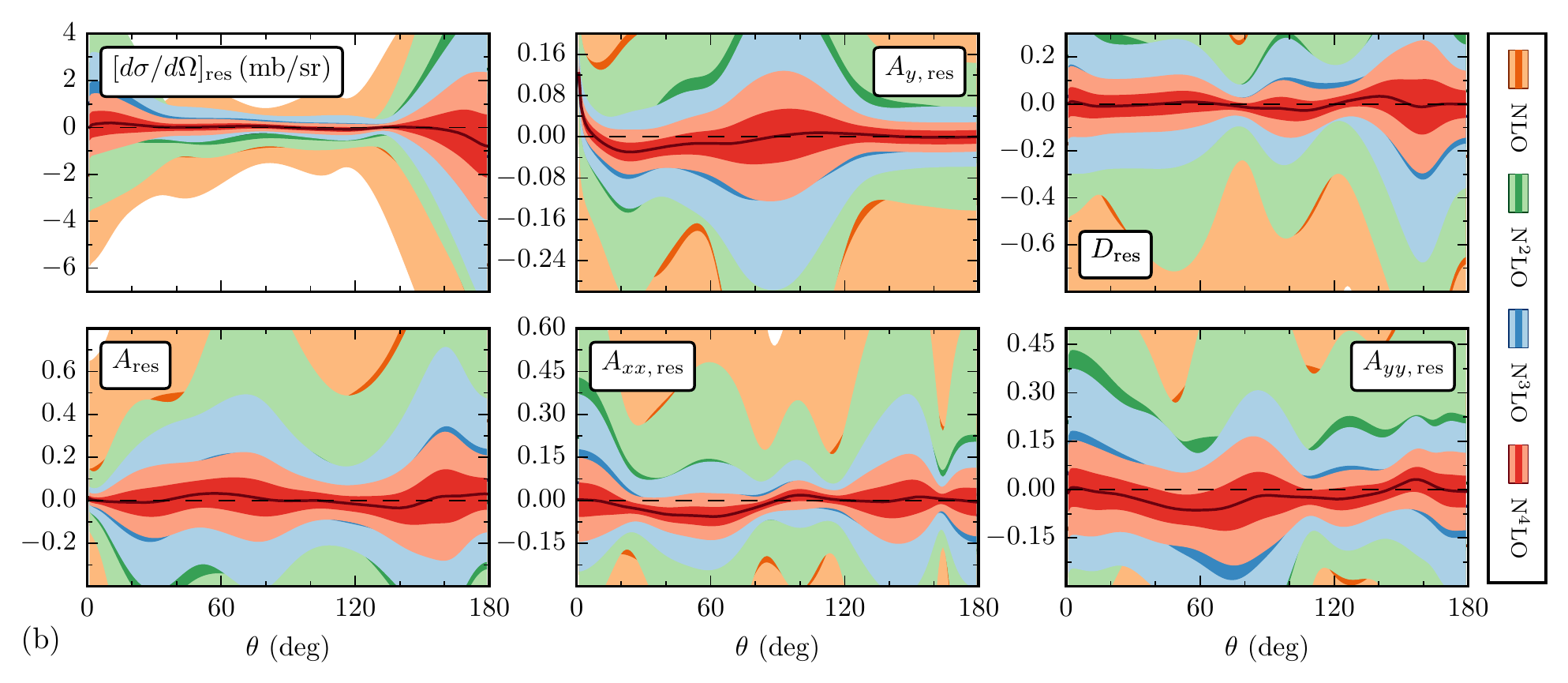}
}
\vspace*{-.42in}\\
\subfloat[\label{fig:observables_coeffs_X0_R0p9_theta_120}]{%
\includegraphics[width=\textwidth]{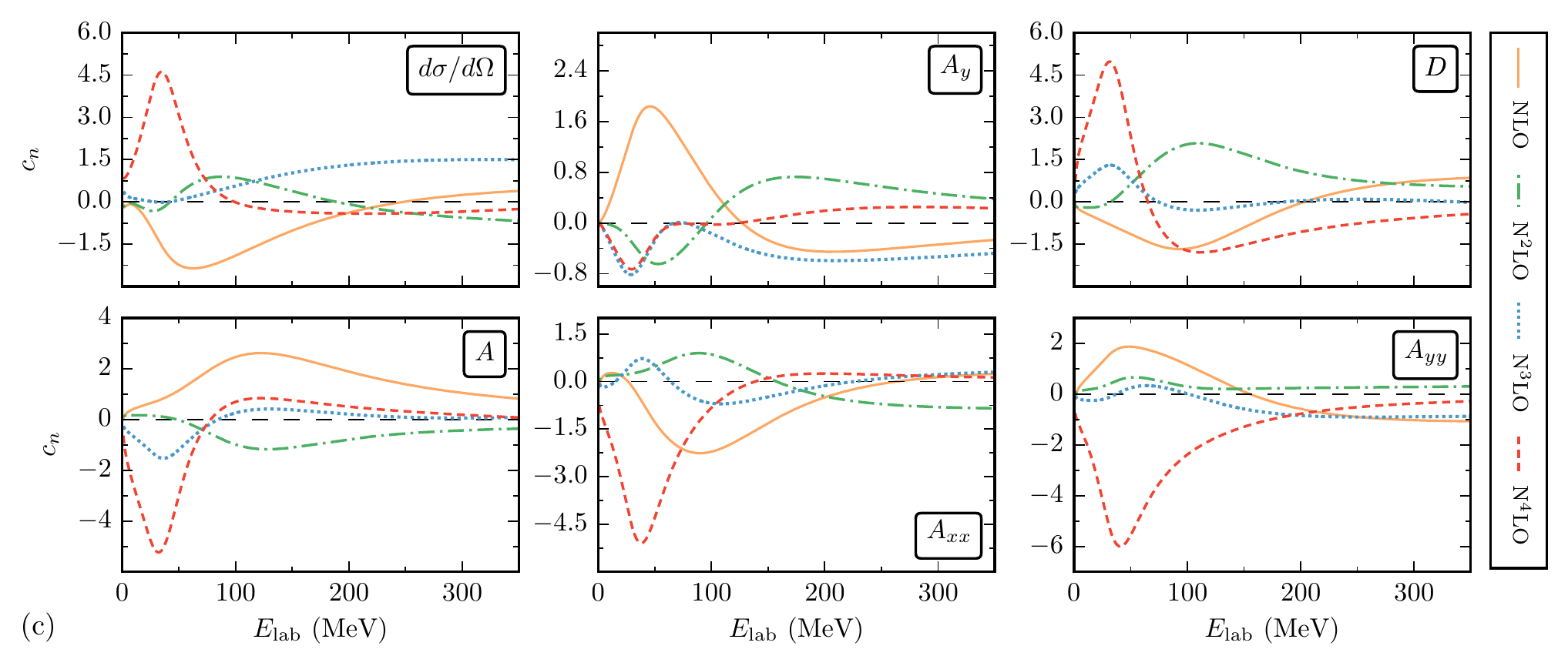}
}
\vspace*{-.23in}
\caption{(a) Dimensionless coefficients as in Eqs.~\eqref{eq:obsexp} and~\eqref{eq:spin_observable_coeffs}, and (b) residuals defined in Eq.~\eqref{eq:residual_def} as a function of $\theta$ at $\Elab = 250$\,MeV.
(c) Dimensionless coefficients
as a function of $\Elab$ at $\theta=120^\circ$.
All use the $R=0.9\,$fm EKM potentials.
$\Xref = X_0$ for the differential cross section and $\Xref=1$ for the spin observables.}
\end{figure*}


For any given spin observable $X_{pqik}$, we assume a natural expansion for 
the full quantity
\begin{align}
  \dsigma X_{pqik} = \Xref \sum_{n=0}^{\infty} c_n Q^n \;,
  \label{eq:spin_observable_full}
\end{align}
which is the probability for a particle to scatter into a solid angle $\dd{\Omega}$, given that the beam and target particles are polarized in the $i$ and $k$ directions and the scattered and recoil particle spins are in the $p$ and $q$ directions, respectively.
The dimensionful scale in Eq.~\eqref{eq:spin_observable_full} is set by the size 
of the differential cross section,
so the natural choice is $\Xref = \dsigma*$.
That means that the expansion for the spin observable itself is
\begin{align}
  X_{pqik} = \sum_{n=0}^{\infty} c_n Q^n \;,
  \label{eq:spin_observable_coeffs}
\end{align}
with no additional prefactor.
Below and in the Supplemental Material~\cite{supplmaterial}
we see that this scaling is consistent with natural ranges for the $c_n$s.

In Fig.~\ref{fig:observable_coeffs_X0_R0p9_Elab_250_MeV} we show the
extracted coefficients as a function of scattering angle for six \npr\ scattering
observables at $\Elab = 250\,$MeV, calculated using the potential with $R=0.9\,$fm.
The corresponding DoB bands for the residuals at this energy, following the
same prescription as applied to the cross sections,
are shown in Fig~\ref{fig:spin_observable_residuals_X0_R0p9_Elab_250_MeV}.   
As already noted, the LO coefficient does not inform our model for truncation errors
and so is not shown and is not used for the truncation error posteriors.
These figures serve as a representative example; figures showing coefficients
and DoB bands for many additional energies are given in the
Supplemental Material~\cite{supplmaterial}.

Except for constraints on some observables at special angles
(e.g., $A_y$ at $\theta = 0^\circ,180^\circ$ and $A$ at $\theta=0^\circ$), the coefficients
truly look like independent bounded
random functions of the angle, which supports our proposition that a 
statistical treatment of their behavior is warranted.
As with the cross sections,
the DoB bands decrease in size systematically and the lower-order bands overlap
the higher-order bands. It also appears that, in general, it is not necessary to
look at observables integrated over all angles to see a natural EFT convergence pattern.
We may therefore apply our statistical model to estimate truncation uncertainties for
NN angular observables.
  
Figure~\ref{fig:observables_coeffs_X0_R0p9_theta_120} shows coefficients
at a fixed angle of $\theta = 120^\circ$ as a function of energy.
The corresponding DoB intervals are shown in the Supplemental Material~\cite{supplmaterial}.
Enhanced \NNNNLO\ coefficients in the double-expansion crossover region,
noted earlier for the total cross section, are visible in most of the observables.
If this region is omitted, 
the behavior of the coefficients
with energies varies with a $\cbar$ scale of about 2, with no other systematic
patterns apparent.

Thus the observed convergence patterns of NN scattering observables for the
EKM interaction with $R=0.9\,$fm, considered as functions
of energy or angle, satisfy the statistical model naturalness assumptions
implied by Fig.~\ref{fig:Bayesian_Network}.
The DoB intervals above 50\,MeV derived using prior set C exhibit reasonable patterns
(as do those using set A; see Supplemental Material~\cite{supplmaterial}), but do not by themselves validate 
the statistical model.
For that purpose we turn to Bayesian model checking to assess the statistical
consistency of all the EKM potentials as well as the sensitivity to the choice 
of prior sets from Table~\ref{tab:priors}.

\clearpage

\section{Model Checking} \label{sec:model_checking}

The predictiveness of our statistical model for EFT truncation errors relies on 
how well our implementation of naturalness aligns with the true convergence pattern 
exhibited by the EFT.
An EFT could fail to exhibit a natural convergence pattern because of regulator artifacts 
or a poorly chosen $\Lambda_b$.
Our prior sets for $\pr(\cbar)$ and $\pr(c_n|\cbar)$, which encode our assumptions
about the
size of the higher-order coefficients, may also be called into question.

The efficacy of our approach for any given EFT or particular observables predicted
by that EFT can be examined using Bayesian model checking~\cite{gelman2013bayesian}.
Here we make use of consistency checks to determine if the DoB intervals behave as advertised.
We also investigate the possibility of determining $\Lambda_b$ solely from the convergence
pattern and the assumption of naturalness.

\subsection{Consistency checks}\label{subsec:consistency}

Once a posterior pdf for $\Delta_k$ is determined via Eq.~\eqref{eq:BayesEpsFull}, 
the probability that the truncation error is in a DoB interval follows directly
from Eq.~\eqref{eq:integralford}.
If our statistical model for the error is valid, a $\ppercent$ DoB interval should 
on average contain the actual next order value of the observable $\ppercent$ of the time
(we use the first-omitted-term approximation in this section).
By applying this test for a range of $p$ values to a sufficiently large set of observables,
we can test for inaccurate models or EFTs with irregular convergence patterns.
Such a consistency check provides us with the statistical toolset to analyze the
sensitivity to our choice of priors and the consistency of the
breakdown scale $\Lambda_b$ taken from Ref.~\cite{Epelbaum:2014efa}.

The procedure for creating consistency plots\footnote{Calibration plots or curves are other common names for such tools.}
to implement model checking is as follows~\cite{Furnstahl:2015rha}:
\be
  \I Choose a set of \emph{independent}
     observables for which the next-order calculation is available (not including LO).%
     \footnote{In Ref.~\cite{Furnstahl:2015rha} the LO to NLO success rates were included as part of the 
     consistency checks.  Because we want to test the convergence pattern only, the LO to NLO success 
     rate is not relevant here, as in the previous sections where we omit $c_0$.}

  \I Select a grid of $\ppercent$ DoB intervals with $p$ ranging from 0 to 1.

  \I Compute the $\ppercent$ DoB interval for each observable in the set, using the
    same priors throughout.

  \I For each next-order calculation that is within the DoB
  interval of the previous order, count one success.

  \I Take the number of successes $n$ and divide by the total number
  of observables $N$ to get the actual success rate.

  \I Plot the success rates versus DoB interval percentage 
  and compare to the ideal result
   given by a $45^\circ$ line.
\ee
Because we will have a finite number $N$ of observables, we expect fluctuations
away from the ideal result for a true $\ppercent$ success rate, as given by the binomial
posterior
\beq
   \pr(n|p,N) = \frac{N!}{n!(N-n)!} p^n (1-p)^{N-n} \;.
   \label{eq:binomial}
\eeq
We apply Bayes theorem with a uniform prior on $p$ to convert to a posterior
for $p$:
\beq
  \pr(p|n,N) \propto \pr(n|p,N) \pr(p) \propto \pr(n|p,N)
  \;,
  \label{eq:p_binomial}
\eeq
and generalize Eq.~\eqref{eq:binomial} to continuous $n$ to calculate horizontal
68\% and 95\% confidence intervals for the DoB percentage, using the HPD prescription (see Sec.~\ref{sec:formulas}).
These become shaded bands in the consistency plots.%
\footnote{In Ref.~\cite{Furnstahl:2015rha} a different procedure 
yielded bands for given $N$, $n$, and $p$ that are reflected about the 45 degree
line from the ones here. Additionally, the bands in~\cite{Furnstahl:2015rha} 
were calculated
using equal-tailed credible intervals for DoBs rather than the HPD prescription.
Both procedures approach symmetric bands for large $N$.}

We can easily evaluate DoB intervals for choices of $\Lambda_b$
different from those identified by EKM, which we have adopted so far.
We follow Refs.~\cite{Bagnaschi:2014eva,Bagnaschi:2014wea,Furnstahl:2015rha} in
doing this by introducing a scaling factor $\lambda$ to
generalize Eq.~\eqref{eq:obsexp} as
\begin{align} \label{eq:lambda_expansion}
  X = \Xref \sum_{n=0}^\infty (c_n \lambda^n) \times \left( \frac{Q}{\lambda} \right)^n
  \;.
\end{align}
Varying $\lambda$ about unity shifts $\Lambda_b$; in the consistency plots here
we consider 20\% variations, namely $\lambda = 0.8$ and $1.2$, with respect to the
EKM choice with $\lambda = 1.0$.

\begin{figure}[tbh!]
  \centering
  \includegraphics[width=.8\columnwidth]{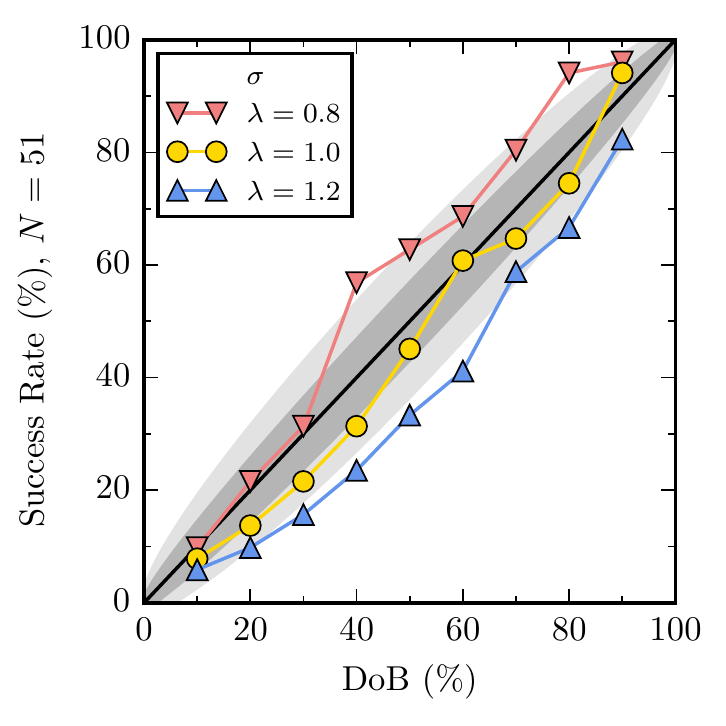}
  \caption{Consistency plot for the total cross section using the $R=0.9$\,fm EKM potential evaluated at $E=20,40,\ldots,340$\,MeV\@.
  Results were obtained using prior set $\Cfone{0.25}{10}$ and are averaged over \NLO, \NNLO, and \NNNLO.
  The shaded bands represent 68\% and 95\% confidence
  intervals for the success rates (see text).}
  \label{fig:consistency_sigma_combined_lambda_dependence_Cp25_E_96-300}
\end{figure}

\begin{figure}[tbh!]
  \centering
  \includegraphics[width=.8\columnwidth]{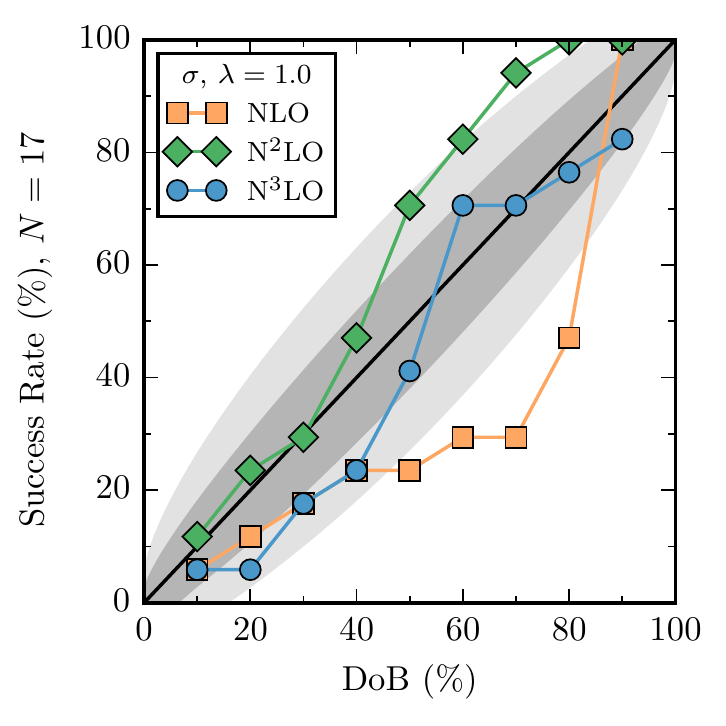}
  \caption{Consistency plot for the total cross section using the $R=0.9$\,fm EKM potential at the recommended $\Lambda_b=600$\,MeV ($\lambda=1$) and separated order-by-order.
  The DoBs were generated using $\Cfone{0.25}{10}$ applied at energies $\Elab=20,40,\ldots,340$\,MeV.}
  \label{fig:consistency_prior_dependence_Cp25}
\end{figure}

\begin{figure}[tb]
  \centering
  \includegraphics[width=.8\columnwidth]{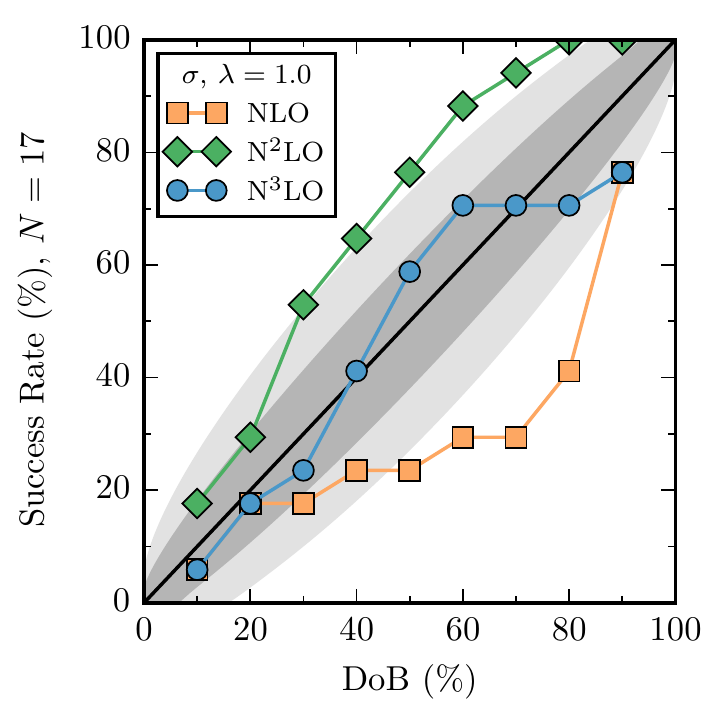}
  \caption{Consistency plot as in Fig.~\ref{fig:consistency_prior_dependence_Cp25} but generated using $\Aepsone$.}
  \label{fig:consistency_prior_dependence_Aeps}
\end{figure}

\begin{figure}[tb]
  \centering
  \includegraphics[width=.8\columnwidth]{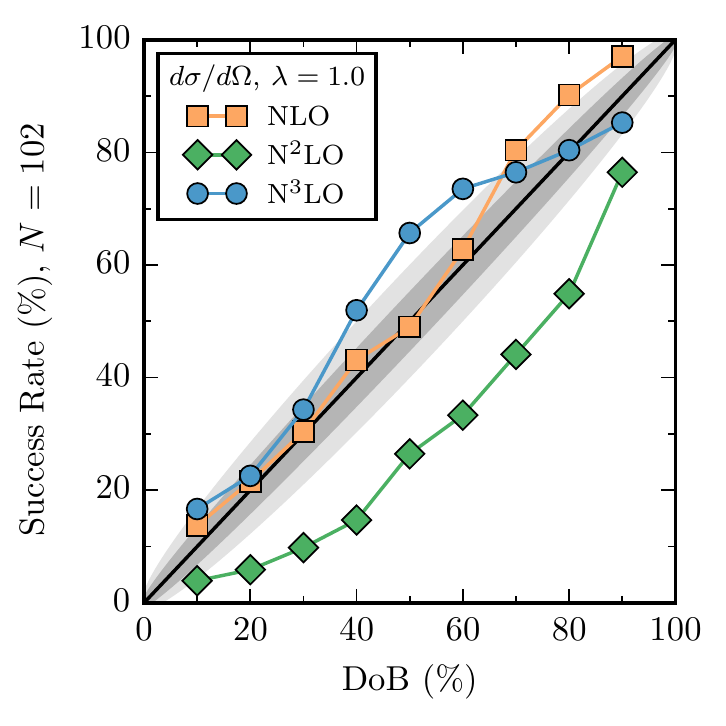}
  \caption{Consistency plot as in Fig.~\ref{fig:consistency_prior_dependence_Cp25} but using $\protect\dsigma*$.
  The observable is evaluated at $\Elab=20,40,\ldots,340$\,MeV and $\theta=40^\circ,60^\circ,\ldots,140^\circ$.}
  \label{fig:consistency_dsdO_order_dependence_Cp25-10_theta-range-40-141-20_energy-range-20-341-20}
\end{figure}

\begin{figure*}[p] 
  \centering
  \includegraphics[width=.34\textwidth]{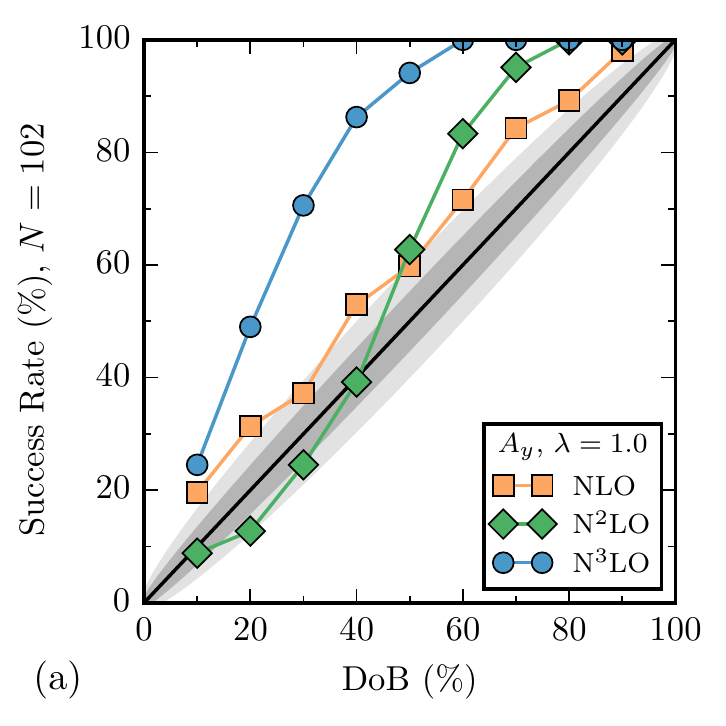}
  \hspace{-.15in}
  \includegraphics[width=.34\textwidth]{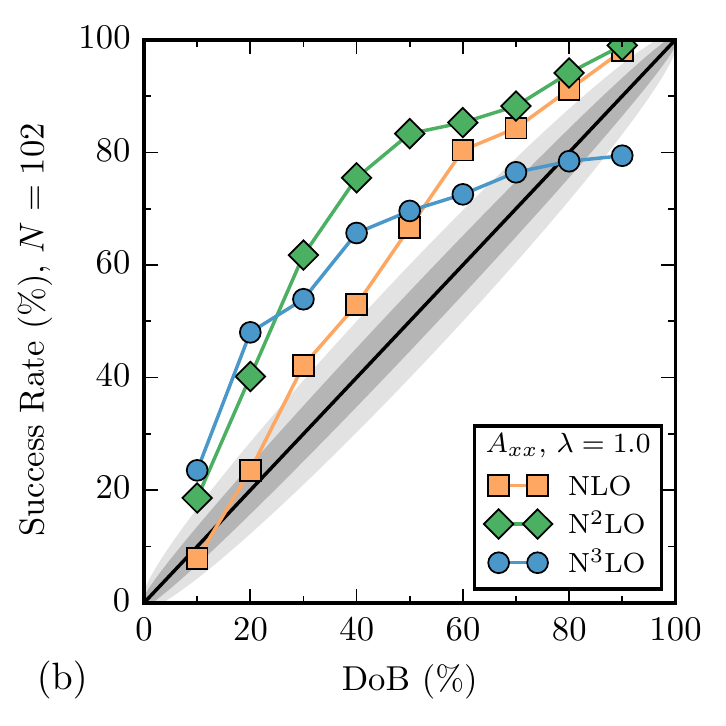}
  \hspace{-.15in}
  \includegraphics[width=.34\textwidth]{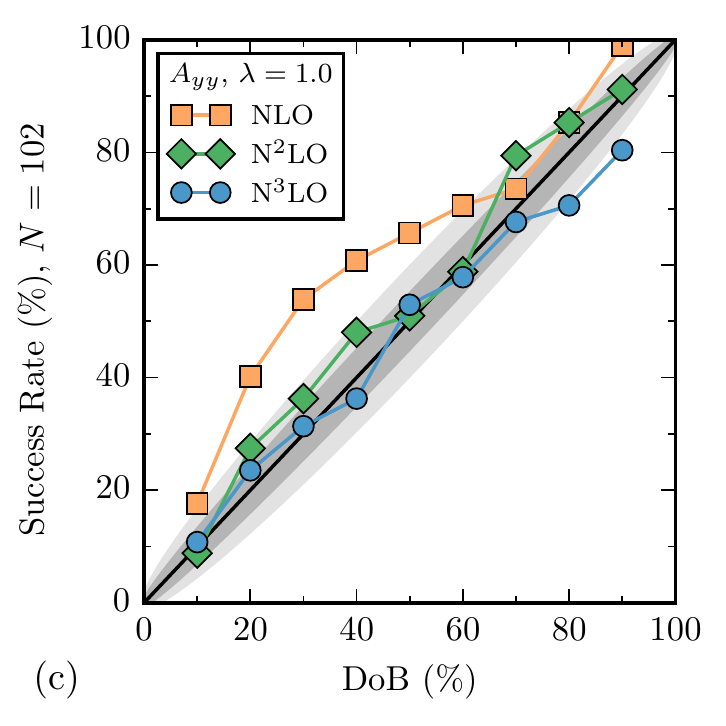}
  \caption{Consistency plots using $\Cfone{0.25}{10}$ for the individual spin observables (a) $A_y$, (b) $A_{xx}$, and (c) $A_{yy}$ with $R=0.9$\,fm and separated order-by-order.
  The observables are evaluated at $\Elab=20,40,\ldots,340$\,MeV and $\theta=40^\circ,60^\circ,\ldots,140^\circ$.}
  \label{fig:consistency_individual_spin_obs_Cp25_10_theta-40-140_energy-20-340}
\end{figure*}

\begin{figure*}[p]
  \centering
  \includegraphics[width=.34\textwidth]{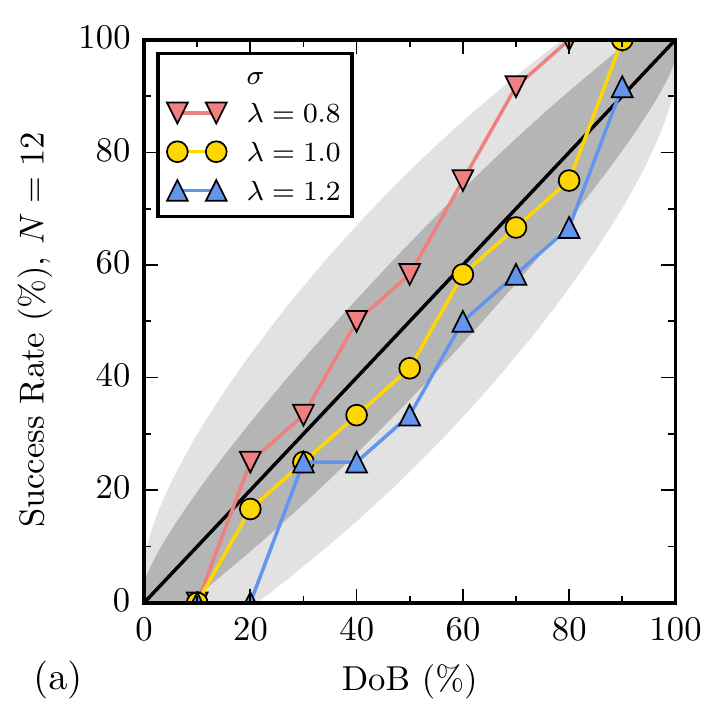}
  \hspace{-.15in}
  \includegraphics[width=.34\textwidth]{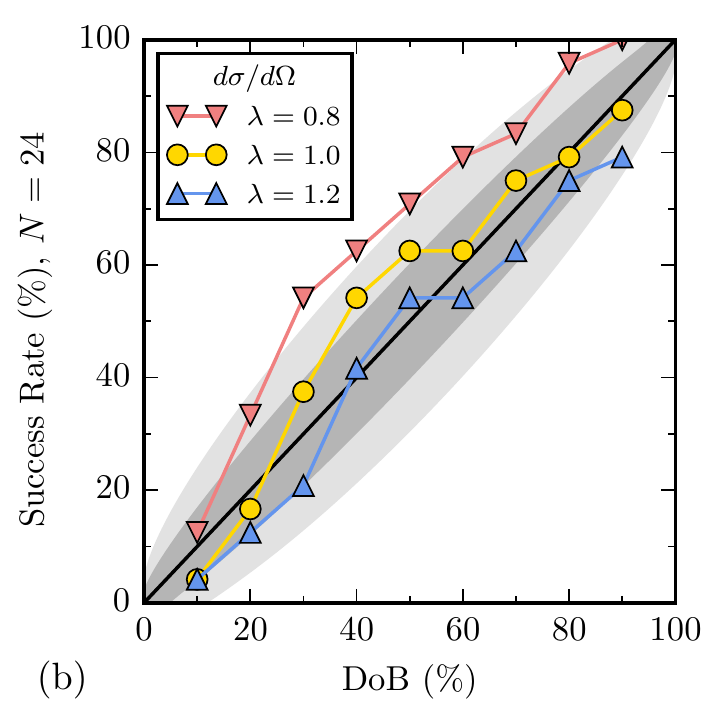}
  \hspace{-.15in}
  \includegraphics[width=.34\textwidth]{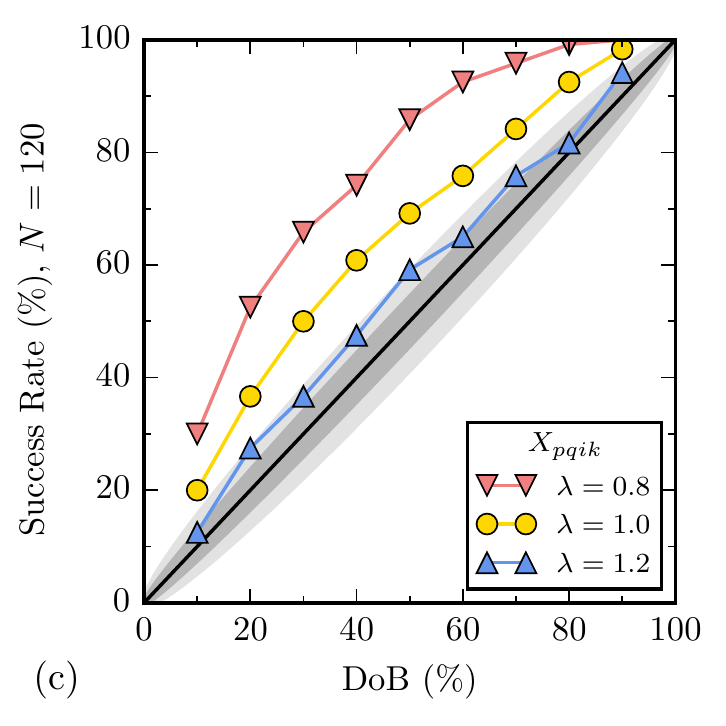}
  \caption{Consistency plots using $\Cfone{0.25}{10}$ averaged over \NLO--\NNNLO\ results for (a) $\sigma$, (b) $\protect\dsigma*$, and (c) the five selected spin observables with $R=0.9$\,fm.
  The observables are evaluated at $\Elab = 96,143,200,300$\,MeV and $\theta=60^\circ,120^\circ$ if applicable.}
  \label{fig:consistency_lambda_dependence_Cp25-10_R0p9_E_96-300}
\end{figure*}

\begin{figure*}[p]
  \centering
  \includegraphics[width=.34\textwidth]{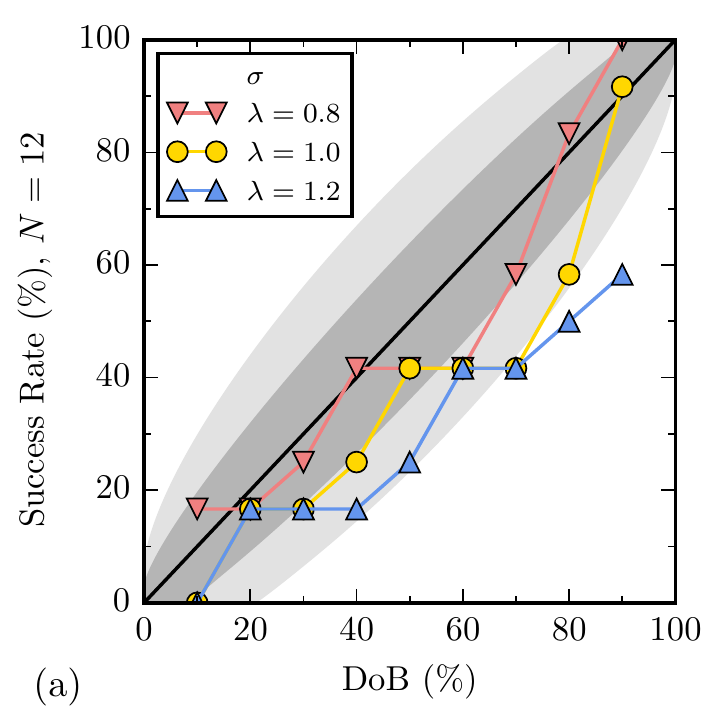}
  \hspace{-.15in}
  \includegraphics[width=.34\textwidth]{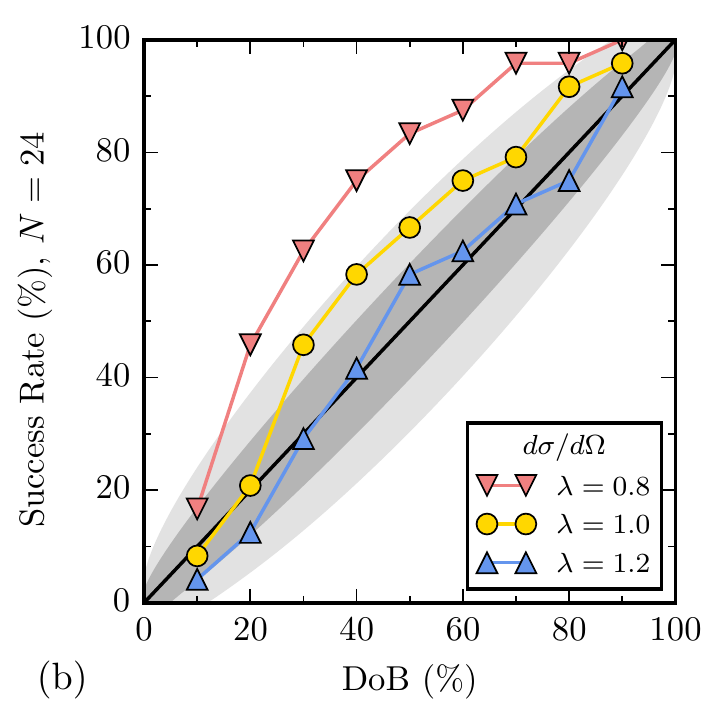}
  \hspace{-.15in}
  \includegraphics[width=.34\textwidth]{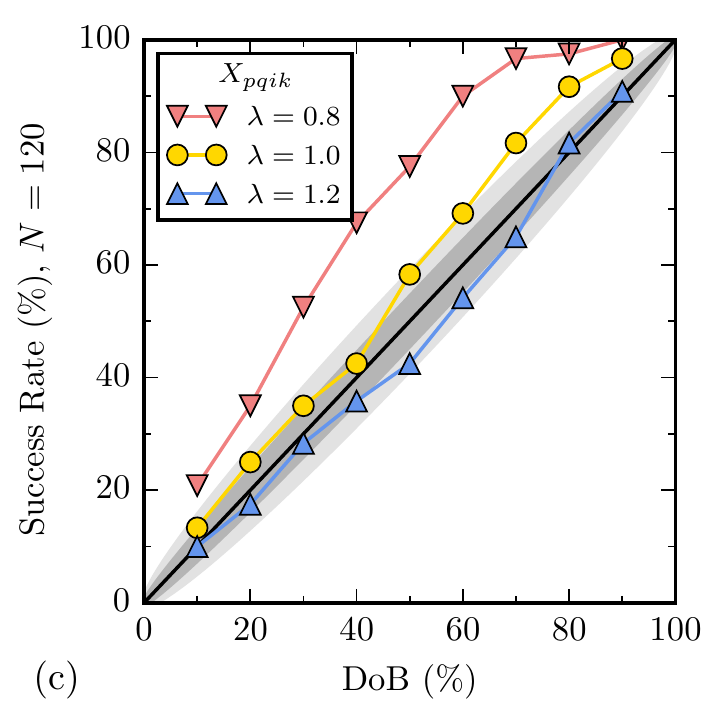}
  \caption{Consistency plots as in Fig.~\ref{fig:consistency_lambda_dependence_Cp25-10_R0p9_E_96-300}, but with $R=0.8$\,fm.}
  \label{fig:consistency_lambda_dependence_Cp25-10_R0p8_E_96-300}
\end{figure*}

\begin{figure*}[p]
  \centering
  \includegraphics[width=.34\textwidth]{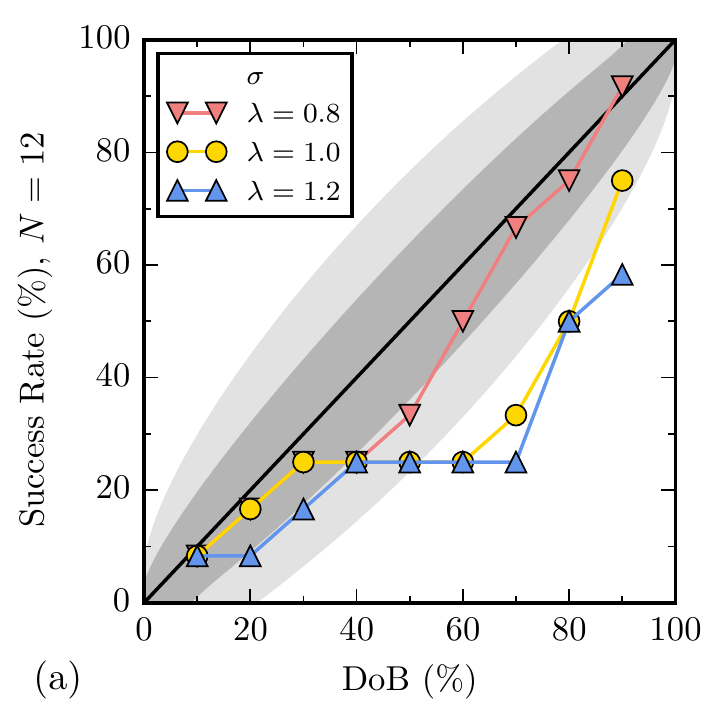}
  \hspace{-.15in}
  \includegraphics[width=.34\textwidth]{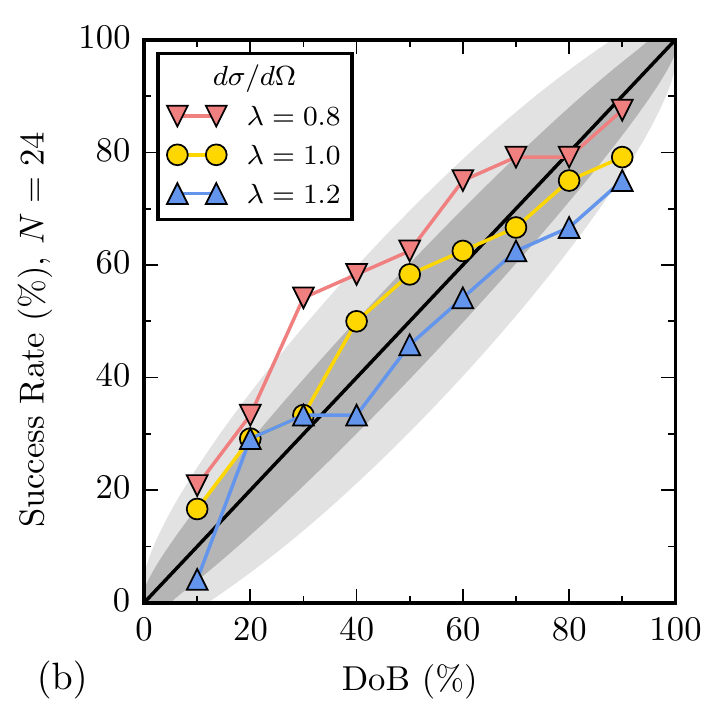}
  \hspace{-.15in}
  \includegraphics[width=.34\textwidth]{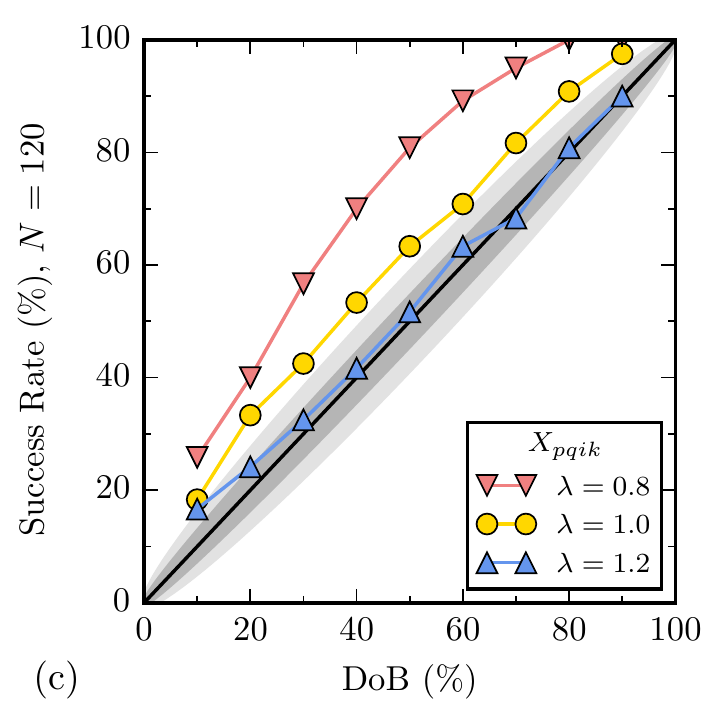}
  \caption{Consistency plots as in Fig.~\ref{fig:consistency_lambda_dependence_Cp25-10_R0p9_E_96-300}, but with $R=1.0$\,fm.}
  \label{fig:consistency_lambda_dependence_Cp25-10_R1p0_E_96-300}
\end{figure*}

\begin{figure*}[p]
  \centering
  \includegraphics[width=.34\textwidth]{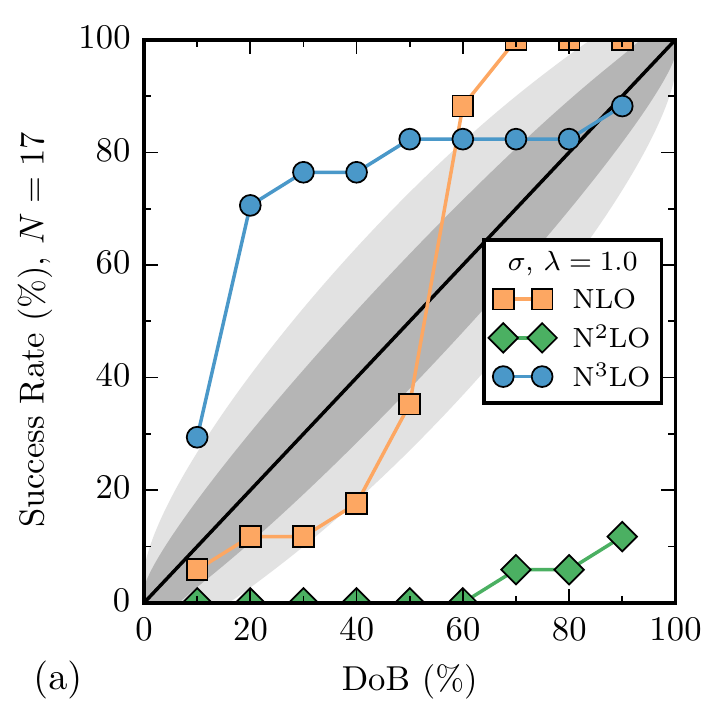}
  \hspace{-.15in}
  \includegraphics[width=.34\textwidth]{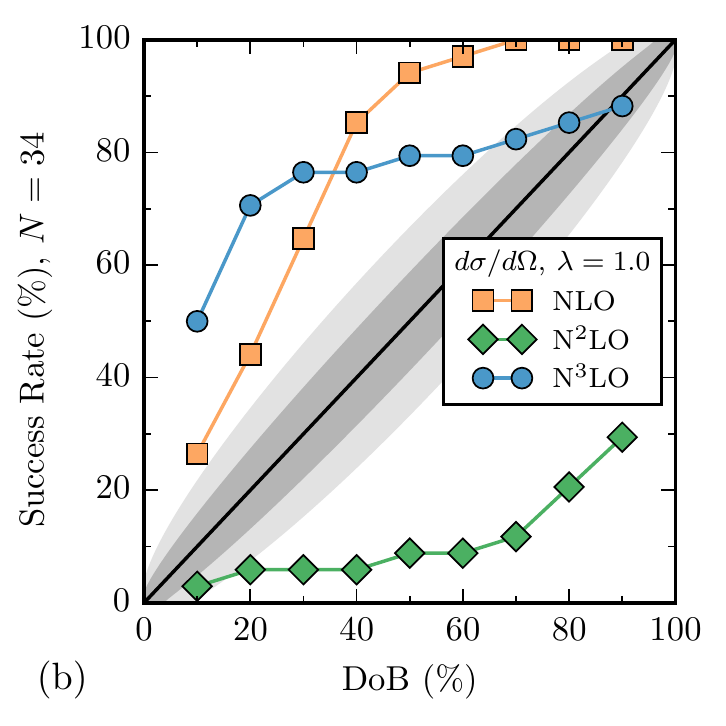}
  \hspace{-.15in}
  \includegraphics[width=.34\textwidth]{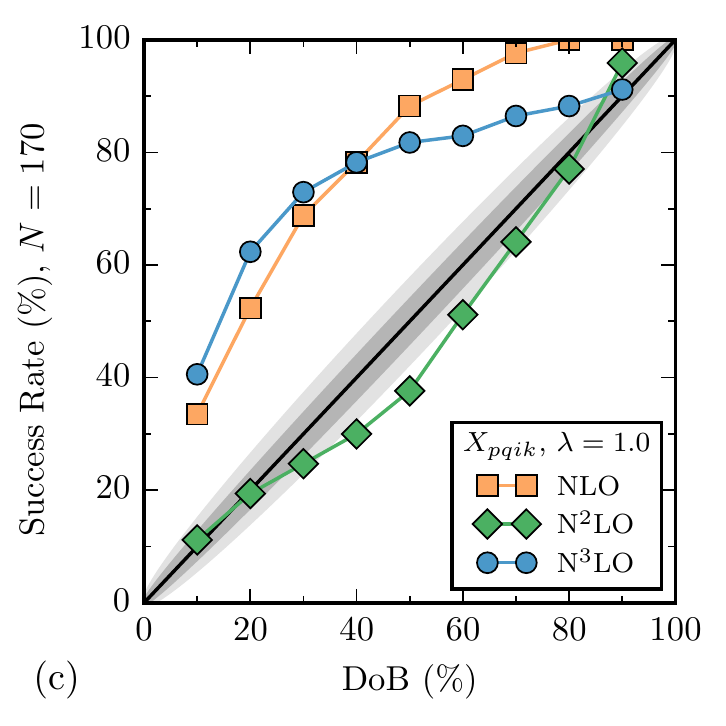}
  \caption{Consistency plots with $\Cfone{0.25}{10}$ and the $R=1.1$\,fm EKM potential showing order-by-order results for (a) $\sigma$, (b) $\protect\dsigma*$, and (c) the five considered spin observables evaluated at $\Elab = 20,40,\ldots,340$\,MeV and $\theta = 60^\circ, 120^\circ$ if applicable.}
  \label{fig:consistency_lambda_dependence_Cp25-10_R1p1_E_20-340_t_60_120}
\end{figure*}

\begin{figure*}[p]
  \centering
  \includegraphics[width=.34\textwidth]{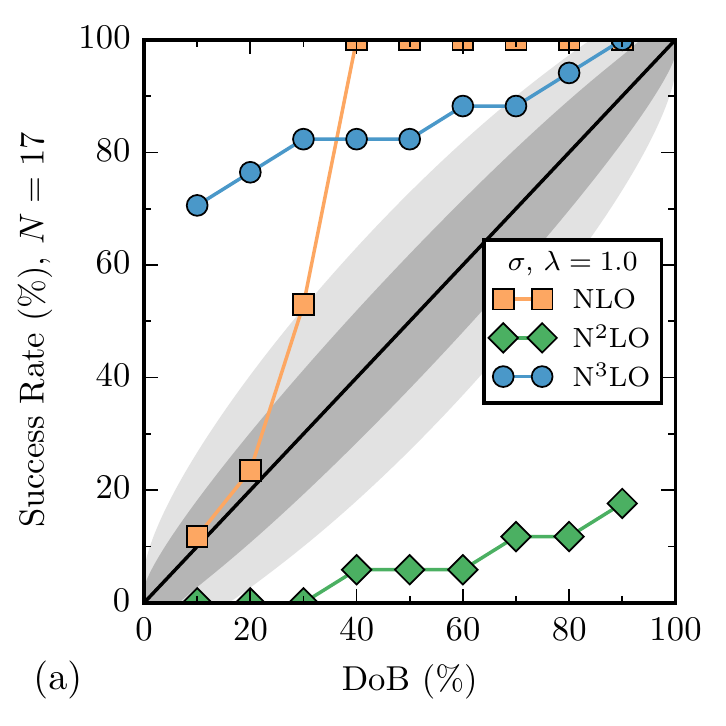}
  \hspace{-.15in}
  \includegraphics[width=.34\textwidth]{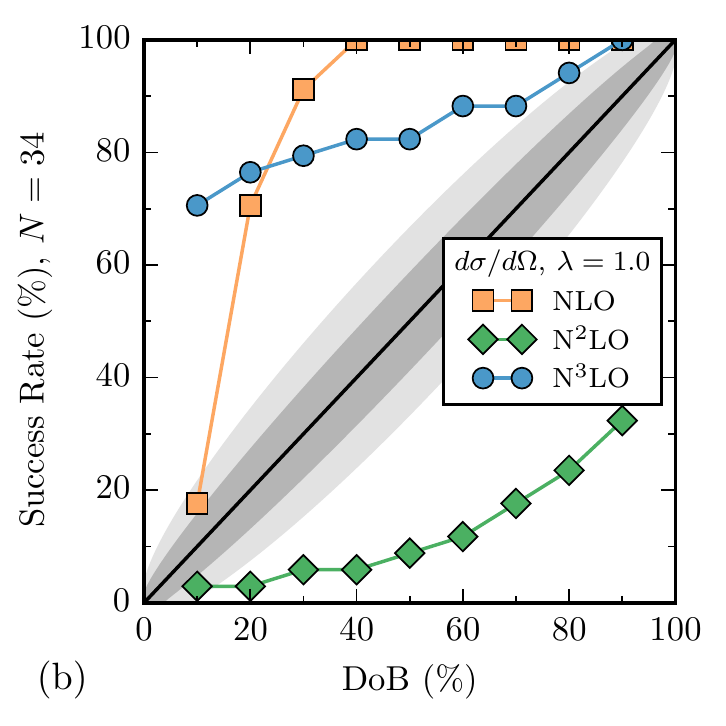}
  \hspace{-.15in}
  \includegraphics[width=.34\textwidth]{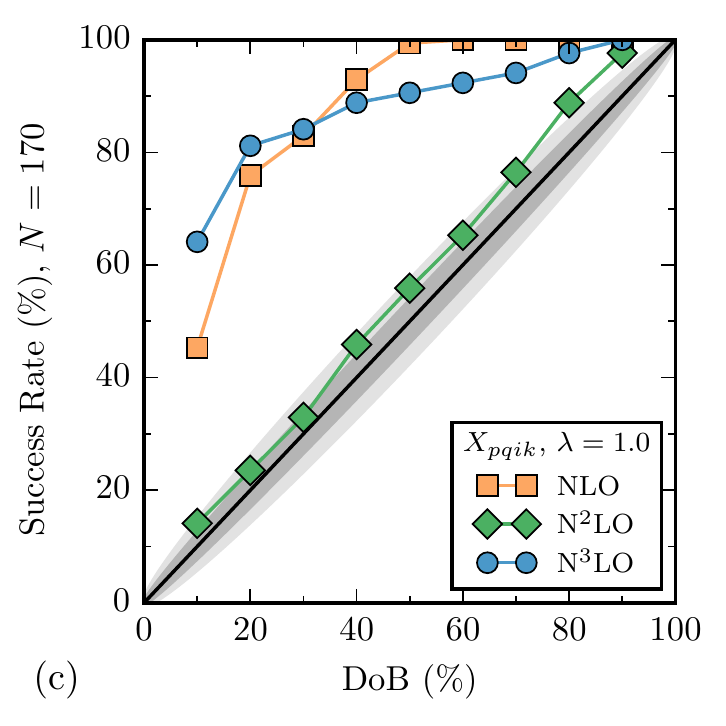}
  \caption{Consistency plots as in Fig.~\ref{fig:consistency_lambda_dependence_Cp25-10_R1p1_E_20-340_t_60_120}, but with $R=1.2$\,fm.}
  \label{fig:consistency_lambda_dependence_Cp25-10_R1p2_E_20-340_t_60_120}
\end{figure*}

The logic of the remainder of this section is as follows.
We begin with a reexamination of the consistency plots for the total cross section,
as begun in Figs.~10 and 11 from Ref.~\cite{Furnstahl:2015rha}, exploring more
energies and stability under prior choice (Figs.~\ref{fig:consistency_sigma_combined_lambda_dependence_Cp25_E_96-300}--\ref{fig:consistency_prior_dependence_Aeps}).
Next we extend the previous analysis with results from the differential cross section
and our selected spin observables (Figs.~\ref{fig:consistency_dsdO_order_dependence_Cp25-10_theta-range-40-141-20_energy-range-20-341-20} and~\ref{fig:consistency_individual_spin_obs_Cp25_10_theta-40-140_energy-20-340}).
Finally, consistency plots of EKM potentials with different regulators are examined
(Figs.~\ref{fig:consistency_lambda_dependence_Cp25-10_R0p9_E_96-300}--\ref{fig:consistency_lambda_dependence_Cp25-10_R1p0_E_96-300}), 
including examples of potentials that fail our analysis 
(Figs.~\ref{fig:consistency_lambda_dependence_Cp25-10_R1p1_E_20-340_t_60_120} and~\ref{fig:consistency_lambda_dependence_Cp25-10_R1p2_E_20-340_t_60_120}).
For a more extensive survey of our results, see the Supplemental Material~\cite{supplmaterial}.

In Fig.~\ref{fig:consistency_sigma_combined_lambda_dependence_Cp25_E_96-300}
we show consistency plots for the total cross section calculated with the $R=0.9\,$fm
EKM potential for prior set $\Cfone{0.25}{10}$
(recall that the superscript indicates that the truncated error
is assumed to be given by the first omitted term).
Here, each line averages over the success rate of the $\NLO$, $\NNLO$, and $\NNNLO$ error bands 
in predicting the corresponding next-order contributions at energies $\Elab=20,40,\ldots, 340$\,MeV\@.
The trends show that for $\lambda = 1$, i.e.\ $\Lambda_b=600$\,MeV,
the predicted $\ppercent$ DoB aligns with the measured success rate to within
the uncertainty predicted by Eq.~\eqref{eq:p_binomial}.
While using data at many $\Elab$ values
improves the statistics, the independence of the results may
be questionable if calculated for too closely spaced kinematic variables.
Dependent measurements would cause the gray error bands in Fig.~\ref{fig:consistency_sigma_combined_lambda_dependence_Cp25_E_96-300}
to be too restrictive, so the  $\lambda=0.8$ and $\lambda=1.2$ lines may be consistent even 
though they are generally outside the 68\% bands (cf.~the leftmost plot in Fig.~\ref{fig:consistency_lambda_dependence_Cp25-10_R0p9_E_96-300}).
In future work we will model the correlation length in energy using 
Gaussian processes~\cite{rasmussen2006gaussian,Gelman03,Jones1998} (GPs) to draw more robust conclusions about independence.

In Fig.~\ref{fig:consistency_prior_dependence_Cp25} we decompose the $\lambda=1$
line of Fig.~\ref{fig:consistency_sigma_combined_lambda_dependence_Cp25_E_96-300}
into the contribution from each individual order, while
Fig.~\ref{fig:consistency_prior_dependence_Aeps} shows the same decomposition
but using prior set $\Aepsone$.
Given the slight changes between
Figs.~\ref{fig:consistency_prior_dependence_Cp25}
and~\ref{fig:consistency_prior_dependence_Aeps}, as well as similar examples not shown,
we conclude that prior choice has little effect on the predictions of EFTs
with good convergence patterns.
For such an EFT, we expect the predictions to improve with the
order of the prediction, because the higher orders contain more
information about the pattern of the observable coefficients.
This is what we see, with the \NNNLO\ predictions being fully consistent within the
gray bands.

Next we consider angle-dependent observables, which were not analyzed in Ref.~\cite{Furnstahl:2015rha}.
Each observable is generated using the $R=0.9$\,fm EKM potential with $\Lambda_b=600$\,MeV ($\lambda = 1$).
Each is evaluated at $N=102$ kinematic points: 17 energies ($20,40,\ldots,340$\,MeV, as for $\sigma$) with 6 angles ($40^\circ,60^\circ,\ldots,140^\circ$) for each energy.
The prior set used in the analysis is $\Cfone{0.25}{10}$ and the plots are decomposed order-by-order.
Figure~\ref{fig:consistency_dsdO_order_dependence_Cp25-10_theta-range-40-141-20_energy-range-20-341-20} shows a consistency plot for the differential cross section, while Fig.~\ref{fig:consistency_individual_spin_obs_Cp25_10_theta-40-140_energy-20-340} shows, as an example, the individual spin observables $A_y$, $A_{xx}$, and $A_{yy}$.
The \NNLO\ DoBs consistently underestimate the size of the \NNNLO\ correction for the differential cross section, but the \NLO\ and \NNNLO\ DoBs do fairly well.
Some of the DoBs for spin observables, such as $A_y$ and $A_{xx}$, overestimate the higher-order
corrections, while the $A_{yy}$ DoB performs well.
The \NLO\ and \NNLO\ coefficients of $A_y$ and $A_{xx}$ are generally larger than \NNNLO\ and particularly the \NNNNLO\ coefficients at $\Elab\gtrsim 100$\,MeV, while $A_{yy}$ tends to have coefficients that each take turns being the largest.

The spacing in angle and energy may be close enough that the calculations used
for the consistency plots are significantly correlated, which will constrain the gray
error bands unnecessarily due to the large number of non-independent
points.
The true impact of this correlation has not yet been quantified and is a topic for future investigation.
From the aforementioned plots, we can conclude that although integrating over angles
is not necessarily required to ensure a natural convergence pattern of coefficients, some observables do show notable patterns that adversely affect the predictive power of their respective DoBs.

Finally, we return to the topic of EKM potentials with varying regulators, first raised in Sec.~\ref{sec:total_sigma}.
Thus far we have mainly focused on the $R=0.9$\,fm EKM potential due to its
natural convergence pattern compared to the other potentials, as evidenced by
Figs.~\ref{fig:cross_sections_EKM_R0p9_R0p9_Lambdab_600_600_Xzero_coeffs}--\ref{fig:cross_sections_EKM_R1p1_R1p2_Lambdab_500_400_Xzero_coeffs}.
Now we relax this focus to gain insight into the effects that regulator choices and their
consequent convergence patterns have on the reliability of the error bands generated by this analysis.
We also test the proposed breakdown scale $\Lambda_b$ for each regulator by varying $\lambda$
defined in Eq.~\eqref{eq:lambda_expansion} about unity.
In an attempt to ensure independent results for the chosen kinematic points,
in Figs.~\ref{fig:consistency_lambda_dependence_Cp25-10_R0p9_E_96-300}--\ref{fig:consistency_lambda_dependence_Cp25-10_R1p0_E_96-300} we use $\Elab = 96,143,200,300$\,MeV and $\theta=60^\circ,120^\circ$ (if applicable).
The choices of separation length in $\Elab$ and $\theta$ are based on a rough analysis of the coefficient curves,
which suggests that energies spaced by about $70$--$80$\,MeV
and $\theta$ spaced by $30^\circ$--$40^\circ$ can be taken as independent for
evaluating DoB successes.

We find in Figs.~\ref{fig:consistency_lambda_dependence_Cp25-10_R0p9_E_96-300}--\ref{fig:consistency_lambda_dependence_Cp25-10_R1p0_E_96-300}, which show results averaged over orders for
$R = 0.9\,$fm, $0.8\,$fm, and $1.0\,$fm, respectively, that our statistical model for truncation errors
is generally successful for these parameters.
For $R=0.9\,$fm, both $\sigma$ and $\dsigma*$ show strong consistency with $\lambda = 1$,  meaning
$\Lambda_b \approx 600$\,MeV, but a wider range of $\Lambda_b$ is not ruled out.  In contrast,
the spin observables are more consistent with somewhat larger $\Lambda_b$, particularly if we
accept the limits of the gray bands. 
The three sets of observables for $R=0.8\,$fm remain fairly consistent with a
single choice for $\Lambda_b$ and overall this potential passes the
test of a natural convergence pattern based on the expected level of consistency.
Although the order-averaged consistency plots for the $R=1.0$\,fm potential
are reasonable, the order-by-order convergence pattern and plausibility of a
single $\Lambda_b$ become suspect;
see the Supplemental Material~\cite{supplmaterial} for more information.

The failure of our statistical model for truncation errors
when applied to $R=1.1$\,fm and $R=1.2$\,fm, which was anticipated by the pattern of
coefficients in Fig.~\ref{fig:cross_sections_EKM_R1p1_R1p2_Lambdab_500_400_Xzero_coeffs},
is best observed in the order-by-order consistency plots, where the impact of fluctuations 
in coefficient size becomes clear.
For better statistics (larger $N$), we use $\theta=60^\circ,120^\circ$ and $\Elab=20,40,\ldots,340$\,MeV\@; since the chosen angles are fairly representative, using more angles does not greatly affect the conclusions.
Figures~\ref{fig:consistency_lambda_dependence_Cp25-10_R1p1_E_20-340_t_60_120}
and~\ref{fig:consistency_lambda_dependence_Cp25-10_R1p2_E_20-340_t_60_120}
explicitly show the unequal nature of the coefficient magnitude for these regulator values.
Because much of the physics content at \NNLO\ and \NNNNLO\ is moved to \NLO\ and \NNNLO, the \NNLO\ 
DoBs tend to underestimate the contribution due to \NNNLO, while the \NLO\ and \NNNLO\ DoBs 
overshoot the error estimates due to \NNLO\ and \NNNNLO, respectively.
Because the trade-off of large and small coefficients causes 
the error bands to be overestimated and then underestimated at alternating orders
in the expansion, this effect could on average cancel out when comparing to the actual 
data, as that comparison highlights the size of \emph{all} left-out higher-order terms.

\clearpage

\subsection{Posterior for \texorpdfstring{$\Lambda_b$}{Lambdab}} \label{sec:posterior}

So far we have assumed that the EFT breakdown scale $\Lambda_b$ was a given quantity,
and then calculated posteriors for EFT truncation errors 
contingent on the known coefficients $c_n$.
We have also checked whether this posterior is statistically consistent with particular
fixed choices for $\Lambda_b$.  
Here we explore whether we can extract a plausible range for $\Lambda_b$ by
calculating a posterior pdf for $\Lambda_b$, contingent only on the order-by-order
results.  
We combine results from
different momenta and angles that are far enough apart that it is reasonable to
assume the EFT calculations are uncorrelated, but also compare to much more closely
spaced kinematics to improve the statistics.
The eventual goal is to be able to use modeled correlations
between observable calculations to calculate $\Lambda_b$ based on the
calculations at many different momenta $p$ and angles.

\begin{figure}[tb]
  \centering
  \includegraphics{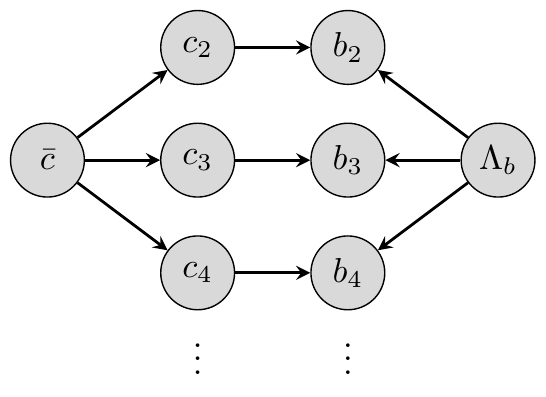}
  \caption{A Bayesian network that outlines the causal relationships between random variables
  when determining $\Lambda_b$.
  For simplicity, only nodes for one $\kinparvec_i$ are shown.}
  \label{fig:Lambdab_BN}
\end{figure}

We first rewrite Eq.~\eqref{eq:obsexp} in terms of powers of $p$ instead of $Q$
(recall that this should not be interpreted as the explicit $p$ dependence of
the observable):
\beq
  X \equiv \Xref \sum_{n=0}^\infty b_n\, p^n \;, 
  \label{eq:X_expansion_full}
\eeq
which defines the dimensionful coefficients $b_n$.
The $b_n$ are trivially related to the $c_n$ from Eq.~\eqref{eq:obsexp} by
\beq
    c_n = \Lambda_b^n b_n \;.
  \label{eq:dn_cn_relate}
\eeq
We proceed based on two independent assumptions: 
(1) the details of the chiral EFT description of low-energy QCD (e.g., renormalization
scale and scheme) dictates a well-defined breakdown scale $\Lambda_b$, and 
(2) a well-formulated EFT implementation will lead to natural expansion coefficients
for observables. 
Although any given $b_n$ can be extracted from order-by-order calculations without any
reference to a breakdown scale or naturalness, our assumptions imply that the value 
of $b_n$ manifests the interplay between the underlying $\Lambda_b$ and the natural 
$c_n$ required by Eq.~\eqref{eq:dn_cn_relate}.
This relationship is represented graphically as a Bayesian
network in Fig.~\ref{fig:Lambdab_BN}.

We note that $b_0$ will not give any information on the expansion
parameter, because it will not modify the convergence pattern.
Because it is known that $b_1 = 0$ in chiral EFT,
it too is omitted from our analysis.

In general, we want to use order-by-order calculations of several observables
at several kinematic points to inform our estimate of $\Lambda_b$.
The full quantity of interest is therefore
\beq
    \pr(\Lambda_b| \bkvec(\kinparvec_1), \ldots, \bkvec(\kinparvec_m)) \;,
\eeq
where the given information is $m$ sets of coefficients
$\bkvec(\kinparvec_i) \equiv (b_2(\kinparvec_i),\cdots,b_k(\kinparvec_i))$ labeled by $\kinparvec_i = ({\Elab}_{\,i}, \theta_i, X_i)$: the set of kinematic parameters and observable $X_i$ from which the $\bkvec$ were calculated.
Using Bayes theorem, we can express the posterior for $\Lambda_b$ as
\beq
  \begin{split}
     \pr(\Lambda_b|\bkvec(&\kinparvec_1), \ldots, \bkvec(\kinparvec_m))  \\&  =
       \frac{ \pr(\bkvec(\kinparvec_1), \ldots, \bkvec(\kinparvec_m)|\Lambda_b) \pr(\Lambda_b) } 
            { \pr(\bkvec(\kinparvec_1), \ldots, \bkvec(\kinparvec_m)) } 
  \;.
  \end{split}
  \label{eq:lambda_b_bayes_chiEFT_full}
\eeq
We have assumed statistical independence of coefficients at different orders,
but it is possible for $b_n(\kinparvec_i)$ to be correlated with $b_n(\kinparvec_j)$.
The coefficients can be correlated for multiple reasons: for a given observable, the kinematic parameters may be close to one another; two distinct observables could themselves be correlated; or a combination of both.
Assuming statistical independence in $\kinparvec_i$, we can factor the likelihood in Eq.~\eqref{eq:lambda_b_bayes_chiEFT_full} into
\beq
	\pr(\bkvec(\kinparvec_1), \ldots, \bkvec(\kinparvec_m)|\Lambda_b) =
	\prod_{i=1}^m \pr(\bkvec(\kinparvec_i)|\Lambda_b) \;.
\eeq
Therefore, the posterior is given by
\beq
  \begin{split}
   \pr(\Lambda_b|\bkvec(&\kinparvec_1), \ldots, \bkvec(\kinparvec_m))    \\ &   = 
     \frac{ \pr(\Lambda_b) \prod_{i=1}^m \pr(\bkvec(\kinparvec_i)|\Lambda_b) } 
       { \pr(\bkvec(\kinparvec_1), \ldots, \bkvec(\kinparvec_m)) } 
  \;.
  \end{split}
  \label{eq:lambda_b_bayes_chiEFT}
\eeq
The denominator of Eq.~\eqref{eq:lambda_b_bayes_chiEFT} is simply a normalization constant and the prior $\pr(\Lambda_b)$ can be chosen later on, leaving only
$\pr(\bkvec(\kinparvec_i)|\Lambda_b)$ to evaluate.
For simplicity, we will refer to this as
$\pr(\bkvec|\Lambda_b)$, noting that the likelihoods for all $\bkvec(\kinparvec_i)$
simply need to be multiplied together to get the final posterior pdf 
in Eq.~\eqref{eq:lambda_b_bayes_chiEFT}.

To express $\pr(\bkvec|\Lambda_b)$ in terms of the prior assumptions
of naturalness, we first use marginalization~\cite{Sivia:2006} to introduce as auxiliary parameters
the dimensionless coefficients $\ckvec$ [see Eq.~\eqref{eq:ckvec}]:
\begin{align}
  \pr(\bkvec|\Lambda_b) = \int \dd{\ckvec}
  \pr(\bkvec|\ckvec,\Lambda_b) \pr(\ckvec|\Lambda_b) \;.
\end{align}
Next, to express the prior pdf for the coefficients $c_n$, we integrate in
the naturalness parameter $\cbar$:
\begin{align}
  \pr (\bkvec|\Lambda_b) = \int \dd{\cbar} \dd{\ckvec}
   \pr(\bkvec|\ckvec,\Lambda_b) \pr(\ckvec|\cbar,\Lambda_b) \pr(\cbar|\Lambda_b) \;.
  \label{eq:lambda_b_cn_cbar_marg}
\end{align}

To simplify Eq.~\eqref{eq:lambda_b_cn_cbar_marg} we use
independence as reflected in the causal relationship outlined in Fig.~\ref{fig:Lambdab_BN}. Neither the $c_n$s nor $\cbar$ depend on $\Lambda_b$ if they are not mediated
by $b_n$. We adopt a prior of independence between the $c_n$s as before. Thus 
\beq
  \pr(\ckvec|\cbar,\Lambda_b)
  =  \prod_{n=2}^k  \pr(c_n|\cbar) 
   \;.
\eeq 
The $b_n$s also only
depend on their corresponding $c_n$ and $\Lambda_b$, and are independent of
one another. This means that 
\begin{align}
  \pr(\bkvec|\ckvec,\Lambda_b)
    & =  \prod_{n=2}^k \pr(b_n|c_n,\Lambda_b) 
  \;.
\end{align}
Therefore,
Eq.~\eqref{eq:lambda_b_cn_cbar_marg} can be written as
\begin{align}
  \pr & (\bkvec|\Lambda_b) = \int \dd{\cbar} \pr(\cbar) \prod_{n=2}^k \int \dd{c_n}
  \pr(b_n|c_n,\Lambda_b) \pr(c_n|\cbar) \;.
\end{align}
The pdf for $b_n$ contingent on $c_n$ and $\Lambda_b$ is simply
\beq
  \pr(b_n|c_n,\Lambda_b) = \delta{\left(b_n - \frac{c_n}{\Lambda_b^n}\right)} \;,
\eeq
which enables us to perform the $c_n$ integrations directly.
Thus,
\begin{align} \label{eq:Lambda_b_cbar_integral}
  \pr & (\bkvec|\Lambda_b) = \Lambda_b^{k(k+1)/2 - 1} \int \dd{\cbar} \pr(\cbar) \prod_{n=2}^k \pr(c_n|\cbar)\;,
\end{align}
where we have used $\prod_{n=2}^k \Lambda_b^n = \Lambda_b^{k(k+1)/2 - 1}$ and have set $c_n = b_n \Lambda_b^n$ from now on.

To evaluate Eq.~\eqref{eq:Lambda_b_cbar_integral}, we must make choices for the priors, such as those from Table~\ref{tab:priors}.
Analytic expressions of Eq.~\eqref{eq:Lambda_b_cbar_integral} can be found for sets $\Aeps$ and 
$\Ceps$, which we will consider here.
It is reasonable to assume no prior knowledge of the scale of $\cbar$, i.e.\ allow $\cbarmin \to 0$ and $\cbarmax \to \infty$, because the scale can vary wildly with a changing $\Lambda_b$.
For set $\Ceps$,
\begin{align} \label{eq:Lambda_b_cbar_integral_Ceps}
  \pr & (\bkvec|\Lambda_b) \propto {\left( \frac{\Lambda_b^{k+2}}{\ckvecsq} \right)}^{(k-1)/2}  \;.
\end{align}
The result for set $\Aeps$ is similar to Eq.~\eqref{eq:Lambda_b_cbar_integral_Ceps},
with the replacement $\ckvecsq \to \cbark^2$, where
$\cbark = \max\{|c_i|: c_i \in \ckvec \}$.
These likelihoods are maximized for values of $\Lambda_b$ 
where the individual $c_n$s are about the same size.

The final step in specifying the posterior is to make a choice of $\pr(\Lambda_b)$.
Here we employ a non-informative log-uniform prior as we did for $\cbar$:
\begin{align}
  \pr(\Lambda_b) = \frac{1}{\ln(\Lambda_>/\Lambda_<)} \frac{1}{\Lambda_b}
    \, \theta(\Lambda_b-\Lambda_<)
    \, \theta(\Lambda_>-\Lambda_b) \;,
  \label{eq:lambda_b_prior}
\end{align}
which assumes we know only limits on the scale of $\Lambda_b$. Then for set $\Ceps$,
\begin{align} \label{eq:pdf_lambda_b_setc_uninformative}
  \pr(\Lambda_b|\bkvec(&\kinparvec_1), \ldots, \bkvec(\kinparvec_m)) 
    \propto \frac{1}{\Lambda_b} \prod_{i=1}^m 
       {\left( \frac{\Lambda_b^{k+2}}{\ckvecsq(\kinparvec_i)} \right)}^{(k-1)/2}
       ,
\end{align}
where the $\theta$ functions on $\Lambda_b$ are implicit. 
A more probable region in $\Lambda_b$ is singled out in Eq.~\eqref{eq:pdf_lambda_b_setc_uninformative} by the interplay of
the $\pr(\bkvec(\kinparvec_i)|\Lambda_b)$ factors, 
which individually favor $\Lambda_b$s
that make the order-by-order $c_n$s
for each $\kinparvec_i$ about the same size.

The lower limit $\Lambda_<$
in the prior of Eq.~\eqref{eq:lambda_b_prior} requires comment:
if $\Lambda_<$ is set less than the momentum scale $p_i$ corresponding to the lab energy
where $\bkvec(\kinparvec_i)$ is calculated, the expansion parameter $Q = p_i/\Lambda_b$ may be
greater than one. If we have an EFT that converges according to our statistical model, $Q>1$ for
the relevant kinematic points should be excluded by Eq.~\eqref{eq:lambda_b_bayes_chiEFT}. 
If it instead favors values of $\Lambda_b$ for which $Q>1$, 
this would signal an inconsistency between the truncation error model and the EFT as implemented.

As already noted, we assume that because of their separations in energy or angle, 
the chosen sets of kinematic parameters can be
treated as independent from one another and their probability densities multiplied.
We make a similar assumption for the observables themselves, i.e., the set labeled $X_{pqik}$ includes $\bkvec$ sets for each of the spin observables 
$A_y$, $A$, $D$, $A_{xx}$ and $A_{yy}$.
The assumption of independence, particularly for observables at the same
energy or angle, may be questioned.
The exploration of methods to combine data from all kinematic parameters, such as through 
GPs~\cite{rasmussen2006gaussian,Gelman03,Jones1998}, and assessments of observable independence, are currently in progress.

\begin{figure}[tb]
  \centering
  \includegraphics[width=0.95\columnwidth]{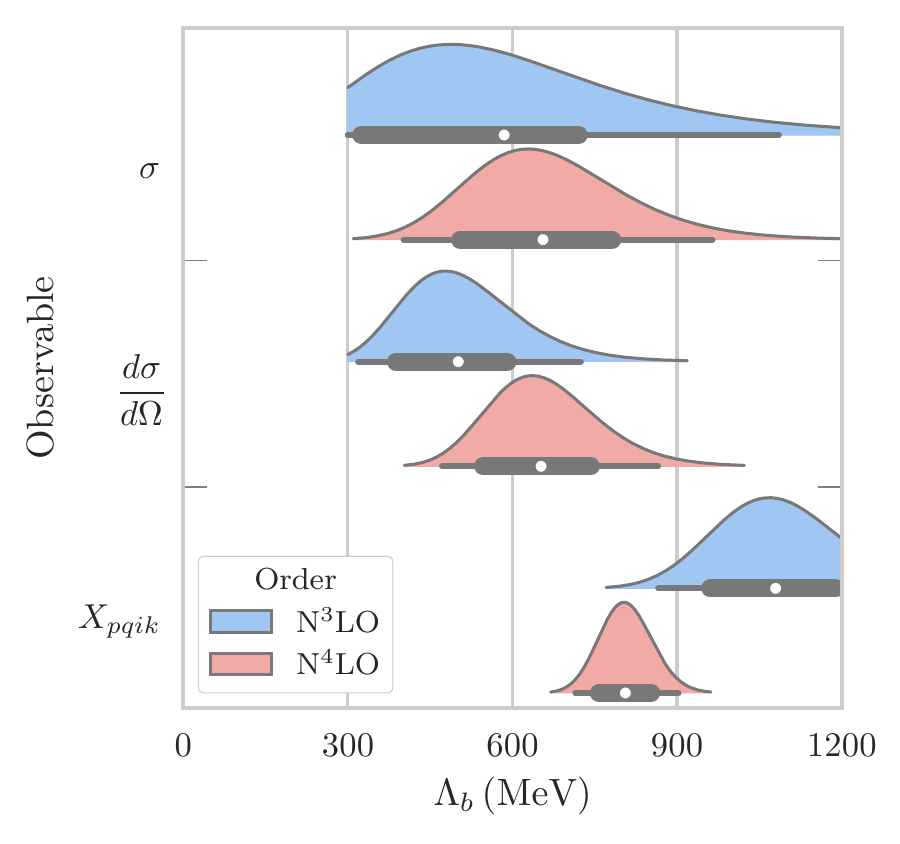}
  \caption{Posterior pdfs $\pr(\Lambda_b|\protect\bkvec)$ for NN observables
  using the $R=0.9$\,fm potential, at the kinematic points
  $\Elab = 96, 143, 200, 300$\,MeV and
  $\theta = 60^\circ, 120^\circ$.
  We use prior set $\Ceps$,
  and assume that $\Lambda_< = 300$\,MeV and $\Lambda_> = 1500$\,MeV\@.
  Thick and thin horizontal lines represent $68\%$ and $95\%$ DoBs, respectively, while the white dot signifies the median.
  $X_{pqik}$ stands for the combination of the 5 considered spin observables $A_y$, $A$, $D$, $A_{xx}$ and $A_{yy}$ treated as independent from one another.
  For $\sigma$ and $\protect\dsigma*$, $\Xref = X_0$, while $\Xref = 1$ otherwise.
  For aesthetic purposes, each plot is scaled to the same height.
  }
  \label{fig:LB_violin_pdfs_cbar_0p001_to_1000_SetC_R0p9_E_96_143_200_300}
\end{figure}

\begin{figure}[tb]
  \centering
  \includegraphics[width=0.95\columnwidth]{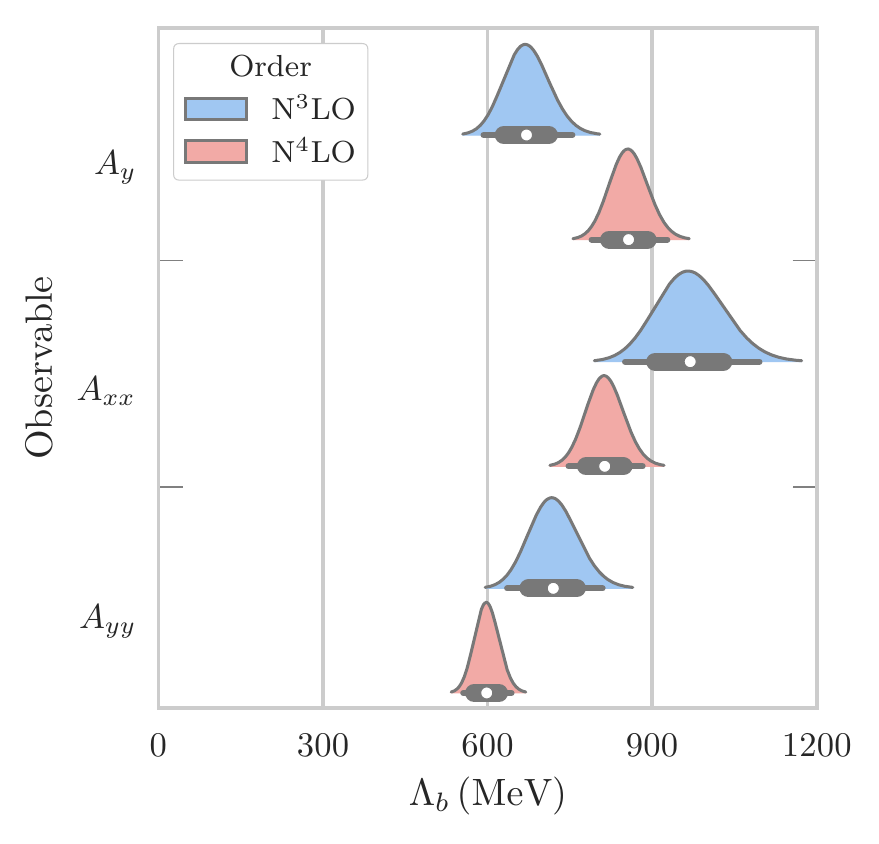}
  \caption{Posterior pdfs for $\Lambda_b$ as in 
  Fig.~\ref{fig:LB_violin_pdfs_cbar_0p001_to_1000_SetC_R0p9_E_96_143_200_300}, except the observables $A_y$, $A_{xx}$, and $A_{yy}$ are considered individually at $\Elab = 20,40,\ldots,340$\,MeV and $\theta = 40^\circ,60^\circ,\ldots,140^\circ$.
  }
  \label{fig:LB_violin_pdfs_cbar_0p001_to_1000_SetC_R0p9_E_20-340_spin_obs}
\end{figure}

\begin{figure}[tb]
  \centering
  \includegraphics[width=0.95\columnwidth]{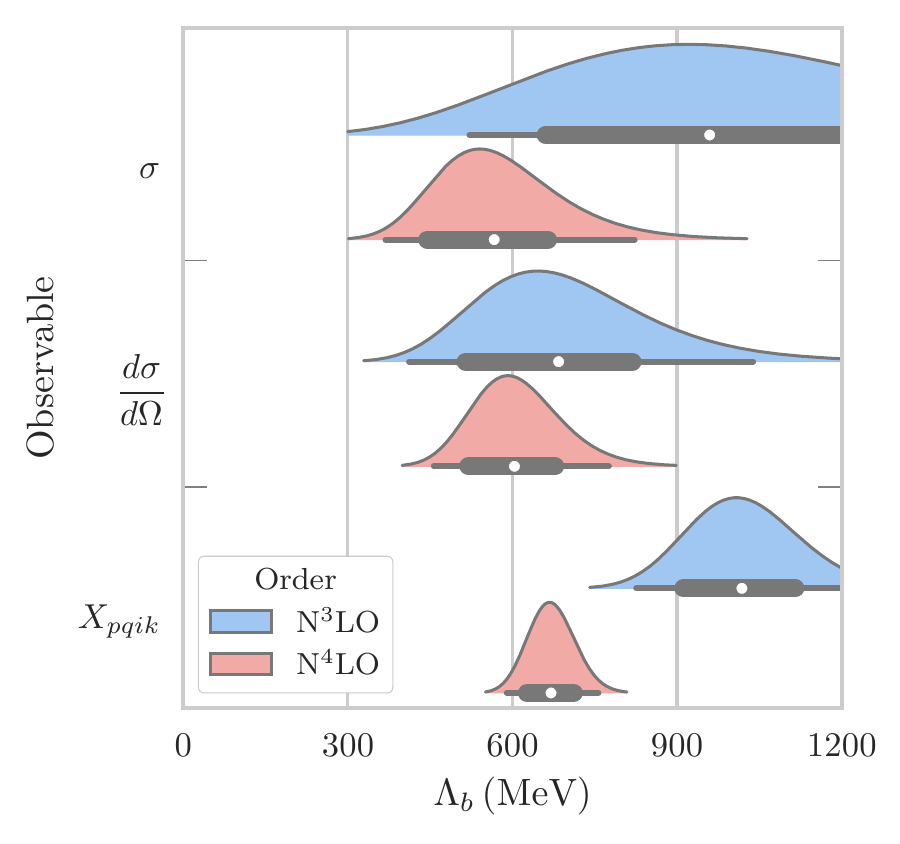}
  \caption{
  Posterior pdfs for $\Lambda_b$ as in 
  Fig.~\ref{fig:LB_violin_pdfs_cbar_0p001_to_1000_SetC_R0p9_E_96_143_200_300}, except with 
  $\Elab = 50, 96, 143, 200$\,MeV.
  }
  \label{fig:LB_violin_pdfs_cbar_0p001_to_1000_SetC_R0p9_E_50_96_143_200}
\end{figure}

\begin{figure}[bt!]
  \centering
  \includegraphics[width=0.95\columnwidth]{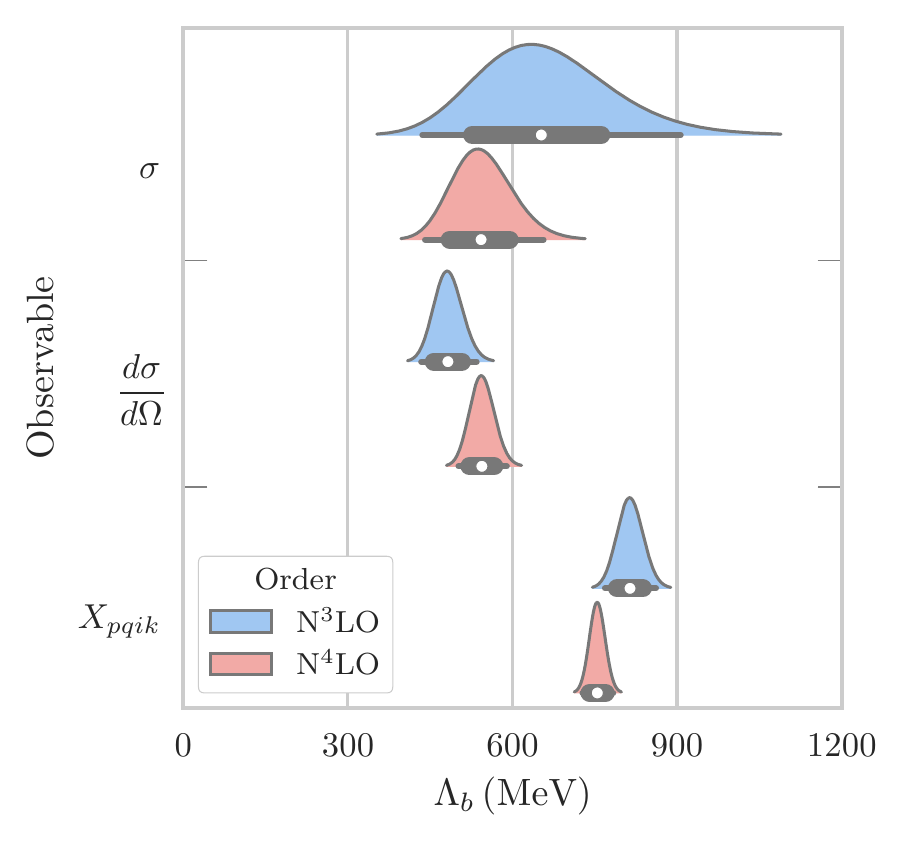}
  \caption{Posterior pdfs for $\Lambda_b$ as in 
  Fig.~\ref{fig:LB_violin_pdfs_cbar_0p001_to_1000_SetC_R0p9_E_96_143_200_300}, except with 
  $\Elab = 20,40,\ldots,340$\,MeV and $\theta=40^\circ,60^\circ,\ldots,140^\circ$.
  }
  \label{fig:LB_violin_pdfs_cbar_0p001_to_1000_SetC_R0p9_E_20-340}
\end{figure}

\begin{figure}[tbh!]
  \centering
  \includegraphics[width=0.95\columnwidth]{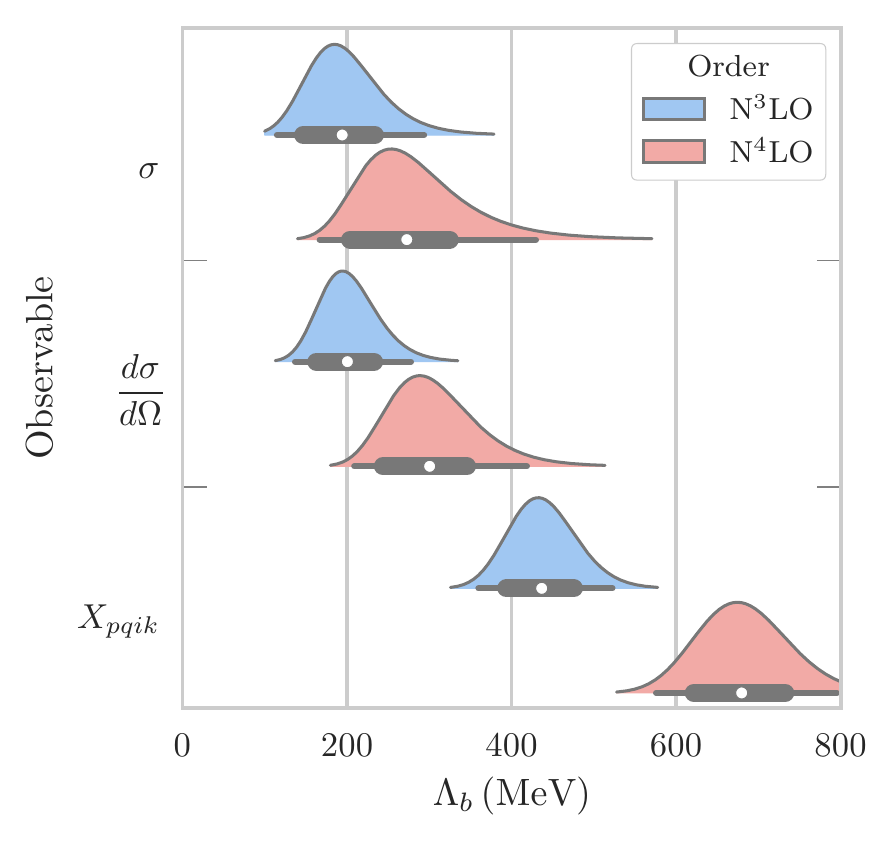}
  \caption{Posterior pdfs for $\Lambda_b$ as in 
  Fig.~\ref{fig:LB_violin_pdfs_cbar_0p001_to_1000_SetC_R0p9_E_96_143_200_300}, except for
  $R=1.2$\,fm
  using $\Lambda_< = 100$\,MeV and $\Lambda_> = 900$\,MeV.
  }
  \label{fig:LB_violin_pdfs_cbar_0p001_to_1000_SetC_R1p2_E_96_143_200_300}
\end{figure}

Given the above assumptions, we have applied Eq.~\eqref{eq:pdf_lambda_b_setc_uninformative} to various potentials, observable sets,
and kinematic parameters; the resulting pdfs and DoB intervals, using the HPD prescription (see Sec.~\ref{sec:formulas}), are presented in 
Figs.~\ref{fig:LB_violin_pdfs_cbar_0p001_to_1000_SetC_R0p9_E_96_143_200_300}--\ref{fig:LB_violin_pdfs_cbar_0p001_to_1000_SetC_R1p2_E_96_143_200_300}.
In contrast to central credibility intervals, the HPD intervals ensure that massive extremes, such as the \NNNLO\ posterior for $\sigma$ near its lower boundary in
Fig.~\ref{fig:LB_violin_pdfs_cbar_0p001_to_1000_SetC_R0p9_E_96_143_200_300},
are not necessarily excluded from our DoB intervals~\cite{kruschke2011doing}.
The posteriors of Fig.~\ref{fig:LB_violin_pdfs_cbar_0p001_to_1000_SetC_R0p9_E_96_143_200_300}
mirror the conclusions drawn from
Fig.~\ref{fig:consistency_lambda_dependence_Cp25-10_R0p9_E_96-300}---both
$\sigma$ and $\dsigma*$ predict $\Lambda_b \approx 600$\,MeV, while
the set of spin observables, taken together, prefers $\Lambda_b > 600$\,MeV\@.
The relatively small amount of data used from $\sigma$ and $\dsigma*$
do not allow for a very precise determination of $\Lambda_b$.

Figure~\ref{fig:LB_violin_pdfs_cbar_0p001_to_1000_SetC_R0p9_E_20-340_spin_obs} explores this 
result for spin observables by splitting out the posteriors for $A_y$, $A_{xx}$, and $A_{yy}$
separately.  These posteriors can be qualitatively 
predicted from the order-by-order consistency plots for these observables
given in Fig.~\ref{fig:consistency_individual_spin_obs_Cp25_10_theta-40-140_energy-20-340}.
In general, the strength of the $\Lambda_b$ posterior at \NNNNLO\ should be highly correlated 
with the pattern in the consistency plot at \NNNLO\ (e.g., do the points lie above or below
the 45~degree line, which imply that $\lambda < 1$ and $\lambda > 1$ are more probable
than $\lambda = 1$, respectively).
Similarly, the $\Lambda_b$ posterior at \NNNLO\ correlates
with the pattern in the consistency plot at \NNLO.
For $A_y$, this rule predicts that the \NNNNLO\ posterior should have its strength concentrated
well above 600\,MeV, and that the \NNNLO\ posterior should be located to its left.
For $A_{xx}$, the \NNNNLO\ posterior should also be well above 600\,MeV, but the \NNNLO\ posterior
should be to its right.  Finally, for $A_{yy}$, the consistency plots predict the \NNNNLO\
posterior will be concentrated near 600\,MeV (i.e., $\lambda = 1$), with the \NNNLO\ posterior
shifted somewhat to the right.
All of these expectations are realized in Fig.~\ref{fig:LB_violin_pdfs_cbar_0p001_to_1000_SetC_R0p9_E_20-340_spin_obs}.

One may wonder to what extent the $\Lambda_b$ posteriors are stable under 
different choices of kinematic parameter sets.
Figure~\ref{fig:LB_violin_pdfs_cbar_0p001_to_1000_SetC_R0p9_E_50_96_143_200} shows the 
posteriors as in Fig.~\ref{fig:LB_violin_pdfs_cbar_0p001_to_1000_SetC_R0p9_E_96_143_200_300},
but with a different (lower) range of energies.
Note that $\Elab = 50$\,MeV is near the crossover region $p \sim m_\pi$, 
where the interpretation of the expansion parameter is unclear.
However, while there are systematic shifts, both sets of posteriors for $\sigma$ and $\dsigma*$
are consistent with the EKM value of $\Lambda_b = 600$\,MeV, with the ensemble spin observable
posteriors favoring significantly higher values. 
In all cases there are wide posteriors at \NNNLO\ and more stability at \NNNNLO.
The only major shift in the median between these energy sets is for the \NNNLO\ cross section result. 
If we use larger sets of observables more closely spaced in both energy and angle 
(see Fig.~\ref{fig:LB_violin_pdfs_cbar_0p001_to_1000_SetC_R0p9_E_20-340}), 
neglecting the possible danger from correlations, the $\Lambda_b$ posteriors become more narrow and
more Gaussian, but are systematically in accord with 
Fig.~\ref{fig:LB_violin_pdfs_cbar_0p001_to_1000_SetC_R0p9_E_96_143_200_300}.

Figure~\ref{fig:LB_violin_pdfs_cbar_0p001_to_1000_SetC_R1p2_E_96_143_200_300} applies 
Eq.~\eqref{eq:pdf_lambda_b_setc_uninformative} to the $R=1.2$\,fm EKM potential, 
for which EKM assumed that $\Lambda_b = 400$\,MeV\@.
Most of the posteriors imply even smaller values of $\Lambda_b$, except for the \NNNNLO\ 
posterior for the spin observables, which is completely inconsistent.
Following the earlier discussion of
the lower limit on $\Lambda_b$ in the prior of Eq.~\eqref{eq:lambda_b_prior}, we
see that the posteriors are not only indicating much lower values of $\Lambda_b$,
but they also weight areas where the expansion parameter $Q>1$. The cross section
and differential cross section posteriors exhibit this behavior most, while
the spin observable posteriors are maximized in regions where $Q<1$.
 The weighting of $Q>1$ regions of the posterior is another indication
that the EFT convergence for this regulator is not
well-described by the statistical model.



\section{Summary and outlook}
\label{sec:summary}

In this work, we extend the analysis from Ref.~\cite{Furnstahl:2015rha} that applies
Bayesian statistics to the quantification of theoretical uncertainties in chiral EFT.
Our approach makes testable predictions of DoB error bands based upon
assumptions about the convergence pattern of EFT observables
and an implementation of naturalness.
In particular, we assume that the scaled observable coefficients $c_n$
defined in Eq.~\eqref{eq:obsexp}
are effectively random functions of natural size whose magnitude provides
an estimate of the error incurred by truncating the EFT expansion.

We apply this model to a set of
\npr\ scattering observables predicted by the semi-local chiral EFT potentials 
of EKM~\cite{Epelbaum:2014efa,Epelbaum:2014sza},
who also proposed a non-statistical protocol for uncertainty quantification.
The EKM error estimates in~\cite{Epelbaum:2014efa} correspond most closely to
the leading approximation of set~\Aeps\ (see Table~\ref{tab:priors}). 
In particular their error bands at
N$^k$LO are $k/(k+1)*100\%$ DoB intervals \cite{Furnstahl:2015rha}; 
i.e., they do not correspond to the same DoB at each order. Additionally, if the known
next-order result does not lie in that $k/(k+1)*100\%$ DoB interval, EKM extends the
interval to the next-order result. Therefore, at some orders it is possible to 
interpret the EKM intervals according to our truncation error model using set~\Aeps,
but not always.
To calculate consistent statistical DoBs at each order, we follow the statistical model
outlined in Fig.~\ref{fig:Bayesian_Network}, which assumes a
natural convergence pattern for the EFT.

\begin{figure*}[tbh!]
  \centering
  \includegraphics[width=.34\textwidth]{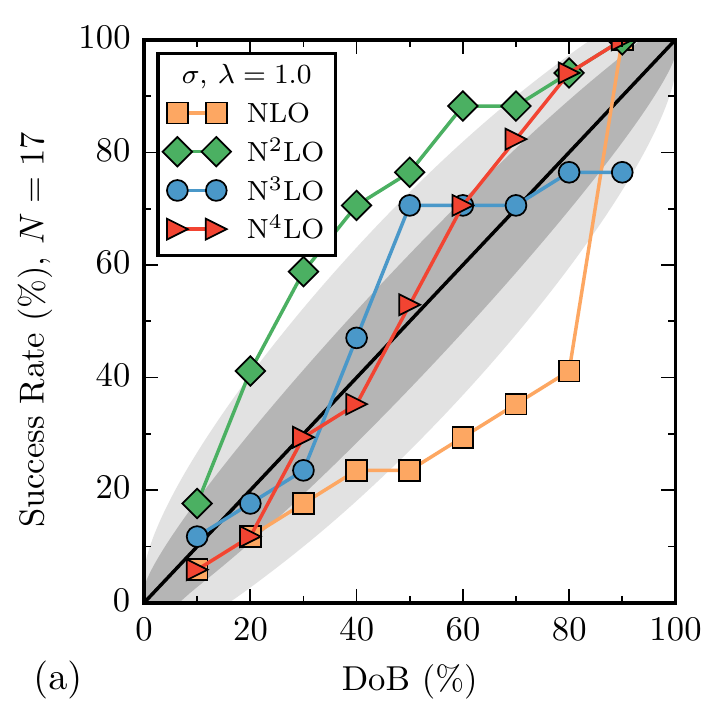}
  \hspace{-.15in}
  \includegraphics[width=.34\textwidth]{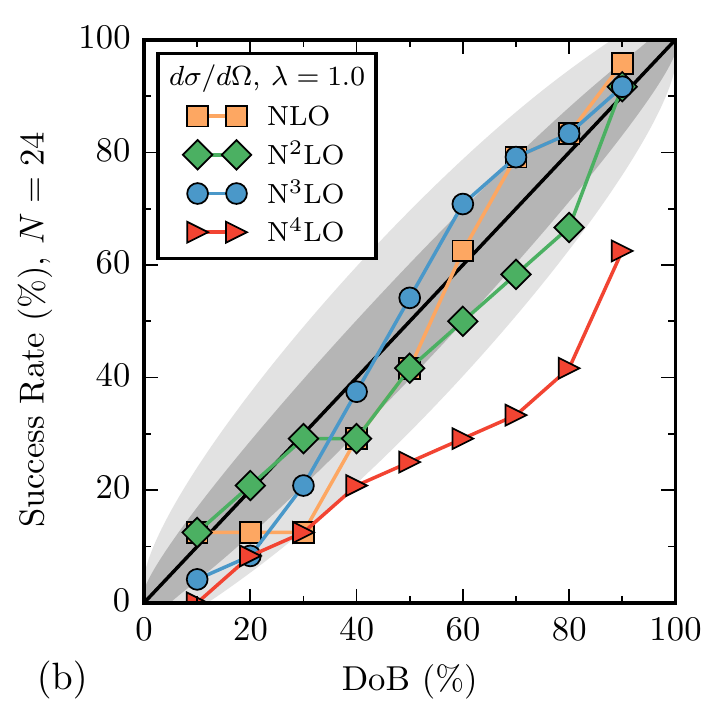}
  \hspace{-.15in}
  \includegraphics[width=.34\textwidth]{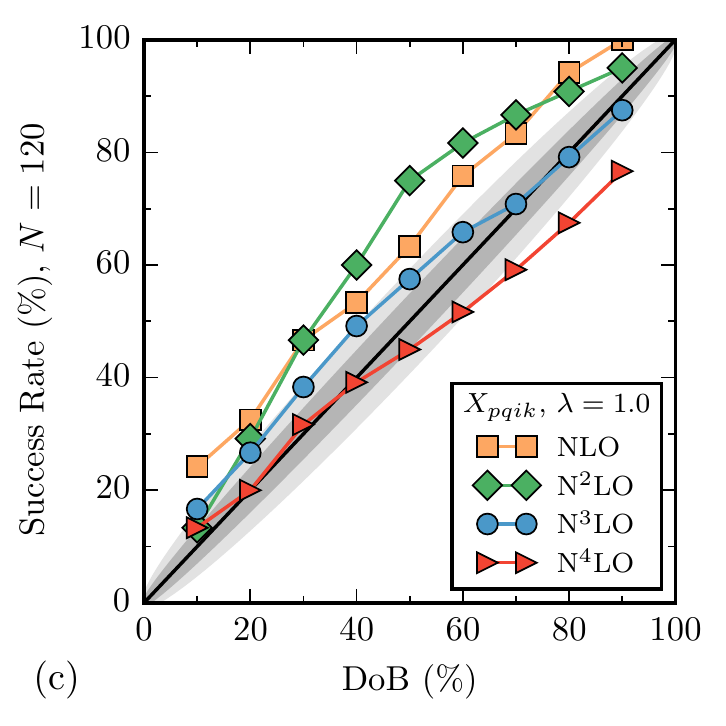}
  \caption{Consistency plots comparing the error band success rate when compared to NPWA data for $R=0.9$\,fm and prior set $\Cf{0.25}{10}$.
  (a) The total cross section is evaluated using $\Elab = 20,40,\ldots,340$\,MeV, while (b) the differential cross section and (c) set of selected spin observables use $\Elab = 96, 143, 200, 300$\,MeV and $\theta=40^\circ,60^\circ,\ldots,140^\circ$.}
  \label{fig:cons_sigma_npwa_Cp25-10_orders}
\end{figure*}

We begin by arguing that appropriate physical choices of scale in Eq.~\eqref{eq:obsexp} are
$\Xref \approx X_0$ for $\sigma$ and $\dsigma*$,
while $\Xref \approx 1$ for any spin observable $X_{pqik}$.
We then validate \emph{a posteriori} for these choices of $\Xref$ the natural distribution 
of observable coefficients for the total cross section, differential cross section, 
and a selection of spin observables for the $R=0.9$\,fm potential.
The $R = 0.8$\,fm potential is also consistent with our statistical model. 
In contrast, the convergence patterns of the $R = 1.0$\,fm, 1.1\,fm and 1.2\,fm potentials
become increasingly unsystematic, and hence they become less well described
by this model.
The expansion parameter for these softer potentials is dominated by regulator artifacts, 
which EKM account for by adopting a smaller value for $\Lambda_b$. This reflects the
effective cutoff in momentum space instead of the intrinsic breakdown scale of chiral
EFT~\cite{Furnstahl:2014xsa}.
But the order-by-order convergence pattern is also modified as the long-range pion contributions
at odd orders in $Q$ are significantly shifted to short-range contributions at even orders.
This is manifested in the coefficients extracted in 
Fig.~\ref{fig:cross_sections_EKM_R1p1_R1p2_Lambdab_500_400_Xzero_coeffs}
and the failed validation of DoBs in Figs.~\ref{fig:consistency_lambda_dependence_Cp25-10_R1p1_E_20-340_t_60_120} and~\ref{fig:consistency_lambda_dependence_Cp25-10_R1p2_E_20-340_t_60_120}.

Our higher-order results are generally insensitive to the specific prior choice;
we compare sets A and C in
Figs.~\ref{fig:consistency_prior_dependence_Cp25} and~\ref{fig:consistency_prior_dependence_Aeps}, and provide more examples
in the Supplemental Material~\cite{supplmaterial}.
The results for potentials with good 
convergence patterns (we focus on $R=0.9\,$fm) 
can be summarized in terms of our 
progress on the questions raised in Sec.~\ref{sec:introduction}:
    \be
      \I The observable coefficients of the total cross section
        vary smoothly with energy, typically changing sign
        once over the energy range from 0 to 350\,MeV\@.  There is no apparent
        order-by-order pattern at any given energy, which supports our
        model of a random distribution characterized by a size $\cbar$.
        The unnatural size of the \NNNNLO\ $\sigma$ coefficient at $\Elab = 50$\,MeV 
        was noted in Ref.~\cite{Furnstahl:2015rha}.
        At low energies, higher-ordered coefficients become more sensitive to the value assigned to the
        expansion parameter $Q$ around the crossover region $p \sim m_\pi$, for which we do not have
        a model.
        Hence we cannot make strong statements about the coefficient spectrum and its 
        implication for naturalness in that energy region. We plan to test alternative
        schemes for $Q$ in the crossover region and to validate the presence of the
        crossover in the $c_n$s using a change-point analysis of the correlations modeled using
        a GP model~\cite{Gramacy2008,rasmussen2006gaussian,Jones1998,Gelman03}.

      \I The observable coefficients for both the differential cross section
        and the chosen spin observables vary smoothly in both $\Elab$ and
        $\theta$ with characteristic sizes between about 1 and 5 for $R = 0.9\,$fm, 
        which validates the assumption that naturalness propagates to these 
        observables for this potential.
        The functional dependences show no obvious patterns, supporting the model
        of effectively random functions.
        As with the cross section, the interpretation of the coefficients is ambiguous for low
        energies.

      \I Because each $c_n$ is a smooth function when plotted against 
      both $\Elab$ and $\theta$,
      the values of the observable coefficients at one value of the kinematic parameters are
      correlated within some neighborhood (a correlation length) of $\Elab$ and $\theta$.
      Through a rough estimation, we find that the correlation length in energy is about 80\,MeV,
       while the correlation length in $\theta$ is approximately $40^\circ$. These values were
       estimated visually here, but in the future we will determine them directly using a GP
       model for the $c_n$s~\cite{rasmussen2006gaussian,Jones1998,Gelman03}. This additional information will then
       be incorporated into our statistical model for truncation uncertainties.

      \I 
       The checks in Sec.~\ref{subsec:consistency} show that taking $\Lambda_b$ to be the same scale for
       both $\sigma$ and $\dsigma*$ is statistically consistent for the $R = 0.9\,$fm
       EKM potential.
       While the spin observable $A_{yy}$ is also consistent with that same scale,
       the ensemble of
       spin observables are more consistent with a somewhat larger value.
       The $R=0.8$\,fm EKM potential also shows promise, but extracting $\Lambda_b$
       becomes questionable for $R=1.0$\,fm and worse for larger $R$.
       The posteriors for $\Lambda_b$, shown in Sec.~\ref{sec:posterior} and the Supplemental Material~\cite{supplmaterial},
       lead to the same conclusions---probable ranges of $\Lambda_b$ consistent with
       the values proposed by EKM can be extracted for $R=0.8$ and $0.9$\,fm, 
       identifying probable ranges for $R=1.0\,$fm is questionable at best, and
       the other EKM potentials ($R = 1.1$ and $1.2\,$fm) are not well described by
       our statistical model.
       Conclusions about $\Lambda_b$ are not warranted for the poorly behaving potentials.
    \ee

An overall validation of our truncation error model as applied to the EKM potential with
$R = 0.9\,$fm is provided in Fig.~\ref{fig:cons_sigma_npwa_Cp25-10_orders}, which shows
order-by-order consistency plots for observables compared with the NPWA data.  These plots
are made by modifying step 4.\ of the procedure laid out in Sec.~\ref{subsec:consistency}
to count a success when the actual NPWA result is within the DoB interval at that order (as
opposed to comparison to the next-order calculation).
In general, the $\ppercent$ error bands work as advertised, predicting the
discrepancy with the NPWA data at the $\ppercent$ level to within expected fluctuations.

The general success of the model for chiral EFT truncation errors motivates 
additional applications, further development of the model (e.g., GP models),
and its full integration into parameter estimation of LECs.
We plan to apply our truncation error model to other chiral interactions
that are available order-by-order, 
such as the recent potential of Entem, Machleidt, and Nosyk
in~\cite{Entem:2017gor}. 
Our error model and Bayesian model checking diagnostics
can be applied not only for other chiral interactions
but also for other EFTs in general. They also apply generically
to any observable calculation that fulfills the
expansion model in Eq.~\eqref{eq:obsexp}, including calculations in perturbation theory.
A Bayesian-type Lepage plot analysis~\cite{Lepage:2001ym,Furnstahl:2014xsa,Wesolowski:2015fqa}
of the power-law behavior of residuals as a function of energy/momentum
will complement the statistically motivated model checks of Sec.~\ref{sec:model_checking}.
Now that we have a framework of testable assumptions
for treating $\Lambda_b$ as a random variable in the posterior pdf calculations of
Sec.~\ref{sec:model_checking}, that information can be used
to marginalize over $\Lambda_b$ as an auxiliary parameter in truncation error estimates rather
than using a fixed $\Lambda_b$ value.
Work in these areas is in progress.


\acknowledgments{We thank E.~Epelbaum, H.~Griesshammer, N.~Klco, and D.~Phillips for useful discussions.
Useful feedback on the manuscript was provided by N.~Klco and D.~Phillips.
This work was supported in part by the National Science Foundation
under Grant Nos.~PHY--1306250 and PHY--1614460 and the NUCLEI SciDAC Collaboration under
Department of Energy Grant DE-SC0008533.}


\appendix

\section{Derivations of \texorpdfstring{$\Delta_k$}{Deltak} Posteriors}
\label{app:derivations_of_posteriors}

Here we continue the analysis started in Sec.~\ref{sec:formulas} by giving
explicit forms of posteriors $\pr_h(\Delta|\ckvec)$ for various prior sets.
The most non-informative case of set A follows if we take $\cbarmin = \epsilon$,
$\cbarmax = 1/\epsilon$ and then take the limit $\epsilon \rightarrow 0$ at the end.
 We designate
this as set~$\Aeps$, and the results for this set were first worked out 
in~\cite{Cacciari:2011ze}. If we further adopt the first-omitted-term approximation, designated
$\Aepsone$, we have analytic expressions for Eq.~\eqref{eq:BayesEpsFull}, 
\begin{align}
      \pr_1 & (\Delta | \ckvec) = \left(\frac{n_c}{n_c+1} \right) \frac{1}{2\cbark\Q^{k+1}} \notag \\
      & ~~~\times
      \begin{cases}
        1 & \mbox{if }|\Delta| \leq \cbark\Q^{k+1} \;, \\
        \displaystyle\left(\frac{\cbark\Q^{k+1}}{|\Delta|}\right)^{n_c+1}
        & \mbox{if }|\Delta| > \cbark\Q^{k+1} \;,
      \end{cases}
\end{align}
and for $d_k^{(p)}$ from Eq.~\eqref{eq:integralford},
\begin{align}
    d_k^{(p)} = \cbark \Q^{k+1}\!
    \times \!
    \begin{cases}
    \displaystyle\frac{ n_c+1}{ n_c}\, p
    & \displaystyle\mbox{if } p\ \leq \frac{ n_c}{ n_c+1} \;,  \\
    \displaystyle\Bigl[ \frac{ 1}{ (n_c+1)(1-p)} \Bigr]^{\frac{1}{n_c}}\!\!\!
      & \displaystyle\mbox{if } p\ > \frac{ n_c}{n_c+1} \;,
    \end{cases}
    \label{eq:dcoeffs}
\end{align}
where
\beq
   \cbar_{(j)} \equiv \max(|c_2|,\cdots,|c_j|)
   \;,
   \label{eq:cbarkDef}
\eeq
and $n_c$ is the number of relevant known coefficients---here, $n_c = k-1$ 
since $c_0$ and $c_1$ do not contribute to our analysis, 
but equations are given in a general form for the reader.

Relaxing the first-omitted-term approximation can pose a numerical challenge,
since the integration volume grows quickly with increasing $h$.
Luckily, by following Ref.~\cite{Perez:2015ufa}, whose results we reproduce in 
Eqs.~\eqref{eq:higher_orders_pdf_hardwall}--\eqref{eq:defintion_of_qsq}, 
Eq.~\eqref{eq:higher_orders_pdf} can be reduced to
one integral for the hard-wall (hw) prior $\pr(c_n|\cbar)$ in sets
A and B, and exactly evaluated for the Gaussian (G) prior in set C\@.
For sets A and B,
\begin{align} \label{eq:higher_orders_pdf_hardwall}
  \!\pr_h^{\text{(hw)}}(\Delta|\cbar) = \frac{1}{2\pi} 
  \int_{-\infty}^\infty \dd{t} \cos(\Delta t) 
  \prod_{i=k+1}^{k+h} \frac{\sin(\cbar Q^i t)}{\cbar Q^i t}
  \;,
\end{align}
which, for $h\to\infty$, becomes the atomic function $h_a$~\cite{victor2009adaptive}.
For set C,
\begin{align} \label{eq:higher_orders_pdf_gaussian}
  \pr_h^{\text{(G)}}(\Delta|\cbar) = \frac{1}{\sqrt{2\pi} q\cbar} e^{-\Delta^2/2q^2\cbar^2}
  \;,
\end{align}
where
\begin{align} \label{eq:defintion_of_qsq}
  q^2 \equiv \sum_{n=k+1}^{k+h} Q^{2n} = Q^{2k+2} \frac{1 - Q^{2h}}{1 - Q^2}
  \;.
\end{align}
Equation~\eqref{eq:higher_orders_pdf_gaussian} is easily evaluated for all $h$;
we use $h=10$ unless otherwise specified, at which point the posteriors have
converged numerically (see Ref.~\cite{Furnstahl:2015rha}).

With Eq.~\eqref{eq:higher_orders_pdf_gaussian}, we can exactly evaluate
Eq.~\eqref{eq:BayesEpsFull} for set C in terms of special functions.
By inserting the priors and making the variable substitution $x = 1/\cbar$,
\begin{align} \label{eq:posterior_set_C_xint}
  \pr_h^{(C)}(\Delta|\ckvec) = 
   \frac
    {\ds\int_{1/\cbarmax}^{1/\cbarmin} \dd{x} x^{n_c} e^{-(\ckvecsq + \Delta^2/q^2) x^2/2}}
    {\ds\sqrt{2\pi} q \int_{1/\cbarmax}^{1/\cbarmin} \dd{x} x^{n_c-1} e^{-\ckvecsq x^2/2}}
    \;,
\end{align}
where, of course,
\begin{align} \label{eq:gamma_def}
  \ckvecsq = \sum_{n=2}^k c_n^2
  \;.
\end{align}
Equation~\eqref{eq:posterior_set_C_xint} can be evaluated in terms of 
the incomplete $\Gamma$ functions via
\begin{align} \label{eq:posterior_setC_analytic}
  & \pr_h^{(C)}  (\Delta|\ckvec) = \frac{1}{\sqrt{\pi q^2 \ckvecsq}} \left( \frac{\ckvecsq}{\ckvecsq + \Delta^2/q^2} \right)^{(1+n_c)/2} \notag \\
  & \quad \null \times \frac{\Gamma \!\left[ \frac{1+n_c}{2}, 
   \frac{1}{2\cbarmax^2}(\ckvecsq + \frac{\Delta^2}{q^2}) \right] 
   - \Gamma \!\left[ \frac{1+n_c}{2}, \frac{1}{2\cbarmin^2}(\ckvecsq 
   + \frac{\Delta^2}{q^2}) \right] }{\Gamma \!\left[ \frac{1}{2} n_c, \ckvecsq/2\cbarmax^2 \right] 
           - \Gamma \!\left[ \frac{1}{2} n_c,  \ckvecsq/2\cbarmin^2 \right]}
           \;,
\end{align}
using the definition
\begin{align}
  \Gamma(s, x) = \int_x^\infty \dd{t} t^{s-1} e^{-t} \;.
\end{align}
Of all sets in Table~\ref{tab:priors}, the posterior as given by set C, 
via Eq.~\eqref{eq:posterior_setC_analytic},
is the only one for which we have found a closed-form expression for all $h$
and ranges of $\cbar$.

For the non-informative set $\Ceps$, where $\cbarmin \to 0$ and $\cbarmax \to \infty$, 
Eq.~\eqref{eq:posterior_setC_analytic} simplifies to a $t$-distribution:
\begin{align} \label{eq:posterior_setC_analytic_eps}
  \pr_h^{(C)} (\Delta|\ckvec) & = \frac{1}{\sqrt{\pi q^2 \ckvecsq}} 
   \frac{\Gamma \!\left( \frac{1+n_c}{2} \right) }{\Gamma \!\left( \frac{1}{2} n_c \right)} \left( \frac{\ckvecsq}{\ckvecsq + \Delta^2/q^2} \right)^{(1+n_c)/2}
   .
\end{align}
Rather than integrating Eq.~\eqref{eq:posterior_setC_analytic_eps}, $d_k^{(p)}$ can be found by numerically solving the transcendental equation 
\begin{align} \label{eq:Ceps_d_transcendental}
  p = \frac{2d_k^{(p)}}{\sqrt{\pi q^2 \ckvecsq}} \frac{\Gamma(\frac{n_c+1}{2})}{\Gamma(\frac{n_c}{2})} \,_2 F_1{\left[ \frac{1}{2}, \frac{n_c+1}{2}; \frac{3}{2}; - \frac{(d_k^{(p)})^2}{q^2\ckvecsq} \right]}
  \;,
\end{align}
where $_2 F_1$ is the hypergeometric function.


\section{Details on NN observables}
\label{app:observables}

For convenience of the reader and because of the multitude of different conventions
in the literature, we have gathered in this appendix the formulas 
used here in the calculation of NN observables~\cite{doi:10.1146/annurev.ns.10.120160.001451,Bystricky:1976jr,lafrance:jpa-00208966,moravcsik:jpa-00210988,Stoks:1990us,Stoks:1993tb,Carlsson:2015vda}.

\subsection{Kinematics}

\begin{figure}
\includegraphics{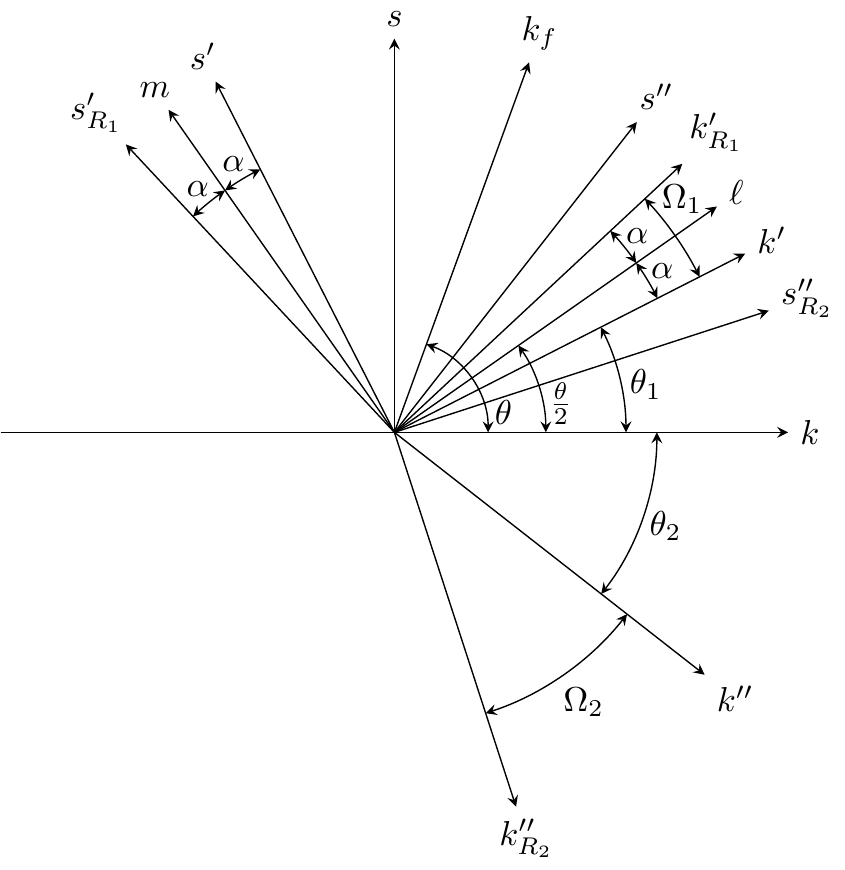}
\caption{The kinematics for nucleon-nucleon scattering~\cite{Bystricky:1976jr}.}
\label{fig:scattering_kinematics}
\end{figure}

In the context of NN scattering, one particle (the beam), with kinetic energy $\Elab$, 
is incident on a stationary particle (the target). 
For \npr\ scattering, the lab system (\ls) is often taken to be the rest frame of the initial proton.
In the center-of-momentum (\cms) system, each particle has a relative momentum of $\prel$.
It is convenient to relate these quantities for each NN experiment:
\begin{align}
    \text{Proton-proton:} \quad \prel^2 & = \frac{1}{2} M_p \Elab \;, \\
    \text{Neutron-neutron:} \quad \prel^2 & = \frac{1}{2} M_n \Elab \;, \\
    \text{Neutron-proton:} \quad \prel^2 & = \frac{\Elab M_p^2 
                       (\Elab + 2 M_n)}{(M_p+M_n)^2 + 2 M_p \Elab} \;,
\end{align}
where relativistic kinematics is used~\cite{Epelbaum:2014efa}.
Unless otherwise stated, $\theta$ is the \cms\ polar scattering angle while $\phi$ denotes the azimuthal scattering angle.
For our purposes, $\phi$ can be set to zero because all observables can be defined relative to the scattering plane.

The spin states of the initial and final states can be expressed in the uncoupled basis $|i\ra_{\text{spin}} = |m_1'm_2'\ra$ and $|f\ra_{\text{spin}} = |m_1m_2\ra$, respectively, where we have suppressed $s_1 = s_2 = 1/2$.
We can also use the coupled singlet-triplet basis, where $|i\ra_{\text{spin}} = |s'm'\ra$ and $|f\ra_{\text{spin}} = |sm\ra$.

\subsection{Observables}

Because nucleons have nonzero intrinsic spin, observables in general are dependent not only on kinematic variables ($\Elab, \theta, \phi$), but also on the relative orientation of the particles' spin.
A generic spin observable can be written as
\begin{align} \label{eq:masterobservable}
    \dsigma X_{pqik} = \frac{1}{4} \text{Tr}\, \sigma^{(1)}_p \sigma^{(2)}_q M \sigma^{(1)}_i \sigma^{(2)}_k M^\dagger \;,
\end{align}
where $\dsigma*$ is the (unpolarized) differential cross section, $M(k_f, k_i)$ is the spin-scattering matrix and $\sigma_v = \boldsymbol{\sigma} \cdot \boldvec{v}$.
The subscripts $p$, $q$, $i$, and $k$ refer to the polarization directions of the scattered, recoil, beam, and target particles, respectively.
If a final-state subscript is zero, its polarization is not analyzed.
If a initial-state subscript is zero, the corresponding particle was unpolarized.

When an observable is considered in the \cms\ system, the polarization of each particle is often decomposed in a common basis using the unit vectors $\ellvec$, $\mvec$, $\nvec$ defined as
\begin{align}
  \ellvec & =
    \begin{pmatrix}
      \ds \sin\frac{\theta}{2} \cos\phi, & \ds\sin\frac{\theta}{2} \sin\phi, & \ds\cos\frac{\theta}{2}
    \end{pmatrix} 
    ,
  \\
  \mvec & =
    \begin{pmatrix}
      \ds\cos\frac{\theta}{2} \cos\phi, & \ds\cos\frac{\theta}{2} \sin\phi, & \ds-\sin\frac{\theta}{2}
    \end{pmatrix}
    ,
  \\
  \nvec & =
    \begin{pmatrix}
      \ds-\sin\phi, & \ds\cos\phi, & 0
    \end{pmatrix}
    ,
\end{align}
and shown in Fig.~\ref{fig:scattering_kinematics}.
Here we consider \textit{pure experiments}, where the spin projections are solely along the basis vectors.
Hence, for a \cms\ observable, the subscripts $p$, $q$, $i$, and $k$ are some combination of $\ell$, $m$, $n$, and 0.

It is often convenient to express spin observables in the \ls, where the scattered and recoil particles deflect at angles $\theta_1$ and $\theta_2$, respectively.
Lab system observables often use three sets of bases to define spin observables, defined by the beam, scattered, and recoil particle directions.
The beam (scattered, recoil) frame aligns $\kvec$ ($\kpvec$, $\kppvec$) with the lab particle momentum and defines $\nvec$ ($= \nvec' = \nvec''$) to be normal to the scattering plane, which leaves $\svec$ ($\spvec$, $\sppvec$) in the scattering plane such that $\svec = \nvec \times \kvec$ ($\spvec = \nvec \times \kpvec$, $\sppvec = \nvec \times \kppvec$).
The initial-state subscripts $i$ and $k$ are then chosen to be $k$, $s$, $n$, or $0$.
Similarly, the scattered-state subscript $p$ is $k'$, $s'$, $n$ or $0$, and the recoil-state subscript $q$ is $k''$, $s''$, $n$ or $0$.

One added complication of calculating \ls\ observables involves accounting for the relativistic spin rotation angles
\begin{align}
  \Omega_1 & = \theta - 2\theta_1 = 2\alpha \;, \\ 
  \Omega_2 & = -\pi + \theta + 2\theta_2 = -\pi + 2\beta \;, 
\end{align}
which rotate the primed and double primed vectors about $\nvec$, respectively.
It is the rotated vectors, denoted with subscripts $R_1$ and $R_2$, that correspond to the \ls\ subscripts for the scattered and recoil particles.
In the nonrelativistic case, $\alpha = 0$ and $\beta = \pi/2$, which implies that
\begin{align}
  \ellvec & \sim \kpvec \sim \kprvec \sim \sppvec \sim \spprvec \;, \\
  \mvec & \sim \spvec \sim \sprvec \sim -\kppvec \sim -\kpprvec  \;.
\end{align}
This too is illustrated in Fig.~\ref{fig:scattering_kinematics}.
Whether an observable is defined in the \cms\ system or the \ls\ should be clear from the chosen subscripts.
All of the observables considered here use the \ls\ notation.

Notational inconsistencies abound in the literature.
While the subscripts of Eq.~\eqref{eq:masterobservable} completely determine a 
given spin observable, often the $X$ is changed to match historical usage.
Other times, the subscript notation is abandoned completely for a nondescript letter.
Table~\ref{tab:obs_lookup} attempts to reconcile some differences by matching a 
common name with Eq.~\eqref{eq:masterobservable} and other popular notations found 
in literature.

\begin{table}[tb]
\centering
\caption{Comparison of notations for selected NN scattering observables.}
\label{tab:obs_lookup}
\begin{ruledtabular}
\begin{tabular}{SlSlSl}
  Name                        & $X_{pqik}$ & Others                      \\
  \colrule
  Differential Cross Section  & $I_{0000}$ & $\sigma$, $\dsigma*$        \\
  Vector Analyzing Power      & $A_{00n0}$ & $A_y$, $P_b$                \\
  Polarization Transfer       & $D_{s'0k0}$ & $A$                        \\
                              & $D_{n0n0}$ & $D$                         \\
  Spin Correlation Parameters & $A_{00ss}$ & $A_{xx}$                    \\
                              & $A_{00nn}$ & $A_{yy}$                    \\
\end{tabular}
\end{ruledtabular}
\end{table}

The spin-scattering matrix is the part of the scattering $S$ matrix that is due to interactions.
In our convention, they are related via $M = (2\pi/ip)(S - 1)$.
To evaluate $M$, it is useful to write it in singlet-triplet space,
and then make a partial-wave expansion
\begin{align} \label{eq:Mwaves}
    M_{m'm}^{s's}(\theta, \phi) &=  \frac{\sqrt{4\pi}}{2ip} 
      \sum_{j, \ell, \ell'}^{\infty} (-1)^{s-s'} i^{\ell-\ell'} 
      \hat J^2 \hat L Y_{m-m'}^{\ell'}(\theta, \phi) \nonumber \\
    & \quad\null\times
    \begin{pmatrix}
        \ell' & s' & j \\
        m - m' & m' & -m
    \end{pmatrix}
    \begin{pmatrix}
        \ell & s & j \\
        0 & m & -m
    \end{pmatrix} 
     \nonumber \\
    & \quad\null\times
    \la \ell' s' | S^j - 1 | \ell s\ra
    \;,
\end{align}
where $\hat J \equiv \sqrt{2j + 1}$ and $\hat L \equiv \sqrt{2\ell + 1}$, 
the second line contains two Wigner $3j$ symbols, and $\mathbf{J} = \mathbf{L} + \mathbf{S}$ 
is the total angular momentum decomposed into orbital and intrinsic angular momentum.
The nuclear potential conserves the total angular momentum $j$, but generally mixes the states of $\ell$ and $s$.
Equation~\eqref{eq:Mwaves} becomes useful only if a small number of $j$ waves are needed to accurately determine $M_{m'm}^{s's}$.

When a partial wave state is uncoupled, such as when $j = 0$, then $S^j$ can be
parameterized by a real phase $\phase{s}{\ell}{j}$ such that $S^j \equiv e^{2i\phase{s}{\ell}{j}}$.
For $j > 0$, $S^j$ is 4-dimensional in angular momentum space and can be written compactly using the triplet submatrix $S_T^j$ and the singlet-triplet
submatrix $S_{ST}^j$ via
\begin{align}
    S^{j\neq 0} =
    \begin{pmatrix}
        S_T^j & 0 \\
        0 & S_{ST}^j
    \end{pmatrix}
    \;.
\end{align}
The triplet submatrix $S_T^j$ can be parameterized by
introducing another real parameter, the mixing angle $\epsmix{j}$.
Using the common notation that subscripts $+$ and $-$ refer to
$\ell = j+1$ and $j-1$, respectively,
then
\begin{align} \label{eq:Tcoupledmatrix}
    S_T^j =
    \begin{pmatrix}
        \cos 2\epsmix{j} e^{2i\phase{1}{-}{j}} & i\sin 2\epsmix{j} e^{i(\phase{1}{-}{j} + \phase{1}{+}{j})} \\
        i\sin 2\epsmix{j} e^{i(\phase{1}{-}{j} + \phase{1}{+}{j})} & \cos 2\epsmix{j} e^{2i\phase{1}{+}{j}}
    \end{pmatrix}
\end{align}
and similarly
\begin{align} \label{eq:STcoupledmatrix}
    S_{ST}^j =
    \begin{pmatrix}
        \cos 2\gammix{j} e^{2i\phase{0}{j}{j}} & i\sin 2\gammix{j} e^{i(\phase{0}{j}{j} 
            + \phase{1}{j}{j})} \\
        i\sin 2\gammix{j} e^{i(\phase{0}{j}{j} + \phase{1}{j}{j})} & \cos 2\gammix{j} 
             e^{2i\phase{1}{j}{j}}
    \end{pmatrix}
    \;.
\end{align}
In the present work, $\gammix{j} = 0$ for all $j$, leaving the singlet-triplet submatrix uncoupled and thus $s' = s$.

Equations~\eqref{eq:Tcoupledmatrix} and~\eqref{eq:STcoupledmatrix} employ the ``Stapp''- or ``bar''-phase shift parameterization.
Another parameterization, with phases and mixing angle denoted here by $\blattphase{s}{\ell}{j}$ and $\blattepsmix{j}$, was made by Blatt and Biedenharn~\cite{Blatt:1952zza}:
\begin{align}
  S_T^j =
    U^{-1}
    \begin{pmatrix}
      e^{2i\blattphase{1}{-}{j}} & 0 \\
      0 & e^{2i\blattphase{1}{+}{j}}
    \end{pmatrix}
    U \;,
\end{align}
where
\begin{align}
  U =
    \begin{pmatrix}
      \cos\blattepsmix{j} & \sin\blattepsmix{j} \\
      -\sin\blattepsmix{j} & \cos\blattepsmix{j}
    \end{pmatrix} \;.
\end{align}
The Blatt eigenphases are related to the Stapp phases via
\begin{align}
    \phase{s}{-}{j} + \phase{s}{+}{j} & = \blattphase{s}{-}{j} 
         + \blattphase{s}{+}{j} \label{eq:stapp-blatt1}  \;, \\
    \sin(\phase{s}{-}{j} - \phase{s}{+}{j}) & = 
      \frac{\tan2\epsmix{j}}{\tan2\blattepsmix{j}} \label{eq:stapp-blatt2} \;, \\
    \sin(\blattphase{s}{-}{j} - \blattphase{s}{+}{j}) & = 
      \frac{\sin2\epsmix{j}}{\sin2\blattepsmix{j}} \label{eq:stapp-blatt3}
      \;.
\end{align}

Given the partial-wave-projected potential $V_{\ell'\ell}^{sj}(p', p)$, it is convenient
and numerically accurate to solve the Lippmann-Schwinger (LS) equation with standing
wave boundary conditions,
\begin{align} \label{eq:LS}
   & R_{\ell'\ell}^{sj}(p', p; \prel^2) = V_{\ell'\ell}^{sj}(p', p) 
    \notag \\
    & \quad \null    + \sum_{\ell''} \frac{2}{\pi} \mathcal{P} 
    \int_0^\infty \dd{q} \frac{q^2 V_{\ell'\ell''}^{sj}(p',q) R_{\ell''\ell}^{sj}(q,p;\prel^2)}
      {\prel^2 - q^2}
     \;, 
\end{align}
for the partial-wave-projected $R$ matrix (known as the $K$-matrix in other contexts).
In the present work we use Gaussian quadrature 
to reduce the LS equation to a system of linear equations, 
from which $R_{\ell'\ell}^{sj}$ is extracted~\cite{Landau:1996}.
Having solved Eq.~\eqref{eq:LS}, the on-shell matrix $\Ron{s}{\ell'}{\ell}{j} = R_{\ell'\ell}^{sj}(\prel, \prel; \prel^2)$ then leads directly to the phases and mixing angle.
For the uncoupled channels,
\begin{align}
  \tan\phase{s}{\ell}{j} = -\prel \Ron{s}{\ell}{\ell}{j}
  \;.
\end{align}
For coupled channels, $\Ron{1}{\ell'}{\ell}{j}$ is 2-dimensional.
The Blatt-Biedenharn phases $\blattphase{s}{\ell}{j}$ and $\blattepsmix{j}$ are extractable via
\begin{align}
    \tan 2\blattepsmix{j} & = \frac{2\Ron{1}{-}{+}{j}}{\Ron{1}{-}{-}{j} 
      - \Ron{1}{+}{+}{j}} \;, \\
    \tan\blattphase{1}{-}{j} & = -\prel {\Biggl(\Ron{1}{-}{-}{j} 
       + \Ron{1}{+}{+}{j} + \frac{\Ron{1}{-}{-}{j} 
       - \Ron{1}{+}{+}{j}}{\cos 2\blattepsmix{j}}\Biggr)} \;, \\
    \tan\blattphase{1}{+}{j} & = -\prel {\Biggl(\Ron{1}{-}{-}{j} 
       + \Ron{1}{+}{+}{j} - \frac{\Ron{1}{-}{-}{j} 
       - \Ron{1}{+}{+}{j}}{\cos 2\blattepsmix{j}}\Biggr)}
       \;,
\end{align}
which can then be converted to the Stapp convention using 
Eqs.~\eqref{eq:stapp-blatt1}--\eqref{eq:stapp-blatt3}.


\clearpage
\bibliography{bayesian_refs} 

\end{document}